\documentclass[prc,amsfonts, amssymb, amsmath, nofootinbib,twocolumn,superscriptaddress]{revtex4-1}
\usepackage[utf8]{inputenc}
\usepackage{graphicx}
\usepackage{mathrsfs}
\usepackage[caption=false]{subfig}
\usepackage{color} 
\usepackage[dvipsnames]{xcolor}
\newcommand{\mtc}[1]{\mathcal{#1}}
\usepackage{ulem}

\newcommand{\clb}{\color{black}}
\newcommand{\clg}{\color{black}}

\title{Methods Paper}
\date{May 2024}

\begin{document}
\title{Applications of the Modified Hulthén--Kohn 
Method for Bound and Scattering States}
\date{\today} 

\begin{abstract}
We \color{black} adapt \color{black} the Hulthén--Kohn method 
suggested by Efros [Phys.~Rev.~C~{\bf99},~034620~(2019)] for calculating various observables in  the  continuum and discrete spectrum using 
two-body interactions in single- and coupled-channel systems.  We explore the convergence of phase shifts and wave functions as well as the location of~$S$-matrix poles  which enables obtaining both resonance and bound state parameters.  We find that \color{black} employing a harmonic oscillator basis, together with an interaction smoothing scheme introduced by Gyarmati \textit{et al.} [Nucl. Phys.~A {\bf 326}, 119 (1979)], and  \color{black} adopting \color{black} approximate bound-state solutions \color{black} for the short-range components of basis wave functions \color{black} lead \color{black} to good convergence \color{black} even with restricted oscillator quanta accessible for modern no-core shell model (NCSM) codes.  The adapted Efros method will facilitate \textit{ab initio} many-body nuclear structure applications. \color{black}
\end{abstract}
\author{Mamoon A. Sharaf}
 \email{Corresponding author: msharaf@iastate.edu}
\affiliation{Institute of Modern Physics, Chinese Academy of Sciences, Lanzhou 730000, China}
\affiliation{Department of Physics and Astronomy, Iowa State University, Ames, Iowa 50011, USA}
\author{Andrey M. Shirokov}
\affiliation{Skobeltsyn Institute of Nuclear Physics, Lomonosov Moscow State University, Moscow 119991, Russia}
\author{Weijie Du}
 \email{Corresponding author: duweigy@gmail.com}
\affiliation{Institute of Modern Physics, Chinese Academy of Sciences, Lanzhou 730000, China}
\affiliation{Department of Physics and Astronomy, Iowa State University, Ames, Iowa 50011, USA}
\author{James P. Vary}
\affiliation{Department of Physics and Astronomy, Iowa State University, Ames, Iowa 50011, USA}

\maketitle

\section{Introduction}
\label{I}

Over the past 25 years, various {\it ab initio} (from first principles) methods have been developed and extensively applied with the aim of describing the structure of atomic nuclei, including the Green's function Monte Carlo method \cite{GFMC0, GFMC1, GFMC2, GFMC3,GFMC4}, the coupled cluster method \cite{CoesterOne, CoesterTwo,Kummel, Wloh, Hagen,Hagen2022,ZHSun2023}, and the no-core shell model (NCSM)~\cite{NCSMPRL, NCSMProg}.  In principle, one can choose any basis within the Hilbert space to construct the eigenfunctions.  For practical reasons, the harmonic oscillator basis is conventionally used in nuclear structure theory as it possesses several convenient features for applications to realistic many-body problems.  The utility is such that the many-body problem is reduced within the M-scheme (in which all basis states have a fixed value of total angular momentum projection but not a fixed total angular momentum) to a sparse matrix eigenvalue problem that is solvable with leadership class supercomputers \cite{MFDnMaris2010,Euro-Par, Aktulga, NCSM-NTSE2014, Shao}.

While there are many advantages in applications of those methods to nuclear structure problems, 
there are 
major technical difficulties in applying them to describe scattering states and reactions. 
While bound states can be calculated by {\it ab initio} approaches with a reasonable degree of accuracy for problems up to a  nucleon number of approximately~$A=20$~and for heavier nuclei around closed shells, successes in describing {\it ab initio} scattering in systems with~$A > 4$~are more limited \cite{CWJohnson}.  This is a pressing issue, since developing experimental techniques  will benefit from  accurate calculations of observables in nuclear reactions such as scattering cross sections, resonance energies~$E_{r}$~and widths~$\Gamma$, the boundaries of stability as either the proton or neutron number is increased~\cite{VSoma}, and  properties of exotic systems like the tetraneutron and
trineutron~\mbox{\cite{AMShirokov2016,4nAIP,JLi2019,3narXiv}}.

In light of these needs, several {\it ab initio} methods have been applied to elastic scattering problems.  
For example, the Green's function Monte Carlo has been used to obtain features of the $J^{\pi}=\frac{3}{2}^{-}$ and $ \frac{1}{2}^{-}$ low-lying resonant states in~$n \alpha$ scattering \cite{Nollett}. \color{black}  The quantum Monte Carlo method has been used to study \color{black}  $n- {^3\textrm{H}}$ scattering \cite{FloresOne} and $n\alpha$ scattering \cite{FloresTwo}. \color{black}   Properties of nuclear resonances and reactions with light nuclei can also be studied within the 
no-core Gamow shell model \cite{Papadimitriou}.  In particular, this approach was successfully applied recently to the study of the tetraneutron and trineutron~\cite{JLi2019}.  Another successful approach is the resonating group method (RGM) in combination with the NCSM \cite{QuagOne, QuagTwo, Navratil2010}, which has been used to study nuclear scattering in systems as large as $A=17$~\cite{Navratil2010}, and its more extended version, the NCSM with continuum~\cite{BaroniTwo,BaroniOne,Navr-scr,NCSMCHebborn2022,NCSMCGysbers2023}.

An alternative approach to the studies of  the continuum spectrum is the harmonic oscillator representation of scattering equations (HORSE)
\cite{Zaitsev1998, BangHORSE}.  The HORSE method allows one to calculate the continuum wave function and the scattering phase shifts~$\delta_{l}(E)$.  It has already been successfully applied to the calculations of~$np$~scattering and the~$M1$ transition cross section of the reaction {$p(n,\gamma)d$}~\cite{DuWeijieHORSE}~as well as to calculations of continuum nuclear spectra within various cluster models
with phenomenological interactions in Refs.~\mbox{\cite{monopoleC12,Lurie1,Lurie2,Lurie3,Lurie4}} and within the RGM~\cite{6He6He,VVasil,Lashko}.  For many-body {\it ab initio} applications, 
however, the HORSE method is impractical due to the prohibitive computational cost.  To get around this, a simplified HORSE version, the single-state HORSE (SS-HORSE) method has been developed~\mbox{\cite{AMShirokovHORSE2016, BlokhintsevSSHorseC, AMShirokovHORSE2018}}, 
which makes it possible to extend the NCSM to calculations of phase shifts.   Moreover, the SS-HORSE method can also provide 
resonance energies and widths via~$S$-matrix pole location. The SS-HORSE method in combination with the NCSM was used to describe resonances in various
light nuclei, in particular, in $^{5}$He, $^{5}$Li~\cite{AMShirokovHORSE2016, AMShirokovHORSE2018}, $^{7}$He~\cite{PPN,Igor2022}, $^{9}$Li~\cite{9Li}, to propose the first description of the tetraneutron resonance~\cite{AMShirokov2016,4nAIP}, and to predict resonant states in the trineutron~\cite{3narXiv}.  The price paid for the \mbox{SS-HORSE} simplification is the loss of information on the continuum wave function.  Without the wave functions, one is not able to perform calculations of important physical observables such as, e.g.,  electromagnetic dissociation cross sections.

To address the shortcomings of the HORSE and SS-HORSE approaches in a computationally feasible manner, the Hulthén--Kohn method originally introduced in Refs.~\cite{Hulthen1948, Kohn1948, Moses1953,  Moses} was modified by Efros~\cite{VDEfros} in a way that makes it convenient for extending the NCSM to the continuum spectrum.  \color{black}  Moreover, unlike the \mbox{SS-HORSE} method, the Efros method easily extends to multichannel problems.  To date, this approach has been tested using simple model problems only in Ref.~\cite{VDEfros}.  Recently, Kohn's variational method has also been used to solve the problem of scattering of nucleons by $^{3}$H or  $^{3}$He within the hyperspherical harmonics approach~\cite{Viviani2020} and has also been successfully utilized to solve fission reactions \cite{HaginoBertsch2024}. 

\color{black} In this work, we adapt the \color{black} Efros method~\cite{VDEfros} for solving scattering problems.~\color{black} We introduce additional features that improve the method \color{black} with the goal of utilizing it in future many-body applications.~\color{black} The adapted Efros method incorporates \color{black} achievements of the HORSE formalism~\cite{Zaitsev1998, BangHORSE} which is a particular version of the~$J$-matrix formalism~\cite{Yamani}~in quantum scattering theory \color{black} that utilizes the oscillator basis. We use the HORSE formalism because the harmonic oscillator basis \color{black} is natural and convenient for nuclear physics applications.  \color{black} Working in the harmonic oscillator basis, we construct basis wave functions by using different choices of linearly independent superpositions of oscillator functions as short-range components.  For many-body calculations, we may choose as short-range components low-lying NCSM eigenfunctions that are obtainable from modern NCSM codes.   We also discuss strategies for implementing our approach to solve \textit{ab initio} many-body problems with restricted oscillator quanta accessible for modern NCSM codes.   \color{black} To further improve convergence, we utilize an interaction smoothing scheme introduced in Refs.~\cite{HungOne,HungTwo}.  \color{black}

 We employ this improved formalism to calculate and analyze phase shifts and continuum wave functions for both single and coupled channels using the Daejeon16~$NN$~interaction~\cite{Daejeon16}~and model two-body problems.  We demonstrate the utilization of \color{black} the adapted \color{black} Efros method for locating~$S$-matrix poles to find resonance energies~$E_{r}$ and widths~$\Gamma$ as well as to improve variational results for the bound state energies~$E_{b}$. \color{black} With these calculations, we demonstrate that the adapted Efros method leads to accurate results even with restricted oscillator quanta.  We envision that our improvements to the method will facilitate \textit{ab initio} many-body calculations of nuclear structure and reactions. \color{black}

In Section \ref{II}, we introduce the formalism of \color{black} the adapted \color{black}  Efros method.  We discuss applications of this method to 
single- and coupled-channel~$np$~scattering in Sections~\ref{III} and~\ref{IV}, respectively.  Calculations of~$S$-matrix poles are examined in Section~\ref{V}. 
We conclude in Section~\ref{VI}.   Additional results are presented in Supplemental Material \color{black} (SM) \color{black} \cite{Supplemental}.

\section{Formalism }
\label{II}
We start with presenting \color{black} the adapted \color{black} Efros method~\cite{VDEfros}~in the case of a two-body problem with
$w$ open channels and some modifications and extensions that we suppose will be useful for many-body applications.

\subsection{Ansatz wave function}
The scattering wave function at the energy~$E$ in the center-of-mass system corresponding to the entrance channel~$i$~is parameterized as 
\begin{equation}
\label{eq1}
u_{i}(E;\vec r)=\eta^-_i(\vec{r})
-\sum_{j=1}^{w}\eta^+_j(\vec{r})\, \mathbb{S}_{ji}(E)
+\sum_{q=0}^{v-1}b_{q}^{i}(E)\,\beta_{q}(\vec r).
\end{equation}
Here~$\vec{r}$ is the coordinate of relative separation of colliding fragments, the channel multi-indices $i$ and~$j$ include all quantum numbers distinguishing the channels and those common for all coupled channels
like the orbital momentum of relative motion of colliding particles~$l_{j}$ in the channel~$j$, their total spin~$S_{j}$, and the total angular momentum~$J$~and its projection~$M_{J}$.  $\beta_{q}(\vec r)$ are~$v$  {linearly independent} short-range functions (SRFs) vanishing as~${r \rightarrow \infty}$~which are introduced to describe the wave function at short distances.  The set of~$v$~energy-dependent coefficients~$b_{q}^{i}(E)$, 
$q=0,...\:,v-1$, together with~$w$~$S$-matrix elements~$\mathbb{S}_{ji}(E)$, $j=1,...\:,w$, comprise a set of~$N=v+w$~unknowns in each given entrance channel~$i$~that should be determined within this approach. 

For nuclear physics applications, it is natural and convenient to use the 
expansions in terms of harmonic oscillator functions.  The harmonic oscillator basis has been used in many \textit{ab initio} nuclear physics applications, for example, 
in the NCSM and coupled cluster applications mentioned above as well as to describe clustering from first principles in~$^8\textrm{Be}$,  $^{10}\textrm{Be}$, and $^{12}\textrm{C}$~\cite{KravvarisVolya2017}.  {We set SRFs~$\beta_{q}(\vec r)$ to be equal to finite sums of harmonic oscillator functions, 
\begin{equation}
\label{eq2}
\beta_{q}(\vec r)=\sum_{j=1}^{w}\sum_{n=0}^{M_{j}-1}a_{q n}^{j} \phi_{nj}(\vec r), 
\end{equation}
or just to single oscillator functions~$\phi_{nj}(\vec r)$~in different channels~$j$ supposing that all 
coefficients~$a_{q n}^{j}$ 
in Eq.~\eqref{eq2} with an
exception of one of them for each value of~$q$~are equal to zero.  The oscillator functions~$\phi_{nj}(\vec r)$~are defined as
\begin{gather}
\label{eq3}
\phi_{n j}(\vec r)  \equiv \phi_{nl_{j}}\!(\vec r) = R_{nl_{j}}\!(r)\,\mathscr{Y}_{J M_{J}}^{l_{j} S_{j}}(\Omega_{\hat{r}}),
\end{gather}
where~$\mathscr{Y}_{J M_{J}}^{l_{j} S_{j}}(\Omega_{\hat{r}})$ are generalized spherical harmonics which describe the
spin-angular  degrees  of freedom in the channel~$j$,
\begin{equation}
\label{GSH}
 \mathscr{Y}_{J M_{J}}^{l S}(\Omega_{\hat{r}})=\sum_{ m_{S},m_{l}}( l m_{l} S m_{S}|J M_{J})Y_{l m_{l}}(\Omega_{\hat{r}})\chi_{m_{S}}^{S},
\end{equation}
$(l m_{l} S m_{S} |J M_{J})$ are real-valued Clebsch--Gordan coefficients \cite{Varsha}, $\chi_{m_{S}}^{S}$~is the spinor, and $Y_{l m_{l}}(\Omega_{\hat{r}})$~is the spherical harmonic function \cite{Varsha};
\color{black}
\begin{equation}
\label{eq4}
R_{nl}(r)=(-1)^{n}\sqrt{\frac{2n!}{b^{3}\Gamma(n+l+\frac{3}{2})}}\Big(\frac{r}{b}\Big)^{l} e^{-\frac{r^{2}}{2b^{2}}}L_{n}^{l+\frac{1}{2}}\!\Big(\frac{r^{2}}{ b^{2}}\Big),
\end{equation}
\color{black}
where~$L_{n}^{\alpha}(z)$~is the associated Laguerre polynomial~\cite{Arfken},
\color{black}
\begin{equation}
\label{eq5}
b=\sqrt{\frac{\hbar}{ m \Omega}}
\end{equation}
\color{black}
is the oscillator length parameter, $m$~is the reduced mass of the colliding particles, and~$\Omega$~is the oscillator frequency. 

The long-range behavior of the wave 
functions $u_{i}(E;\vec r)$~is described by the functions
\begin{gather}
\label{eq6}
\eta^\pm_i(\vec{r}) \equiv\eta^\pm_i(k,\vec{r})
=\tilde{h}^{\pm}_{l_i}(kr)\mathscr{Y}_{J M_{J}}^{l_{i} S_{i}}(\Omega_{\hat{r}}).
\end{gather}
Here
\begin{gather}
\label{eq7}
\tilde{h}^{\pm}_{l}(kr)=\tilde{n}_{l}(kr) \pm ij_{l}(kr), \\[1ex]
\label{eq8}
k=\frac{\sqrt{2mE}}{\hbar}
\end{gather}
is the relative momentum of colliding particles, $j_{l}(z)$~is the spherical Bessel function~\cite{Arfken}, and 
the functions~$\tilde{n}_{l}(z)$
and~$\tilde{h}^{\pm}_{l}(z)$ behave asymptotically as
\begin{align}
\tilde{n}_{l}(z) &\mathop{\longrightarrow}_{z\to\infty} -n_{l}(z) 
\mathop{\longrightarrow}_{z\to\infty} \frac{1}{z}\cos\!\Big(z-\frac12\pi l\Big),
\label{eq9} \\
\tilde{h}^{\pm}_{l}(z)&\mathop{\longrightarrow}_{z\to\infty} { \pm ih^{(1,2)}_l(z)}
\mathop{\longrightarrow}_{z\to\infty}  \frac{i^{\mp l}}{z}e^{\pm iz}.
\label{eq10}
\end{align}
The singularity at the origin of the spherical Neumann~$n_{l}(z)$ and Hankel~$h^{(1,2)}_l(z)$ functions~\cite{Arfken} is regularized in the functions~$\tilde{n}_{l}(z)$
and~$\tilde{h}^{\pm}_{l}(z)$ that makes it possible to avoid divergences of integrals calculated with them. 

In the case of the continuum spectrum, the series of oscillator functions~$\phi_{n j}(\vec r)$ should be formally infinite:
\begin{equation}
\label{eq11}
u_{i}(E;\vec r)=\sum_{j=1}^{w}\sum_{n=0}^{\infty}d_{n,ij}(E)\,\phi_{n{j}}(\vec r).
\end{equation}
 Therefore, contrary to Efros~\cite{VDEfros}~who uses regulator functions for  the  regularization of the functions~$n_{l}(z)$ and~$h^{(1,2)}_l(z)$, we utilize here the functions developed within the HORSE formalism~\cite{Zaitsev1998, BangHORSE}  defined through their infinite expansions in 
 a  series of harmonic oscillator functions both for the regularized functions~$\tilde{n}_{l}(kr)$~and~$\tilde{h}^{\pm}_{l}(kr)$ and for the regular spherical Bessel function~$j_{l}(kr)$:
\begin{align}
\label{eq12}
j_{l}(kr)=&\sum_{n=0}^{\infty}S_{nl}(k)R_{nl}(r),\\
\label{eq13}
 \tilde{n}_l(kr)=&\sum_{n=0}^{\infty}C_{nl}(k)R_{nl}(r),\\
 \label{eq14}
\tilde{h}^{\pm}_{l}(kr)=&\sum_{n=0}^{\infty}C^{\pm}_{n l}(k)R_{nl}(r),
\end{align}
where~\cite{BangHORSE}
\color{black}
\begin{align}
\label{eq15}
S_{nl}(k)=&\:\sqrt{\frac{\pi b^{3} n!}{\Gamma(n+l+\frac{3}{2})}}(kb)^{l}e^{-\frac{k^{2}b^{2}}{2}}L_{n}^{l+\frac{1}{2}}(k^{2}b^{2}),
\\[1.5ex]
\label{eq16}
C_{nl}(k)=&\:\frac{(-1)^{l}}{\Gamma(-l+\frac{1}{2})}\sqrt{\frac{\pi b^{3} n!}{\Gamma(n+l+\frac{3}{2})}}(kb)^{-l-1}e^{-\frac{k^{2}b^{2}}{2}} \notag \\
&\phantom{\Gamma(-l)} \times
{_{1}F_{1}}\Big(\!-n-l-\frac{1}{2},-l+\frac{1}{2};k^{2}b^{2}\Big),
\\[1.5ex]
\color{black}
\label{eq17}
\color{black} C^{\pm}_{n l}(k)=&\color{black}\:C_{n l}(k) \pm i S_{n l}(k),
\end{align}
\color{black}
and ${_{1}F_{1}}(a,b;z)$ \color{black} is the confluent hypergeometric function~\cite{Arfken}.

It is convenient sometimes to employ the standing-wave approach where the continuum wave functions are real-valued: 
\begin{multline}
\label{eq18}
{u}_{i}(E;\vec r)=j_{l_{i}}(kr)\,\mathscr{Y}_{J M_{J}}^{l_{i} S_{i}}(\Omega_{\hat{r}})
\\ 
+\sum_{j=1}^{w}K_{ji}(E)\,\tilde{n}_{l_{j}}(kr)\,\mathscr{Y}_{J M_{J}}^{l_{j} S_{j}}(\Omega_{\hat{r}})+\sum_{q=0}^{v-1}
{b}_{q}^{i}(E)\,\beta_{q}(\vec r) , 
\end{multline}
where we  define the  $K$-matrix as
\begin{equation}
\label{eq19}
K=i(I-\mathbb{S})(I+\mathbb{S})^{-1}\!.
\end{equation}

 \color{black}
\color{black} Our method can incorporate Coulomb interactions in a straightforward manner by following Ref.~\cite{BangHORSE}.  In particular, for the scattering of charged particles, we introduce an auxiliary interaction operator $V^{sh}$.  \color{black} The Coulomb interaction is included in $V^{sh}$ and is formally cut at a distance~$r_{c}$ that is greater than the range of the nuclear interaction.~Using the auxiliary interaction, we obtain the scattering wave function~$u_{i}^{sh}(E;\vec r)$.  The solutions corresponding to~$V^{sh}$~are then utilized to calculate the Coulomb-distorted scattering wave function~$u_{i}^{Coul}(E;\vec r)$~that is obtained via trivial recalculations.  \color{black} Additional} \color{black} details are presented in \color{black} Ref.~\cite{BangHORSE} \color{black} and alternative approaches can be found in \color{black} Refs.~\cite{Okhrimenko} and \cite{Yanikov2025}.

\color{black}

\subsection{Formal solutions}
The wave functions~$u_{i}(E;\vec r)$ entering Eq.~\eqref{eq1} fit the Schr\"odinger equation
\begin{gather}
(H-E)\,u_{i}(E;\vec r)=0,
\label{eq20}
\end{gather}
where~$H$~is the Hamiltonian of the system. To find~$N$~unknowns entering the function~$u_{i}(E;\vec r)$~in each entrance channel~$i$, we can obtain a set of~$w$~matrix equations
\begin{gather}
\label{eq21}
\langle \bar{\beta}_{q}|H-E| u_{i}(E;\vec r)\rangle=0 , \hspace{1mm} q=0,...\,,N-1; 
\hspace{1mm}
 i=1,...\,,w.
\end{gather}
Here,  $\bar{\beta}_{q}=\bar{\beta}_{q}(\vec{r})$ are SRFs.
We note that the number of unknowns~$N=v+w$~in Eq.~\eqref{eq1} is larger than~$v$, the number of SRFs~$\beta_{q}(\vec r)$ involved in this equation. Hence we need to use {in Eq.~\eqref{eq21}}~$w$ more 
SRFs~{$\bar{\beta}_{q}(\vec r)$~as compared with  the  SRFs~$\beta_{q}(\vec r)$} used to describe the short-range physics in Eq.~\eqref{eq1}. 
{
As in Eq.~\eqref{eq2}, we can expand~$\bar{\beta}_{q}^{i}(\vec{r})$~in a finite series of oscillator functions,
\begin{equation}
\label{eqBarBeta}
\bar{\beta}_{q}^{i}(\vec{r})=\sum_{j=1}^{w}\sum_{n=0}^{\bar{M}_{j}-1}\bar{a}_{qn;i}^{j}\phi_{nj}(\vec{r}).
\end{equation}

Efros suggests~\cite{VDEfros} to use
the same~$v$ SRFs $\bar{\beta}_{q}(\vec r)=\beta_{q}(\vec r)$, $q=0,...\,,v-1,$ and we shall 
mostly follow this suggestion in what follows.  In this case, 
$\bar{a}_{qn;i}^{j}=a_{qn}^{j}$ for~$q=0,...\,,{v-1}$.  For $q=v,...\,,N$, 
$\bar{\beta}_{q}(\vec{r})$ may be arbitrarily chosen with the only constraint being that the functions~$\bar{\beta}_q(\vec{r}), q=0,... , N$~should be linearly independent.  One of the  reasonable choices, for example, 
may be some oscillator functions 
as we shall see below.

However, generally, the set of SRFs~$\{\bar{\beta}_{q}\}$ can differ completely
from the set of linearly independent SRFs~$\{{\beta}_{q}\}$; it is only important to have a set of~$N$~linearly independent SRFs~$\bar{\beta}_{q}(\vec r)$~which may be nonorthogonal.
That means that, generally, the coefficients $\bar{a}_{qn;i}^{j}$ are 
distinct from~$a_{qn}^{j}$ in Eq.~\eqref{eq2}.  Moreover, $\bar{M}_{j}$, the finite truncations
of oscillator series in each channel~$j$, may differ from the truncations~$M_{j}$ in Eq.~\eqref{eq2}.   }

Let us define a vector~$\bar{c}^i$ of the unknowns in the entrance channel~$i$ with components
\begin{gather}
\bar{c}^i_q=\begin{cases}
\phantom{-}b_{q}^{i}(E),\quad &q=0,...\, ,v-1, \\
-\mathbb{S}_{q-v+1,i}(E),  &q=v,...\, ,N-1. 
\label{eq22}
\end{cases}
\end{gather}
Equation~\eqref{eq21} can be rewritten as
\begin{equation}
\label{eq23}
\bar{A}\bar{c}^{i}=\bar{B}^{i} , \quad  i=1,...\, ,w,
\end{equation}
where the elements of matrix~$\bar{A}$ are  
\begin{subequations}
\begin{align}
\bar{A}_{q q'}=\langle  {\bar{\beta}_{q}}|H-E|\beta_{q'} \rangle, \quad &\begin{cases}
							q=0,...\, ,N-1, \\ 
							q'=0,...\, ,v-1, \end{cases}
\label{eq24a}
\\
\bar{A}_{qq'}= {\langle {\bar{\beta}_{q}}|H-E| \eta^{+}_{{j}} \rangle } ,  \quad &\begin{cases}
							{q=0,...\, ,N-1} , \\ 
							q'=v,...\, ,N-1, \\
							{j=q'-v+1 ,} \end{cases}
\label{eq24b}
\intertext{and the components of vector~$\bar{B}^{i} $ are}
\bar{B}_{q}^{i}=-\langle  {\bar{\beta}_{q}}|H-E| \eta^{-}_{{i}} \rangle  , \quad 
							&{q=0,...\, ,N-1} . 
\label{eq24c}
\end{align}
\end{subequations}

We can combine~$w$ equations~\eqref{eq23}~corresponding to different entrance channels~$i$~into a single matrix equation
\begin{equation}
\label{eq25}
Ac=B,
\end{equation}
where
\begin{equation}
\label{eq26}
A=\begin{pmatrix}
    \bar{A}  & \textbf{0} & \hdots & \textbf{0} \\[0.3em]
    \textbf{0}  & \bar{A} & \hdots & \textbf{0} \\[0.3em]
    \vdots  & \vdots & \ddots & \vdots \\[0.3em]
    \textbf{0}  & \textbf{0} & \hdots & \bar{A} \\[0.3em]
     \end{pmatrix}
\end{equation}
is a~$wN \times wN$~matrix comprising~{$w$}
identical~{$N \times N$} matrices~$\bar{A}$~as blocks on its diagonal, and 
the~$wN$-dimensional vectors~$B$~and~$c$~are
\begin{equation}
\label{eq27}
B=\begin{pmatrix}
    \bar{B}^{1} \\[0.3em]
    \vdots \\[0.3em]
    \bar{B}^{w} \\[0.3em]
     \end{pmatrix}\!
\end{equation}
and
\begin{equation}
\label{eq28}
c=\begin{pmatrix}
    \bar{c}^{1} \\[0.3em]
    \vdots \\[0.3em]
    \bar{c}^{w} \\[0.3em]
     \end{pmatrix}\!. 
\end{equation}

By solving Eq.~\eqref{eq25} we can obtain all $wN$ unknowns entering Eqs.~\eqref{eq1} with different
entrance channels $i=1,...\,,w$.
In the single-channel case, the above equations provide
a complete solution of the problem.  However, not all the~$Nw$~unknowns are independent
in the multichannel case when~$w>1$. In particular, the~$S$~matrix should be symmetric, i.\:e., $\mathbb{S}_{ij}(E) = \mathbb{S}_{ji}(E)$.  So, the $S$~matrix has~$w^2$ matrix elements but only ${\frac12w(w+1)}$~independent matrix elements. 
Therefore, Eq.~\eqref{eq25} is over-complete, providing~${w^2-\frac12w(w+1)}=\frac12w(w-1)$ unneeded extra relations between unknowns, and the number of independent unknowns
is ${\mathbb{\clb N}}{=Nw-\frac12w(w-1)}$.  \color{black} In \color{black} Ref. \color{black} \cite[Section III]{Supplemental}, we discuss \color{black} this situation using an example of a two-channel system.  \color{black}  We show \color{black} that in some cases the over-complete Eq.~\eqref{eq25} 
{provides a correct solution with~$\mathbb{S}_{12}(E) = \mathbb{S}_{21}(E)$.} However,
 in cases of a reduced set of SRFs~$\bar{\beta}_q$ 
most interesting for many-body applications,  solving Eq.~\eqref{eq25}
results in~$\mathbb{S}_{12}(E) \ne \mathbb{S}_{21}(E)$.  \color{black}

In the cases where the symmetry is not guaranteed, we impose the~$S$-matrix symmetry and
replace all~$S$-matrix elements~$\mathbb{S}_{ij}(E)$~with~$i<j$ 
by~$\mathbb{S}_{ji}(E)$ in Eqs.~\eqref{eq1} and~\eqref{eq21} and define a new
vector of unknowns~$c$ by replacing the vectors~$\bar{c}^i$
defined by Eq.~\eqref{eq22} by the vectors~$c^i$ with the components
\begin{gather}
{c}^i_q=\begin{cases}
\phantom{-}b_{q}^{i}(E),\quad &q=0,...\, ,v-1, \\
-\mathbb{S}_{q-v+i,i}(E),  &q=v,...\, ,N-i. 
\label{eq29}
\end{cases}
\end{gather}
We combine the vectors~${c}^i_q$ in a total ${\mathbb{\clb N}}$-dimensional vector of unknowns
\begin{equation}
\label{eq30}
c=\begin{pmatrix}
    c^{1} \\[0.3em]
    \vdots \\[0.3em]
    c^{w} \\[0.3em]
     \end{pmatrix}\! .
\end{equation}

Instead of~$wN$~equations~\eqref{eq21}~we need now to formulate only~$\mathbb{\clb N}$
independent equations of the same type determining the~$\mathbb{\clb N}$ independent unknowns. We can do this by reducing the sets of bra SRFs~$\langle {\bar{\beta}_{q}}|$~in Eq.~\eqref{eq21} in some entrance
channels~$i$. We denote these reduced sets as~$\{\langle\bar{\beta}_q^i|\}$.  We include in the set~$\{\langle\bar{\beta}_q^1|\}$~in the first entrance channel ($i=1$) all~$N$~SRFs~$\langle {\bar{\beta}_{q}}|$, $\{\langle\bar{\beta}_q^1|\}\equiv\{\langle {\bar{\beta}_{q}}|\}$.  We note here that the channel ordering is arbitrary and we can always choose a particular numbering of channels which is optimal from the point of view of providing the best convergence.  For the second entrance channel with~$i=2$, we use~$N-1$~independent SRFs~$\langle\bar{\beta}_q^2|$~which may be obtained from the set~$\{\langle {\bar{\beta}_{q}}|\}$~by excluding from it one of the SRFs~$\langle {\bar{\beta}_{q}}|$.  We can also define some of the SRFs~$\langle\bar{\beta}_q^2|$~as linear combinations of the SRFs~$\langle {\bar{\beta}_{q}}|$: the set~$\{\langle\bar{\beta}_q^2|\}$ should include~$N-1$~independent SRFs which can be nonorthogonal.  We note that any superposition of~$v$~linearly independent SRFs~$\beta_q(\vec{r})$, $q=0,...\, ,v-1$, entering
Eq.~\eqref{eq1} should produce exactly the same results while superpositions involving additional~$w$~SRFs~${\bar{\beta}_q(\vec{r})}$,
$q=v,...\, ,v+w$, involved in Eq.~\eqref{eq21} may affect the convergence. Next, we
obtain in the same manner the set~$\{\langle\bar{\beta}_q^3|\}$~including~$N-2$~independent SRFs~$\langle\bar{\beta}_q^3|$, etc. \color{black} More generally, our prescription for SRF selection is as follows: for the first channel, we use a set of~$N$~SRFs~$\{\langle \bar{\beta}_{q}^{1}|\}$, where~$q=0,...,N-1$.  In the~$i$th channel, where~$i=2,...,w$, we remove the SRF~$\bar{\beta}_{N-i-w+1}^{i-1}(\vec r)$~from the~$(i-1)$th~set to form the set of SRFs~$\{\langle \bar{\beta}_{q}^{i}|\}$, where~$q=0,...,N-i$. \color{black} We note that the set of ket  SRFs~$\{|\beta_{q'}\rangle\}$ is the same in all entrance channels.

As a result, instead of Eq.~\eqref{eq23}, we obtain
\begin{equation}
\label{eq31}
{A}^ic^{i}=B^{i} , \quad  i=1,...\, ,w,
\end{equation}
where the matrices~${A}^i$, contrary to matrices~$\bar{A}$~in
Eq.~\eqref{eq23}, differ in size for different channels~$i$~and their elements  are
\begin{subequations}
\label{eq32}
\begin{align}
\!\!{A}^i_{q q'}&=\langle \bar{\beta}^i_{q}|H-E|\beta_{q'} \rangle ,\;&\begin{cases}
							q=0,...\, ,N-i, \\ 
							q'=0,...\, ,v-1, \end{cases}\label{eq32a}\\
\!{A}^i_{qq'}&=\langle \bar{\beta}^i_{q}|H-E|\eta^{+}_{{j}} \rangle  , \;&  \begin{cases} 
							{q=0,...\, ,N-i}, \\ 
							 q'= v,...,N-i ,\color{black}  \\
						{  j=q'-v+i . }\color{black} \end{cases}
\label{eq32b}
\end{align}
{For the components of vector~$B^{i} $ we define for $i=1$ as}
\begin{align}
B^1_q=-\langle \bar{\beta}^1_{q}|H-E| \eta^{-}_{{i}} \rangle ,\;\qquad 
							&\hspace{-3ex}q=0,...\, ,N-1  , & 
\label{eq32c}
\intertext{and for $i>1$ as}							
B_{q}^{i}=-\langle \bar{\beta}^i_{q}|H-E|{ \eta^{-}_{{i}} }\rangle 
 -\sum_{j=1}^{i-1} & 
               \langle \bar{\beta}^i_{q}|H-E| \eta^{+}_{{j\color{black}}} \rangle c^{j}_{v+i-j\color{black}} 
                                              ,  &\nonumber \\
							&\hspace{-3ex} 
					q=0,...\, ,N-i.
\label{eq32e}
\end{align}
\end{subequations}

Equations~\eqref{eq31} can be combined into the matrix equation~\eqref{eq25}
where
\begin{equation}
\label{eq33}
A=\begin{pmatrix}
     {A}^1  & \textbf{0} & \hdots & \textbf{0} \\[0.3em]
    \textbf{0}  &  {A}^2 & \hdots & \textbf{0} \\[0.3em]
    \vdots  & \vdots & \ddots & \vdots \\[0.3em]
    \textbf{0}  & \textbf{0} & \hdots &  {A}^w \\[0.3em]
     \end{pmatrix}
\end{equation}
is an $\mathbb{\clb N} \times \mathbb{\clb N}$ matrix comprising $(N-i+1) \times (N-i+1)$ matrices~$A^i$~as diagonal blocks,
\begin{equation}
\label{eq34}
B=\begin{pmatrix}
    B^{1} \\[0.3em]
    \vdots \\[0.3em]
    B^{w} \\[0.3em]
     \end{pmatrix}\!,
\end{equation}
and~$c$ given by Eq.~\eqref{eq30} are $\mathbb{\clb N}$-dimensional vectors.  \color{black}  We note that our SRF selection indicates which SRF $\bar{\beta}_{N-i-w+1}(\vec{r})$ to remove from the $i>1$ channel.  Therefore, changing the choices of which functions to remove from the~$i>1$~calculations amounts to replacing one row of~$A^{2}, A^{3},...,A^{w}$~in Eq.~\eqref{eq33}~by another.  See Section~\ref{IV} and \color{black} Ref. \color{black} \cite[Section III]{Supplemental} for the two-channel case.
\color{black}

Equations~\eqref{eq31} can be resolved step-by-step.  First, we solve this matrix equation for the first entrance channel with~$i=1$~and obtain~$v$~coefficients~$b_{q}^{1}(E)={c}^1_q$, $q=0,...\,,v-1$, and~$w$ $S$-matrix elements ${\mathbb{S}_{j1}(E)=-c^1_{j+v-1}}$, $j=1,...\,,w$~at any desired
energy~$E$. Next, we solve Eq.~\eqref{eq31} for~$i=2$. The~$S$-matrix 
element~$\mathbb{S}_{12}(E)=\mathbb{S}_{21}(E)=-c^1_{v+1}$ in the right-hand side of Eq.~\eqref{eq32e} was already obtained at the previous step. With the help of Eq.~\eqref{eq31} we now obtain~$v$ coefficients~${b_{q}^{2}(E)={c}^2_q}$,
$q=0,...\,,v-1$,  and~$w-1$~$S$-matrix elements  ${\mathbb{S}_{j2}(E)=-c^2_{j+v-2}}$, 
{\clb$j=2,...\,,w$}.  Now we have $S$-matrix elements~{$\mathbb{S}_{3j}(E)=-c^{j}_{v+3-j}$}
{\clb with} $j=1$, 2 in the 
right-hand side of Eq.~\eqref{eq31} for 
{\clb$i=3$ and solve it.}
Next we solve Eq.~\eqref{eq31} for~$i=4$, 5, etc.

{
The calculations of matrices~$A^i$ and~$B^i$ can be simplified using the following expressions for the function~$\eta^\pm_i(\vec{r}) $~\cite{Zaitsev1998}:
\begin{equation}
\label{eq35}
(T-E)\eta^\pm_i(\vec{r}) =\frac{\hbar^{2}}{2m}
\frac{\phi_{0 i}(\vec r)}{k\,S_{0 l_{i}}(k)} ,
\end{equation}
where $T$ is the operator of kinetic energy of relative motion of colliding particles. The 
Hamiltonian
\begin{gather}
\label{eq36}
H=T+V
\end{gather} 
 includes also the potential energy operator~$V$. Thus, 
\begin{equation}
\label{eq37}
(H-E)\eta^\pm_i(\vec{r}) 
=V\eta^\pm_i(\vec{r}) 
+\frac{\hbar^{2}}{2m}\frac{\phi_{0 i}(\vec r)}{k\,S_{0 l_{i}}(k)} .
\end{equation}

Following the ideas of the HORSE approach~\cite{Zaitsev1998, BangHORSE}, the potential energy operator~$V$~can be approximated by a finite matrix~$V^{tr}$~in the oscillator basis, 
\begin{multline}
\label{eq38}
V\approx V^{tr}=\sum_{i,j=1}^{w}\sum_{n=0}^{\mtc{N}_{i}-1}\sum_{n'=0}^{\mtc{N}_{j}-1}|\phi_{ni}\rangle\langle\phi_{ni}|V|\phi_{n'j}\rangle\langle\phi_{n'j}| \\
\equiv\sum_{i,j=1}^{w}\sum_{n=0}^{\mtc{N}_{i}-1}\sum_{n'=0}^{\mtc{N}_{j}-1}|\phi_{ni}\rangle V_{ni,n'j}\langle\phi_{n'j}|,
\end{multline}
since the potential energy matrix elements in the oscillator basis~$V_{ni,n'j}$ decrease with~$n$ and/or~$n'$ 
and can be neglected at large~$n$ and/or~$n'$ as compared  to  the nonzero kinetic energy matrix 
elements~$T_{ni,n'j}=\langle\phi_{ni}|T|\phi_{n'j}\rangle=T^{l_i}_{nn'}\delta_{ij}$,
\begin{subequations}
\label{eq39}
\begin{align}
T^{l}_{nn}&=\frac12\hbar\Omega\Big(2n+l+\frac32\Big),
\label{eq39a} \\
T^{l}_{n-1,n}&=-\frac12\hbar\Omega\sqrt{n\Big(n+l+\frac12\Big)},
\label{eq39b} \\
T^{l}_{n+1,n}&=-\frac12\hbar\Omega\sqrt{(n+1)\Big(n+l+\frac32\Big)},
\label{eq39c} \\
T_{nn'}^l&=0 \quad\text{if}\quad {\clg |n-n'|>1.}
\label{eq39d}
\end{align}
\end{subequations}
Note,  the approximation of Eq.~\eqref{eq38} is conventionally employed in many-body {\it ab initio} nuclear structure calculations where the kinetic energy matrix~\eqref{eq39} is usually also  truncated in the same manner.  However, we do not truncate the kinetic energy to allow for scattering.

Using Eq.~\eqref{eq35} together with Eqs.~\eqref{eq2}, \eqref{eq6},  \eqref{eq14}, \eqref{eqBarBeta}, and defining 
\begin{gather}
\label{eq40}
H_{ni,n'j}=\langle\phi_{ni}|H|\phi_{n'j}\rangle,
\end{gather} 
\color{black}
 we rewrite Eqs.~\eqref{eq32} as 
\begin{widetext}
\begin{subequations}
\label{eq41}
\begin{align}
\!\!\!{A}^i_{q q'}    
=&{ \sum_{j,j'=1}^{w}}\sum_{n=0}^{\bar{M}_{j}-1}\sum_{n'=0}^{M_{j'}-1} { \big(\bar{a}_{qn;i}^{j}\big)^{\!*}}a_{q'n'}^{j'} (H_{nj,n'j'}-E\delta_{nn'}\delta_{jj'}) , \quad &&
                                                           \begin{cases}
							q=0,...\, ,N-i, \\ 
							q'=0,...\, ,v-1, \end{cases}\label{eq41a}\\
\label{eq41b}
\!\!\!{A}^i_{qq'} =&
    \sum_{j=1}^{w}\sum_{n=0}^{\bar{\cal M}_{j}-1}\sum_{n'=0}^{\mathcal{N}_{j'}-1}\color{black} \big(\bar{a}_{qn;i}^{j}\big)^{\!*}C^{+}_{n'l_{\!j'}}\!(k)\, V_{nj,n'j'}
       +\frac{\hbar^{2}}{2m}\frac{\big(\bar{a}_{q0;i}^{j'}\big)^{\!*}}{k\,S_{0l_{\!j'}}\!(k)}, \quad &&
							 \begin{cases}
							{q=0,...\, ,N-i} , \\ 
							q'=v,...,N-i, \\
							{j'=q'-v+i , } \end{cases} \! \! \\
 \!\!  B^1_q =&-    \sum_{j=1}^{w}\sum_{n=0}^{\bar{\cal M}_{j}-1}\sum_{n'=0}^{\mathcal{N}_{1}-1}\color{black} \big(\bar{a}_{qn;1}^{j}\big)^{\!*}C^{-}_{n'l_{1}}\!(k)\, V_{nj,n'1}
       -\frac{\hbar^{2}}{2m}\frac{\big(\bar{a}_{q0;1}^{1}\big)^{\!*}}{k\,S_{0l_{1}}\!(k)} , 
							\quad && q=0,...\, ,N-1  , \!
\label{eq41c}\\
 \!\! B_{q}^{i}=&-\sum_{j=1}^{w}\sum_{n=0}^{\bar{\cal M}_{j}-1}\sum_{n'=0}^{\mathcal{N}_{i}-1}\color{black} \big(\bar{a}_{qn;i}^{j}\big)^{\!*}C^{-}_{n'l_{i}}\!(k)\, V_{nj,n'i}
       -\frac{\hbar^{2}}{2m}\frac{\big(\bar{a}_{q0;i}^{i}\big)^{\!*}}{k\,S_{0l_{i}}\!(k)}
       \nonumber
 \\
 &-\sum_{j'=1}^{i-1} \sum_{j=1}^{w}\sum_{n=0}^{\bar{\cal M}_{j}-1} \sum_{n'=0}^{\mathcal{N}_{j'}-1}\big(\bar{a}_{qn;i}^{j}\big)^{\!*}C^{+}_{n'l_{j'}}\!(k)\, V_{nj,n'j'}c^{j'}_{v+i-j'}-\sum_{j'=1}^{i-1} \frac{\hbar^{2}}{2m}\frac{\big(\bar{a}_{q0;i}^{j'}\big)^{\!*}}{kS_{0l_{j'}}(k)}c^{j'}_{v+i-j'}, \hspace{-1.8ex}
							\quad && \begin{cases}i>1,\\ q=0,...\, ,N-i , \end{cases}
\label{eq41d}
\end{align}
\end{subequations}
\end{widetext}
{where~{\clb
$\bar{\mathcal{M}}_{i}=\min\{\bar{M}_{i},\mathcal{N}_{i}\}$,}
$\big(\bar{a}_{qn;i}^{j}\big)^{\!*}$~is a complex conjugate of~$\bar{a}_{qn;i}^{j}$, and, as previously mentioned, we set~$\bar{a}_{qn;i}^{j}=a_{qn}^{j}$ for~$q=0,...,{v-1}$ and arbitrarily chosen coefficients for~$q=v,...,{N-i}$.  }
\color{black}

Efros mentioned~\cite{VDEfros} that, contrary to the conventional 
Hulth\'en--Kohn variational method~\cite{Hulthen1948, Kohn1948, Moses1953,  Moses}, his approach does not make use of ``free-free'' matrix elements like~$\langle\eta^\pm_i |H-E|\eta^\pm_j \rangle$~but, instead, includes only ``bound-bound''~$\langle \bar{\beta}^i_{q}|{H-E}|\beta_{q'} \rangle$~and ``bound-free''
$\langle\bar{\beta}^i_{q} |{H-E}|\eta^\pm_j \rangle$ matrix elements which are easier to calculate.  We note that Eqs.~\eqref{eq41} include matrix elements of the ``bound-bound'' type
only, which are evaluated within a restricted Hilbert space spanned by a limited set of oscillator functions. \\

\vspace{-0.5cm}

\color{black} This is significant because the evaluation of the matrix elements $\langle\bar{\beta}^i_{q} |{H-E}|\eta^\pm_j \rangle$ as ``bound-bound" (i.\,e., within a restricted Hilbert space) facilitates calculations.  \color{black} In particular, as seen in  Eqs.~\eqref{eq41b}--\eqref{eq41d}, the term $\langle\bar{\beta}_{q}^{i}|T-E|\eta_{j}^{\pm}\rangle$ disappears due to Eq.~\eqref{eq37} and coupling to continuum is represented by $\pm \frac{\hbar^{2}}{2m}\frac{\big(\bar{a}_{q0;i}^{j'}\big)^{\!*}}{k\,S_{0l_{j'}}\!(k)}$.~Because the remaining terms that include the interaction and coupling to continuum are in the restricted Hilbert space, we obtain a convenient representation of the matrix elements~$\langle\bar{\beta}_{q}^{i}|H-E|\eta_{j}^{\pm}\rangle$.  This is especially useful for many-body problems.~\color{black} We should note \color{black} however, \color{black} that in many-body applications, Eqs.~\eqref{eq41b}--\eqref{eq41d}
become more complicated.~The problem is that the 
terms~$\pm \frac{\hbar^{2}}{2m}\frac{\big(\bar{a}_{q0;i}^{j'}\big)^{\!*}}{k\,S_{0l_{j'}}\!(k)}$
in these equations account for the kinetic energy of relative motion of colliding particles~$T^{rel}$ only. In the many-body case,
the colliding particles have an internal structure.~Therefore, the Hamiltonian includes the kinetic energy of the internal motion of these particles~$T^{int}$, which enters Eqs.~\eqref{eq41b}--\eqref{eq41d} in the same manner as potential energy terms.  The separation of the total kinetic energy operator~$T=T^{rel}+T^{int}$~into the  relative $T^{rel}$ and internal~$T^{int}$ parts may be nontrivial within NCSM but will be much more natural and simpler in approaches like the RGM. \\

\vspace{-0.4cm}

In the case of the standing-wave approach starting from the wave-function ansatz~\eqref{eq18}, one can use all the above equations of this subsection with replacements~$-\mathbb{S}_{ij}(E) \rightarrow K_{ij}(E)$, $\tilde{h}^{-}_{l_{i}}(kr) \rightarrow j_{l_{i}}(kr)$, ${\tilde{h}^{+}_{l_{i}}(kr) \rightarrow \tilde{n}_{l_{i}}(kr)}$, $C^{-}_{nl_{i}}(k) \rightarrow S_{nl_{i}}(k)$,  
${C^{+}_{nl_{i}}(k) \rightarrow C_{nl_{i}}(k)}$,  having  in mind that with these replacements,  Eqs.~\eqref{eq35} should be substituted by the following relations:
%
\begin{align}
(T-E)j_{l_{i}}(kr)\mathscr{Y}_{J M_{J}}^{l_{i} S_{i}}(\Omega_{\hat{r}})&=0 , 
\tag{\ref{eq35}a}
 \\
(T-E)\tilde{n}_{l_{i}}(kr)\mathscr{Y}_{J M_{J}}^{l_{i} S_{i}}(\Omega_{\hat{r}})&=\frac{\hbar^{2}}{2m}
\frac{
\color{black} \phi_{0i}(\vec r)\color{black}}{k\,S_{0l_{i}}(k)} . \tag{\ref{eq35}b}
\label{eq35rep}
\end{align}
As a result, Eqs.~\eqref{eq41a}--\eqref{eq41b} preserve their structure with the replacement $C^{+}_{nl}(k)\rightarrow C_{nl}(k)$ while in Eqs. \eqref{eq41c}--\eqref{eq41d} the substitution $C^{-}_{nl}(k) \rightarrow S_{nl}(k)$ should be
 accompanied by removing the terms $-\frac{\hbar^{2}}{2m}\frac{\big(\bar{a}_{q0;i}^{j}\big)^{\!*}}{kS_{0l_{j}}\!(k)}$. \color{black} 
 \color{black}

\subsection{$S$-matrix poles}
The matrix formulation of the problem presented by Eqs.~\eqref{eq25},  \eqref{eq30}, \eqref{eq33}, \eqref{eq34} is convenient for searching for the~$S$-matrix poles. By locating the $S$-matrix poles in the complex energy plane,  
{\clb it is possible to} obtain resonance energies~$E_{r}$~and widths~$\Gamma$~{\clb\cite{AMShirokov2016,4nAIP,Lurie3,Lurie4,AMShirokovHORSE2016,AMShirokovHORSE2018, PPN,Igor2022,9Li,3narXiv}} and 
{\clb to improve essentially} the results of variational calculations of bound state energies~\cite{Vasya21,Lurie3,Lurie4,9Li}; moreover, by calculating the residues for poles corresponding to bound states, 
{\clb it is possible to} obtain~\cite{Vasya21,9Li} the asymptotic normalization constants~\cite{Baz}.

The formal solution of Eq.~\eqref{eq25} for the vector of unknowns~$c$~can be written as
\begin{equation}
\label{eq42}
c=A^{-1}B.
\end{equation}
Since~$c$~contains the~$S$-matrix elements, it is natural to search for the~$S$-matrix poles by solving the equation
\begin{equation}
\label{eq43}
\det A(E) =0
\end{equation}
in the complex energy plane.  Due to the diagonal structure of matrix~$A$, see Eq.~\eqref{eq33},  it looks like that the 
solutions of Eq.~\eqref{eq43} can be assured by solving any of the equations
$$
\det A^{i}(E) =0;
$$
however, it is safe to use solutions only of the equation
\begin{gather}
\label{eq44}
\det A^{1}(E) =0.
\end{gather}
Indeed, Eq.~\eqref{eq42} can be split into a set of equations for pieces of the vector of unknowns~$c$:
\begin{gather}
\label{eq45}
c^{i}=(A^i)^{-1}B^{i}.
\end{gather}
The vector~$B^{i}$~at~$i>1$ includes~$S$-matrix elements { [the 
components~{\clb $c^j_{v+i-j}$} 
of the vector~$c$, see Eqs.~\eqref{eq32e} and~\eqref{eq29}]} which have poles; thus searching for poles of the vector~$c^{i}$~by locating the zeros of~$\det A^{i}$ with~$i>1$~is not justified and one can omit some of the~$S$-matrix poles.  Locating the zeros of~$\det A^{1}(E) $ in the complex energy plane can be efficiently performed by the technique  of calculating contour integrals described in Ref.~\cite{AMShirokovHORSE2018}.  \color{black} Other techniques \color{black} for locating $S$-matrix poles \color{black} are \color{black} presented in \color{black} Refs.\color{black}~\cite{Moiseyev1998, Fanto2018}.

Equation~\eqref{eq44} can be used for locating not only resonance~$S$-matrix poles but also those corresponding to bound states. In this case, the pole energies correspond to physical states and 
it is important to calculate the wave functions at these energies.  Contrary to the continuum spectrum where we have~$w$~independent solutions corresponding to different entrance (or exit) channels, in the case of bound states, we have a unique solution no matter how many closed channels are accounted for.  Neglecting the first term~$\eta^-_i(\vec{r}) $ in Eq.~\eqref{eq1} as compared to the second pole term
$-\sum_{j=1}^{w}\eta^+_j(\vec{r})\, \mathbb{S}_{ji}(E)\vphantom{\int^{A^{A}}}$,  we obtain the following equation for the complete set of unknowns~$c^{1}$:
\begin{gather}
\label{eq46}
A^{1}c^{1}=0.
\end{gather}
Note, Eq.~\eqref{eq44} guarantees that Eq.~\eqref{eq46} has a nonzero solution. However, the vector of unknowns~$c^{1}$ is obtained up to an arbitrary multiplier, and the bound state wave function~$u_{i}(E;\vec r)$ should be additionally normalized. {\clb Moreover, the bound state
wave function in this case is expressed as an infinite converging series of oscillator functions
since the function~$\eta^+_i(\vec{r}) $ entering~$u_{i}(E;\vec r)$ is defined through infinite 
oscillator function series; see Eqs.~\eqref{eq6} and~\eqref{eq14}.}

\section{Single-Channel Scattering }
\label{III}
We proceed to discuss the efficiency of \color{black} the adapted \color{black} Efros approach and its advantages and disadvantages using two-body single-channel scattering. As the first example, we consider~$np$~scattering in the~${^1S_{0}}$~channel
using the Daejeon16~$NN$~interaction~\cite{Daejeon16}. \color{black} The Daejeon16 interaction is defined as a finite matrix and therefore, the HORSE method gives the exact solution that can be used for comparison with our calculated results. \color{black} In the single-channel case, it is convenient to utilize the standing-wave approach with real-valued wave functions where the only $K$-matrix element in  
Eq.~\eqref{eq18} is expressed through the scattering phase shifts~$\delta_{l}(E)$: 
\begin{gather}
K_{11}=\tan\delta_{l}(E).
\label{eq47}
\end{gather}
We shall omit the channel index~$i$~in the matrix~$A^{i}$, vectors~{$B^{i}$} and~$c^{i}$, 
coefficients~$b_{q}^{i}(E)$, etc., while discussing single-channel applications.

Various nuclear structure approaches, e.g., the NCSM,  utilize wave function expansions in the 
{\clb finite}
series of many-body harmonic oscillator functions~${\phi_{n}(\vec r)\equiv\phi_{nj}(\vec r)}$ 
{\clb like}  Eqs.~\eqref{eq2} and~\eqref{eq3}, where we {\clb exclude in our case of two-body single-channel scattering
the summation over channels and} can omit now the
channel index~$j$. In this case, the Hamiltonian has a structure 
\begin{gather}
\label{eq48}
H^{tr}=\sum_{n,n'=0}^{\mtc{N}-1}|\phi_{n}\rangle\langle\phi_{n}|H|\phi_{n'}\rangle\langle\phi_{n'}|
\equiv\sum_{n,n'=0}^{\mtc{N}-1}|\phi_{n}\rangle H_{nn'}\langle\phi_{n'}|,
\end{gather}
where~$\mtc{N}$~is the number of harmonic oscillator basis functions used in the truncated 
wave function expansion. The truncation corresponds conventionally to the maximal value of
(excitation) oscillator quanta~$N_{\max}$.  \color{black} If we naively suppose that the Hamiltonian structure conventionally employed in \textit{ab initio} NCSM calculations is the Hamiltonian of the system (i.e., we set $H=H^{tr}$), then we get incorrect results \color{black} as $H^{tr}$ does not include coupling to continuum. \color{black}  That is discussed in \color{black} Ref.~\color{black} \cite[Section I]{Supplemental}.  \color{black}

In the case of single-channel two-body scattering, the relative motion oscillator functions~$\phi_{n}(\vec r)$ are given by Eq.~\eqref{eq3} and~$2n+l \leq N_{\textrm{max}}$, where~$n=0,1,...\,,\mtc{N}-1$.
Equation~\eqref{eq11} takes the form
\begin{equation}
\label{eq49}
u(E;\vec r)=\sum_{n=0}^{\infty}d_{n}(E)\phi_{n}(\vec r),
\end{equation}
where the expansion coefficients are expected to be 
\begin{align}
\label{eq50}
\hspace{-1.3ex}d_{n}(E)&= 
\begin{cases} 
    g_{n}(E)+a^{as}_{n}(E), &  n =0,...\,,\mtc{N}-1 , 
    \\ 
      a^{as}_{n}(E), &  n = \mtc{N},...\,,\infty , 
   \end{cases}\\
   \label{eq51}
\hspace{-1.3ex} a^{as}_{n}(E)&=  S_{nl}(k)+C_{nl}(k)\tan\delta_{l}(E), \quad n=0, ...\,, \infty ,
\end{align}
and 
$\mtc{N}$~energy-dependent coefficients~$g_{n}(E)$~should be found within the approach.  In
our case of~$NN$~scattering in the~$^{1}S_{0}$~partial wave,  the  fixed quantum numbers are~$l=0$, $S=0$, $J=0$, and~$M_{J}=0$; therefore ${\mtc{N}=\frac12N_{\max}+1}$.

The asymptotic expansion 
coefficients~$a^{as}_{n}(E)$ describing the free motion at energy~$E$~and within the HORSE formalism fit the free Schr\"odinger equation~\cite{Zaitsev1998, BangHORSE} 
\begin{gather}
\label{eq63}
\sum_{n'=0}^{\infty}(T^{l}_{nn'}-E\delta_{nn'\!})\,a^{as}_{n'}(E)=0, \qquad n=1,...\,,\,\infty,
\end{gather}
which, due to the tridiagonal structure of the kinetic energy matrix elements, Eqs.~\eqref{eq39}, reduces to the three-term recurrent relation 
\begin{gather}
\label{eq64}
T^{l}_{n,n-1}a^{as}_{n-1}(E) +(T^{l}_{nn}-E)a^{as}_{n}(E)+ T^{l}_{n,n+1}a^{as}_{n+1}(E) =0.
\end{gather}
Therefore, according to our proposition regarding the structure of the wave function expansion coefficients,
Eqs.~\eqref{eq50} and \eqref{eq51}, the Hamiltonian~\eqref{eq48} should be extended to infinity in the oscillator space by the kinetic energy matrix elements,
\begin{multline}
\label{eq65}
H=\sum_{n,n'=0}^{\mtc{N}-1}|\phi_{n}\rangle V_{nn'}\langle\phi_{n'}|
+\sum_{n,n'=0}^{\infty}|\phi_{n}\rangle T^{l}_{nn'}\langle\phi_{n'}| \\
\equiv\sum_{n,n'=0}^{\mtc{N}-1}|\phi_{n}\rangle H^{tr}_{nn'}\langle\phi_{n'}|
+\sum_{n,n'=\mtc{N}}^{\infty}|\phi_{n}\rangle T^{l}_{nn'}\langle\phi_{n'}|\\
+|\phi_{\mtc{N}-1}\rangle T^{l}_{\mtc{N}-1,\mtc{N}}\langle\phi_{\mtc{N}}|
+|\phi_{\mtc{N}}\rangle T^{l}_{\mtc{N},\mtc{N}-1}\langle\phi_{\mtc{N}-1}|,
\end{multline}
where the truncated Hamiltonian~$H^{tr}$~given by Eq.~\eqref{eq48} has the structure conventionally used in variational calculations like NCSM and its matrix elements~$H^{tr}_{nn'}$~include both potential energy matrix elements~$V_{nn'}$~and kinetic energy~matrix elements~$T^{l}_{nn'}$ with~$n,n'=0$, ...\,,\,$\mtc{N}-1$, the term
$\sum_{n,n'=\mtc{N}}^{\infty}|\phi_{n}\rangle T^{l}_{nn'}\langle\phi_{n'}|$ describes the free motion (continuum) of colliding particles, and the last two terms in Eq.~\eqref{eq65} describe the coupling between the continuum and the interaction region in the oscillator space.  The role of these interaction region{\clb--}continuum coupling terms is very important and cannot be neglected. So, for the correct application of the Efros formalism with oscillator basis, one should use in the single-channel case the Hamiltonian~\eqref{eq65}.  

We need to specify the sets of SRFs~$\{\beta_{q}\}$ and~${\{\bar{\beta}_{q}\}\equiv\{\bar{\beta}^{1}_{q}\}}$~entering Eqs.~\eqref{eq32}.  Regarding the SRFs~$\beta_{q}(\vec r)$, one can either use a set of~$N-1$~oscillator functions~$\phi_{n}(\vec r)$ with~$n\leq \mtc{N}-1$, eigenfunctions~$\beta_{q}^{\textrm{Efn}}(\vec r)$ of the Hamiltonian~$H^{tr}$, or a set of~${N-1}$ linearly independent superpositions of these functions which may be nonorthogonal.  
 As for the functions~$\bar\beta_{q}(\vec r)$, we suggest to use~${N-1}$~functions~$\beta_{q}(\vec r)$~plus an additional function~$\bar\beta_{N-1}(\vec r)$~that guarantees the coupling to the continuum. It is most convenient to set it equal to~$\phi_{\mtc{N}}(\vec r)$ which is coupled to a single function~$\phi_{\mtc{N}-1}(\vec r)$~in the interaction region by the analytic matrix element~$T^{l}_{\mtc{N}-1,\mtc{N}}$.  That is, we choose 
\begin{equation}
\label{eq66}
\bar{\beta}_{q}(\vec r)=\ \begin{cases} 
    \beta_{q}(\vec r), & 0 \leq q \leq N-2 , \\ 
     \phi_{\mtc{N} }(\vec r), &  q=N-1. 
   \end{cases}
\end{equation}
 A natural option is to use  $ \color{black} N-1 \color{black} \le \mtc{N}$ eigenfunctions  $\beta_{q}^{\textrm{Efn}}(\vec r)$ of the truncated Hamiltonian \eqref{eq48}, 
 \color{black}	
\color{black}								
which solve the Schr\"odinger equation
\begin{gather}
H^{tr} \beta_{q}^{\textrm{Efn}}(\vec r) = E_{q} \beta_{q}^{\textrm{Efn}}(\vec r), 
\label{eq53}
\end{gather}
and can be expressed as superpositions of oscillator functions~\eqref{eq3}, 
\begin{equation}
\label{eq54}
\beta_{q}^{\textrm{Efn}}(\vec r)=\sum_{n=0}^{\mtc{N}-1}a^{\textrm{Efn}}_{q n}\phi_{n}(\vec r).
\end{equation}
\color{black}
 In this case, $\mtc{N}$~coefficients~$g_{n}(E)=\sum_{q=0}^{N-2}b_{q}(E)\,a^{\textrm{Efn}}_{qn}$
in Eq.~\eqref{eq50} are expressed through~${v=N-1}$~unknown energy-dependent coefficients~$b_{q}(E)$~in the wave function expansion~\eqref{eq1} which should be found within the approach.

The use of the eigenfunctions simplifies the calculation of matrix~$A=A^{1}$:
\begin{equation}
\label{eq55}
A_{q q'}=(E_{q}-E)\delta_{q q'}, \quad q=0,...\,,N-1  ; \quad q'=0,...\,,N-2.
\end{equation}
The choice of eigenfunctions~$\beta_{q}^{\textrm{Efn}}(\vec r)$~as SRFs is also convenient from the point of view of many-body applications in combination with the NCSM or other {\it ab initio} approaches. These approaches conventionally utilize the \mbox{M-scheme}  and thus operate  with a huge number of many-body basis functions (Slater determinants) that do not have a definite angular momentum~$J$.  Therefore, to simplify applications of \color{black} the adapted \color{black}  Efros method, it is preferable to organize superpositions of these basis functions with a definite~$J$~value.  The available NCSM codes like MFDn~\mbox{\cite{MFDnMaris2010,Euro-Par,Aktulga,NCSM-NTSE2014,Shao}} and \mbox{BIGSTICK}~\cite{BIGSTICK} provide the NCSM eigenenergies and eigenfunctions with definite~$J$ values obtained dynamically during the Hamiltonian diagonalization. 
So, the most natural and simple way to arrange definite~$J$~superpositions of \mbox{M-scheme} Slater determinants is to use the NCSM eigenfunctions corresponding to eigenenergies in the energy range of interest.

We note that in the case of~$N=\mtc{N}$ (and, generally speaking, with~$N>\mtc{N}$)
the version of the Efros formalism under discussion is equivalent to the HORSE formalism~\cite{Zaitsev1998, BangHORSE,Yamani},  though these formalisms utilize different equations for calculating phase shifts, wave functions, etc. The advantage of \color{black} the adapted \color{black} Efros approach is that it can use a smaller number~$N$~of SRFs~$\bar{\beta}_{q}(\vec r)$~than the dimensionality of the truncated Hamiltonian matrix~$\mtc{N}$.  The HORSE  formalism provides exact solutions for the Hamiltonian Eq.~\eqref{eq65} and its multichannel extensions including their versions with smoothed interactions [see Eqs.~\eqref{eq67} and \eqref{tilH2channel} below]. The Hamiltonian~\eqref{eq65} is an approximation since the potential energy is truncated in the oscillator space, however the accuracy of this approximation is improved by increasing the truncation boundary~${\cal N}$. \color{black} The adapted \color{black} Efros formalism with~$N<{\cal N}~$~suggests an approximation of the HORSE and generates solutions which converge to the HORSE solutions with increasing~$N$.  We shall see below that one can obtain well-converged results with~$N<\mtc{N}$ and even with~$N\ll \mtc{N}$ in some~cases.  \color{black} This feature is especially important because the HORSE method is impractical for many-body problems because it requires the calculation of all eigenstates with a definite $J$ value and $0s$ center-of-mass motion.  The number of such eigenstates is very large and is unknown.  See Ref.~\cite{PengYinPLB2024} for a discussion of the number of $1^{-}$ $^4\textrm{He}$ eigenstates with $0s$ center-of-mass motion up to energies of 100 MeV above the ground state calculated within NCSM.\color{black}

 In SM \color{black} \cite[Section II]{Supplemental}, we examine first
the $^1S_{0}$~$NN$~phase shifts within the HORSE method utilizing the Hamiltonian~\eqref{eq65}~and investigate their convergence with~$N_{\textrm{max}}$~at various~$\hbar \Omega$.  We find that the
complete convergence is attained at~$N_{\textrm{max}}=24$~in a wide range of energies and that~$\hbar \Omega= 30$~MeV is
reasonable for our studies of~$np$~scattering in what follows, see \color{black} Ref. \color{black} \mbox{\cite[Figs.~S3 and~S4]{Supplemental}}.

To improve convergence, we smooth the potential truncation as suggested in
Refs.~\cite{HungOne,HungTwo} by replacing in Eq.~\eqref{eq65}
the potential energy operator~$\mathop{\sum}\limits_{n,n'=0}^{\mtc{N}-1}|\phi_{n}\rangle V_{nn'}\langle\phi_{n'}|$ by
\begin{gather}
\label{eq67}
\sum_{n,n'=0}^{\mtc{N}-1}|\phi_{n}\rangle \tilde{V}_{n n'}\langle \phi_{n'} |=
\sum_{n,n'=0}^{\mtc{N}-1} |\phi_{n}\rangle \sigma_{\mtc{N}-1}^{n} V_{n n'} \sigma_{\mtc{N}-1}^{n'} \langle \phi_{n'} |,
\end{gather}
where
\begin{equation}
\label{eq68}
\sigma_{\mtc{N}-1}^{n}=
\frac{1-e^{-\left(\!a\frac{n-\mtc{N}}{\mtc{N}}\!\right)^{\hspace{-.7pt}2}}}{1-e^{-a^{2}}}
\end{equation}
and~$a$~is a dimensionless parameter.~The recommended values for the parameter~$a$~are from
the interval~${a=(2.5,10)}$.  We find that~$a=2.5$ is sufficient for our purposes, see \color{black} Ref. \color{black} \cite[Fig.~S5]{Supplemental}. \color{black}
The smoothing results in an excellent convergence of the phase shifts within 
the HORSE method at~$N_{\textrm{max}}=14$ and~$\hbar \Omega=30$~MeV as opposed to~$N_{\textrm{max}}=24$ needed for convergence without smoothing at the same~$\hbar \Omega$,   see \color{black} Ref. \color{black} \cite[Fig.~S6]{Supplemental} \color{black} for more details.  \color{black}

As it was pointed out above, \color{black} the adapted \color{black} Efros approach suggests an approximation of the
HORSE method that seems to be reasonable for many-body applications.  We therefore examine the~$^1S_{0}$~phase shifts within \color{black} the adapted \color{black} Efros method utilizing the Hamiltonian~\eqref{eq65}.  As is demonstrated in \color{black} Ref.  \color{black} \cite[Fig.~S7]{Supplemental}, \color{black}
 the harmonic oscillator SRFs \color{black} \cite[Eq.~(S5)]{Supplemental} \color{black} and the eigenfunction SRFs are comparable in terms of convergence, but { in what follows we shall concentrate mostly on the case of} the eigenfunction SRFs since they are more viable for many-body applications and suggest a more natural connection to the NCSM.  We also note that smoothing improves convergence within \color{black} the adapted \color{black} Efros method}, as seen in \color{black} Ref. \color{black} \cite[Figs.~S8 and~S9]{Supplemental}. \color{black}

\begin{figure}[t!]
{%
  \includegraphics[width=1\columnwidth]{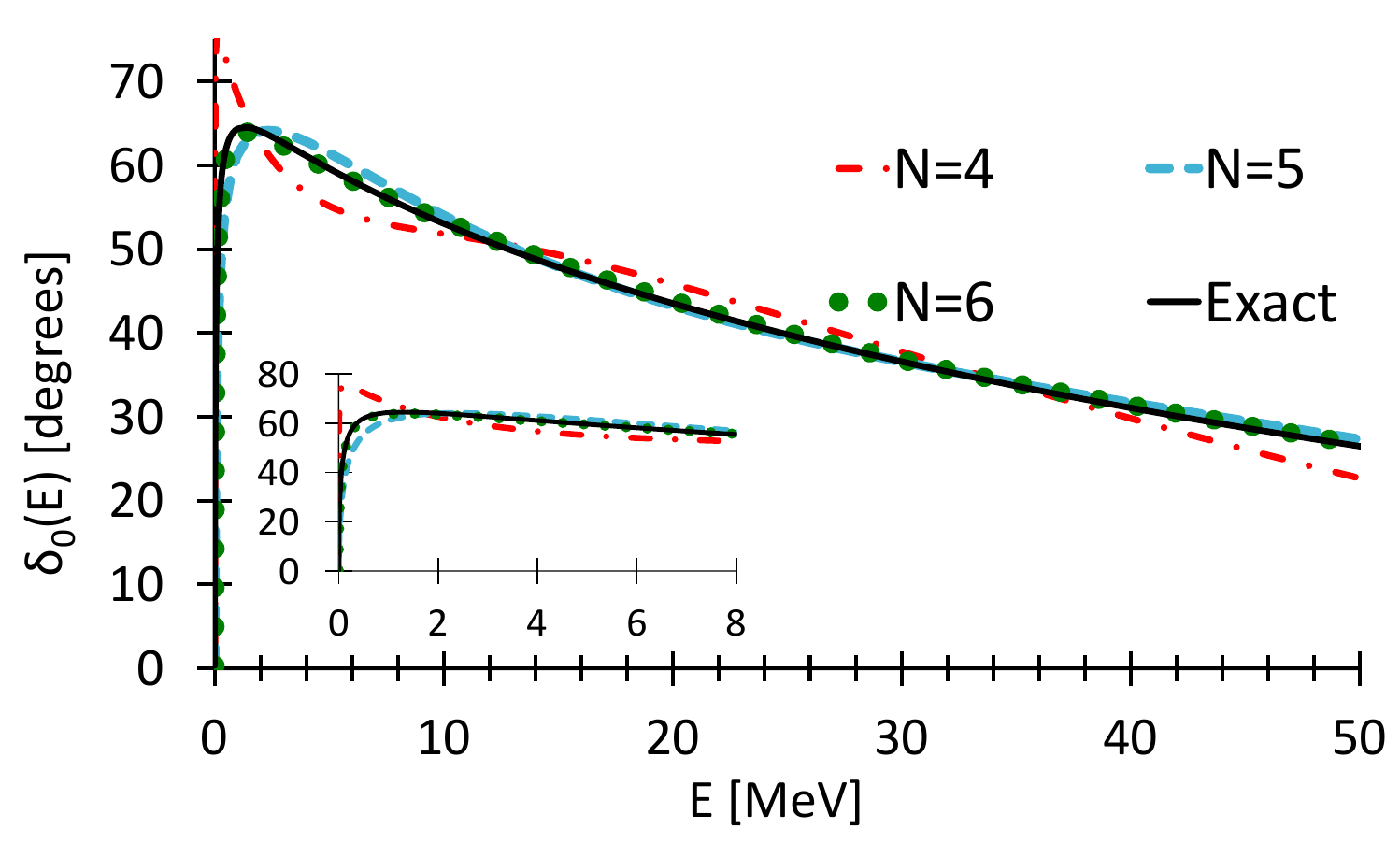}%
}
\caption[]{$^1S_{0}$  phase shifts obtained by \color{black} the adapted \color{black} Efros method with 
various~$N$ at $N_{\textrm{max}}=14$ and~$\hbar \Omega=30$~MeV using the eigenfunction SRFs 
and smoothing with parameter~$a=2.5$.
}
\label{fig9}
\end{figure}

In Fig.~\ref{fig9}, we investigate the phase shift convergence with~$N$~for~$N_{\textrm{max}}=14$~and~$\hbar \Omega=30$~MeV.  We obtain reasonable $^1S_{0}$ phase shifts with~$N=5$  at~$N_{{\max}}=14$ while~$N=6$ provides very accurate results for~$\delta_{0}(E)$
to at least~$E=50$~MeV.  We note also \color{black} (see \color{black} Ref. \color{black}~\mbox{\cite[Fig.~S10]{Supplemental}}) \color{black} that the proposed approach provides the correct scattering wave function at the energy~${E=0.5}$~MeV where the $^1S_{0}$ phase shift obtained with smoothing at~$N_{\max}=14$ 
and~$\hbar \Omega=30$~MeV is close to the exact.  The smoothing results in insignificant changes of the amplitudes~$d_{n}(E)$ of the order of the line thickness
if the phase shifts obtained with smoothing are very close to those obtained without the smoothing, but the
amplitudes~$d_{n}(E)$, of course, differ in the case when the smoothing results in the changes of the
phase shift values.

\color{black}
In some cases, it may be useful to consider a basis that mixes the harmonic oscillator SRFs and the eigenfunction SRFs. For example, in many-body applications, it may happen that only a few lowest eigenstates are in the energy range of interest and those higher in energy do not significantly contribute to the low-energy phase shifts. In this case,
it seems reasonable to introduce a few additional harmonic oscillator SRFs describing the relative motion of colliding fragments in the channel of interest.  We present in \color{black} Ref. \color{black} \cite[Figs.~S11~and~S12]{Supplemental} \color{black} phase shift convergence with a couple of sets of nonorthonormal linearly independent SRFs \color{black} \cite[Eqs.~(S8) and (S9)]{Supplemental}. \color{black}

\quad

\section{Two-Channel Scattering }
\label{IV}

We have shown that \color{black} the adapted \color{black}  Efros method is consistent and leads to a convergence
{\clg of} 
single-channel {\clg scattering calculations using} 
a limited basis space {\clg obtained by diagonalizing the truncated Hamiltonian~$H^{tr}$}.  We now extend our application to a system with two open channels.  In particular, we investigate~$np$~scattering in the coupled~$^3SD_{1}$~partial waves {\clg using the 
Daejeon16 $NN$ potential and eigenfunction SRFs.}

\subsection{\clg\boldmath Hamiltonian, $S$ matrix and Stapp parametrization for~$np$~scattering in coupled partial waves}

{\clg The Hamiltonian of the $np$ system in the coupled partial waves}
can be written as
\begin{widetext}
\begin{multline}
\label{eq72}
H={\clg\sum_{i,j=1,2}}\sum_{n=0}^{\mtc{N}_{i}-1}\sum_{n'=0}^{\mtc{N}_{j}-1}|\phi_{ni}\rangle V_{ni,n'j}\langle\phi_{n'j}|
+{\clg\sum_{i =1,2}}\:\sum_{n,n'=0}^{\infty}|\phi_{n i}\rangle T^{l_{i}}_{nn'}\langle\phi_{n'i}| 
\equiv {\clg\sum_{i,j=1,2}} \sum_{n=0}^{\mtc{N}_{i}-1}\sum_{n'=0}^{\mtc{N}_{j}-1}|\phi_{ni}\rangle H^{tr}_{ni,n'j}\langle\phi_{n'j}|
\\
+\sum_{i=1,2}\:\sum_{n,n'=\mtc{N}_{i}}^{\infty}|\phi_{n i\!}\rangle T^{l_{i}}_{nn'}\langle\phi_{n'i}\!|\!
+\sum_{i=1,2}\!\big[|\phi_{\mtc{N}_{i}-1,i\!}\rangle T^{l_{i}}_{\mtc{N}_{i}-1,\mtc{N}_{i}}\!\langle\phi_{\mtc{N}_{i}i}\!|
+|\phi_{\mtc{N}_{i}i}\rangle T^{l_{i}}_{\mtc{N}_{i},\mtc{N}_{i}-1}\langle\phi_{\mtc{N}_{i}-1,i}|\big]\!.
\end{multline}
{We define {\clg the truncation of  the potential energy matrix by}
the maximal oscillator quanta~{\clg$N_{\max}$,  i.\,e., ${2n+l_{i} }\leq N_{\max}$
and~$\mtc{N}_{i}=\frac12 (N_{\max}-l_{i})+1$, $i=1,2$.} 
}

The 
smoothing {\clg of the interaction can be} 
also applied {\clg in the two-channel case by introducing the potential energy operator}  
\begin{gather}
\label{eq73}
\tilde{V}\equiv \sum_{i,j=1,2}\sum_{n=0}^{\mtc{N}_{i}-1}\sum_{n'=0}^{\mtc{N}_{j}-1}|\phi_{ni}\rangle \tilde{V}_{ni, n'j}\langle \phi_{n'j} |=
\sum_{i,j=1,2}\sum_{n=0}^{\mtc{N}_{i}-1}\sum_{n'=0}^{\mtc{N}_{j}-1} |\phi_{ni}\rangle \sigma_{\mtc{N}_{i}-1}^{n} V_{n i, n'j} \sigma_{\mtc{N}_{j}-1}^{n'} \langle \phi_{n'j} |.
\end{gather}
\end{widetext}
{\clg Smoothing also improves the
convergence in the two-channel case and we present below results of calculations with the smoothed
interactions.} {\clb That means that we use the truncated Hamiltonian
\begin{gather}
H^{tr}\equiv \tilde{H}^{tr}
={\clg\sum_{i,j=1,2}} \sum_{n=0}^{\mtc{N}_{i}-1}\sum_{n'=0}^{\mtc{N}_{j}-1}
|\phi_{ni}\rangle H^{tr}_{ni,n'j}\langle\phi_{n'j}|
\label{Htr2ch}
\end{gather}
with matrix elements
\begin{gather}
H^{tr}_{ni,n'j}\equiv\tilde{H}^{tr}_{ni,n'j}=
\langle\phi_{ni}|T+\tilde{V}|\phi_{n'j}\rangle, \quad n<{\cal N}_{i},\ n'<{\cal N}_{j}.
\label{tilH2channel}
\end{gather}
}

\color{black}
For the two-channel problem, it is convenient to solve Eq.~\eqref{eq21} in the~$S$-matrix representation.  We {\clg define the set of SRFs~$\{\beta_{q}\}$ as 
a set of~$N-2$ eigenfunctions of~$H^{tr}\!$,
\begin{gather}
{\beta}_q(\vec r)=\beta_{q}^{\textrm{Efn}}(\vec r), \quad q=0,...,N-3. 
\label{betaq-2ch}
\end{gather}
{\clb{Note,}} $N=v+2$ is the number of unknowns in {\clb each} 
channel and 
the total number
of independent unknowns  is~${{\mathbb{\clb N}}=2N-1\clb=2v+3}$. 
\color{black}
{\clb We} first solve the full set of equations for the~$i=1$ entrance channel with the 
set~{\clg$\{\bar{\beta}_q^1\}$} constituting a set of~$N-2$ eigenfunctions of~$H^{tr}$ 
{\clg plus} 
two additional oscillator functions that are used exclusively for orthogonalization and guarantee coupling to the continuum but do not appear explicitly in Eq.~\eqref{eq1}: 
\begin{equation}
\label{eq74}
{\clg\bar{\beta}_q^1(\vec r)} 
=\begin{cases} 
    \beta_{q}^{\textrm{Efn}}(\vec r), & q=0,...,N-3,\\ 
     \clg \phi_{\mtc{N}_{1} 1}(\vec r), &  q=N-2, \hspace{1mm} \\
     \clg \phi_{\mtc{N}_{2} 2}(\vec r),  &  q=N-1. \hspace{1mm} 
   \end{cases} 
\end{equation}
Here, 
\begin{multline}
\label{eq75}
\beta_{q}^{\textrm{Efn}}(\vec r)=\sum_{q'=0}^{\mtc{N}_{1}-1}a_{q q'}^{l_{1}}\,{\clg\phi_{q'1}(\vec r)}
	+\sum_{q'=0}^{\mtc{N}_{2}-1}a_{q q'}^{l_{2}}\,\clg\phi_{q'2}(\vec r) .
\end{multline}

Next, we solve the reduced set of equations for the ${i=2}$ entrance channel.  Since there is one less unknown, namely, the off-diagonal~$S$-matrix 
element {\clg${\mathbb{S}_{12}(E)=\mathbb{S}_{21}(E)}$}, we use 
{\clg for the set~$\{\bar{\beta}_q^2\}$}
a subset of~{\clg$\{\bar{\beta}_q^1\}$ that includes} 
one less function. 
The choice of {\clg the} function to be omitted 
{\clg for the orthogonalization in the~$i=2$} channel 
is arbitrary.  We can 
omit{\clg{, e.\,g.,}}~$\beta_{N-3}^{\textrm{Efn}}(\vec r)$,  and 
{\clg define the set~$\{\bar{\beta}_q^2\}$ as}
\begin{equation}
\label{eq76}
{\clg \bar{\beta}_{q}^{2}(\vec r)}=
\begin{cases} 
    \beta_{q}^{\textrm{Efn}}(\vec r), & q=0,...,N-4, \\ 
 \clg     \phi_{\mtc{N}_{1} 1}(\vec r), &  q=N-3, \hspace{1mm} \\
 \clg     \phi_{\mtc{N}_{2} 2}(\vec r), &  q=N-2. \hspace{1mm} \\
   \end{cases}
\end{equation}
We note that the set of SRFs~{\clg$\{\beta_q\}$} appearing explicitly in Eq.~\eqref{eq1} and 
{\clg defined by Eq.~\eqref{betaq-2ch}}
is identical {\clg{in}} 
both 
channels.


Since we {\clg have specified the SRFs,}
we can 
 write {\clg down} the coefficients $d_{n,ij}(E)$ \clg of the continuum
wave function~$u_{i}(E;\vec r)$ expansion in oscillator functions [see Eq.~\eqref{eq11}].~For the~$i=1$~entrance channel, {\clg the wave function~$u_{1}(E;\vec r)$ has}
a ``strong'' 
component $\clg\sum_{n=0}^{\infty}d_{n,11}(E)\,\phi_{n1}(\vec r) $ 
{\clg in the channel~1} and a ``weak'' 
component~$\clg\sum_{n=0}^{\infty}d_{n,12}(E)\,\phi_{n2}(\vec r)$ 
{\clg in the channel~2}
obtained from the set of~$i=1$ linear equations. {\clg The respective expansion
coefficients are} 
\begin{gather}
\label{eq77}
d_{n,11}(E)\!=\!
\begin{cases} 
    a_{n,11}^{as}(E)+g_{n,11}(E), & n=0,...\,, \mtc{N}_{1}-1,  \\ 
      a_{n,11}^{as}(E),&  n=\mtc{N}_{1},...\,,\infty \hspace{1mm} 
   \end{cases}
\end{gather}
and
\begin{gather}
\label{eq80}
d_{n,12}(E)\!=\! 
\begin{cases} 
    a_{n,12}^{as}(E)+g_{n,12}(E), \!\!&  n=0,...\,,\mtc{N}_{2\!}-1,  \\ 
      a_{n,12}^{as}(E),&  n=\mtc{N}_{2},...\,,\infty.
   \end{cases}
\end{gather}
For the~$i=2$ entrance channel,  {\clg the wave function $u_{2}(E;\vec r)$ has} 
a ``weak" 
component $\clg\sum_{n=0}^{\infty}d_{n,21}(E)\,\phi_{n1}(\vec r)$ 
{\clg in the channel~1} and a ``strong" 
component $\clg\sum_{n=0}^{\infty}d_{n,22}(E)\,\phi_{n2}(\vec r)$ 
{\clg in the channel~2 with the~expansion
coefficients}
\begin{gather}
\label{eq82}
d_{n,21}(E)\!=\! 
\begin{cases} 
   a_{n,21}^{as}(E)
    +g_{n,21}(E), \!\!& n=0,...\,,\mtc{N}_{1}-1,  \\ 
      a_{n,21}^{as}(E), &  n=\mtc{N}_{1},...\,,\infty 
   \end{cases}
\end{gather}
and
\begin{gather}
\label{eq84}
d_{n,22}(E)=
\begin{cases} 
    a_{n,22}^{as}(E)
    +g_{n,22}(E), & n=0,..., \mtc{N}_{2}-1,  \\ 
      a_{n,22}^{as}(E),&  n=\mtc{N}_{2},...,\infty.   \hspace{1mm} 
   \end{cases}
\end{gather}
{\clg In Eqs.~\eqref{eq77}--\eqref{eq84},
\begin{align}
\label{anij}
\!\!a_{n,ij}^{as}(E)&=C^{-}_{n l_{i}}(k)\,\delta_{ij}-\mathbb{S}_{ji}(E)\,C^{+}_{n l_{j}}(k),
						\!\!&&i,j=1,2\\
\label{gnij}
\!\!g_{n,ij}(E)&=\sum_{n'=0}^{v-1} b_{n'}^{i}(E)\, a_{n' n}^{l_{j}},
						\!&&i,j=1,2.
\end{align}
}
{\clg In the case of the $np$ scattering, it is conventional to express the $S$ matrix
in the Stapp parametrization~\cite{Stapp} where
the phase shifts~$\delta_{l_{i}}=\delta_{l_{i}}(E)$ are introduced in each of the 
partial waves and 
the mixing parameter~$\epsilon=\epsilon(E)$. These 
parameters are related  
to the $S$ matrix as}  
\begin{equation}
\label{eq71}
\mathbb{S}(E)=
\begin{pmatrix}
   e^{2i \delta_{l_{1}}}\textrm{cos}(2 \epsilon) & i e^{i(\delta_{l_{1}}+\delta_{l_{2}})}\textrm{sin}(2 \epsilon)  \\[0.3em]
   i e^{i(\delta_{l_{1}}+\delta_{l_{2}})}\textrm{sin}(2 \epsilon) & e^{2i \delta_{l_{2}}}\textrm{cos}(2 \epsilon)  
     \end{pmatrix} \!. 
\end{equation}

\subsection{Two-channel results}

We start with discussion of calculations of the phase shifts and mixing parameter~$\epsilon$ 
in the ${^{3}SD}_{1}$ coupled partial waves of $np$ scattering with $N_{\textrm{max}}=12$, $\hbar \Omega=30$~MeV, and smoothing parameter~$a=2.5$.  As is shown in \color{black} SM \color{black}, see \color{black} Ref. \color{black} \cite[Fig.~S13]{Supplemental}, \color{black} the HORSE method provides in this case
reasonably accurate results for a wide range of energies.

\color{black} For \color{black} the ``complete'' set of SRFs~$\{\beta_{q}\}$,
 \color{black} the adapted \color{black} Efros and HORSE methods give the same results. In this case, one can use the
over-complete Eq.~\eqref{eq25} which suggests additional relations for independent
calculations of the unknowns that should be equal 
such as~$\mathbb{S}_{12}(E)=\mathbb{S}_{21}(E)$ in the case of two-channel scattering.  \color{black} See \cite[Section III]{Supplemental} for an example in which we compute the $S$-matrix elements at the energy of $E=0.5$ MeV using the ``complete" set of SRFs and get $\mathbb{S}_{12}(E)=\mathbb{S}_{21}(E)=-0.0022-0.0035i$. 

\color{black} Since the ``complete'' set of SRFs in
current {\it ab initio} nuclear structure applications is extremely large, it is more interesting to investigate the essential simplification suggested within \color{black} the adapted  \color{black} 
Efros method as compared to the HORSE method which makes it possible to use a reduced set of SRFs.  In \color{black} the \color{black} case \color{black} of a reduced set of
SRFs~$\{\beta_{q}\}$, \color{black} Eq.~\eqref{eq74} results in the $S$ matrix with broken unitarity and $\mathbb{S}_{12}(E) \neq \mathbb{S}_{21}(E)$.  \color{black} Therefore, \color{black} one is pushed to use Eq.~\eqref{eq31} and impose the $S$-matrix symmetry either by calculating~$\mathbb{S}_{21}(E)$ and setting~$\mathbb{S}_{12}(E)$ equal to it or vice versa.  \color{black} We present an example in \color{black} Ref.~\cite[Section III]{Supplemental} \color{black} in which we compute the $S$-matrix elements at the energy of $E=0.5$ MeV for $N=12$, where we get $\mathbb{S}_{12}(E)=-0.0021-0.0033i$ and $\mathbb{S}_{21}(E)=-0.0016-0.0025i$.
\color{black}

\begin{figure}[t!]
\subfloat[]{%
  \includegraphics[width=1\columnwidth]{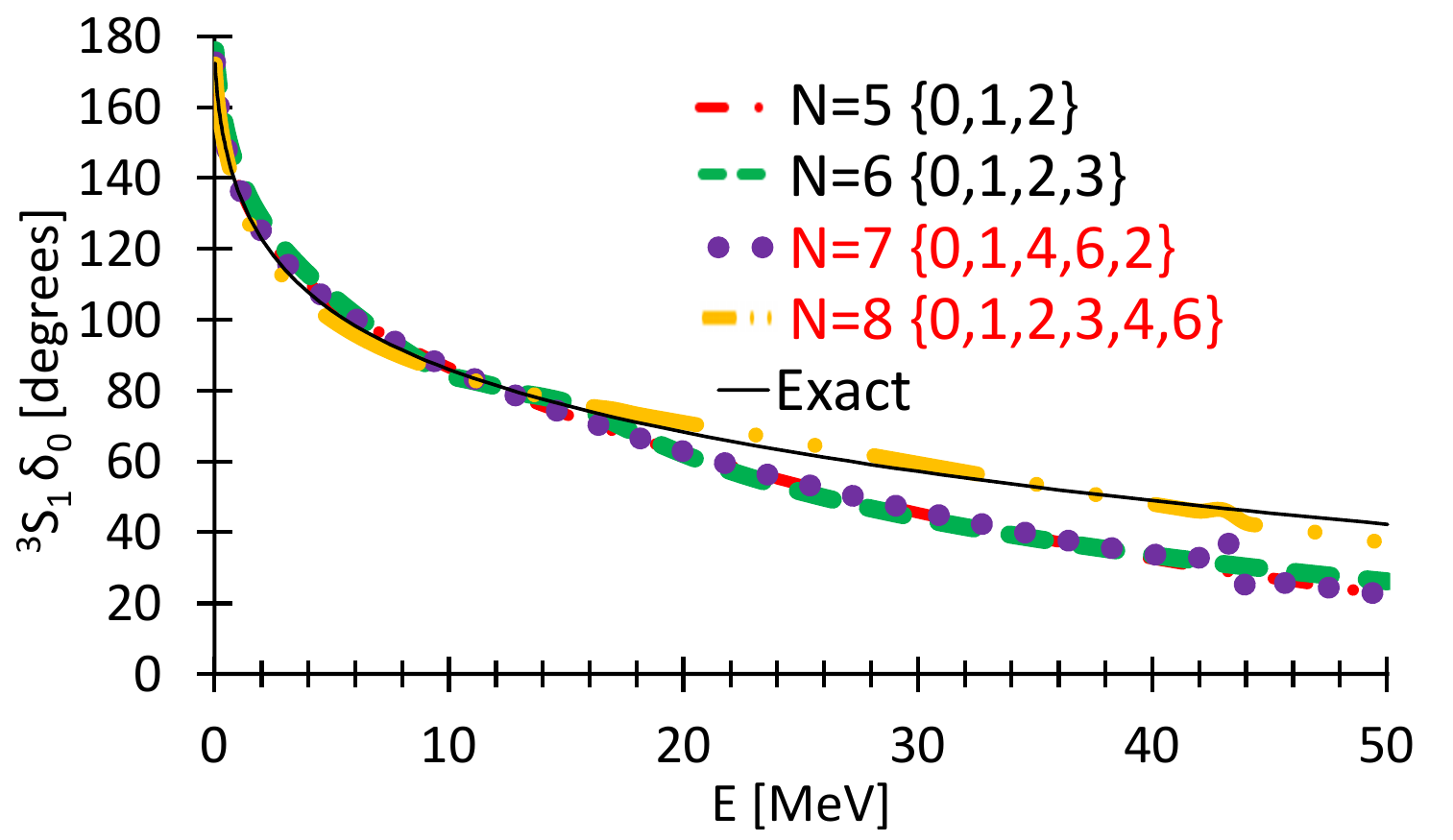}%
}\hfill
\subfloat[]{%
  \includegraphics[width=1\columnwidth]{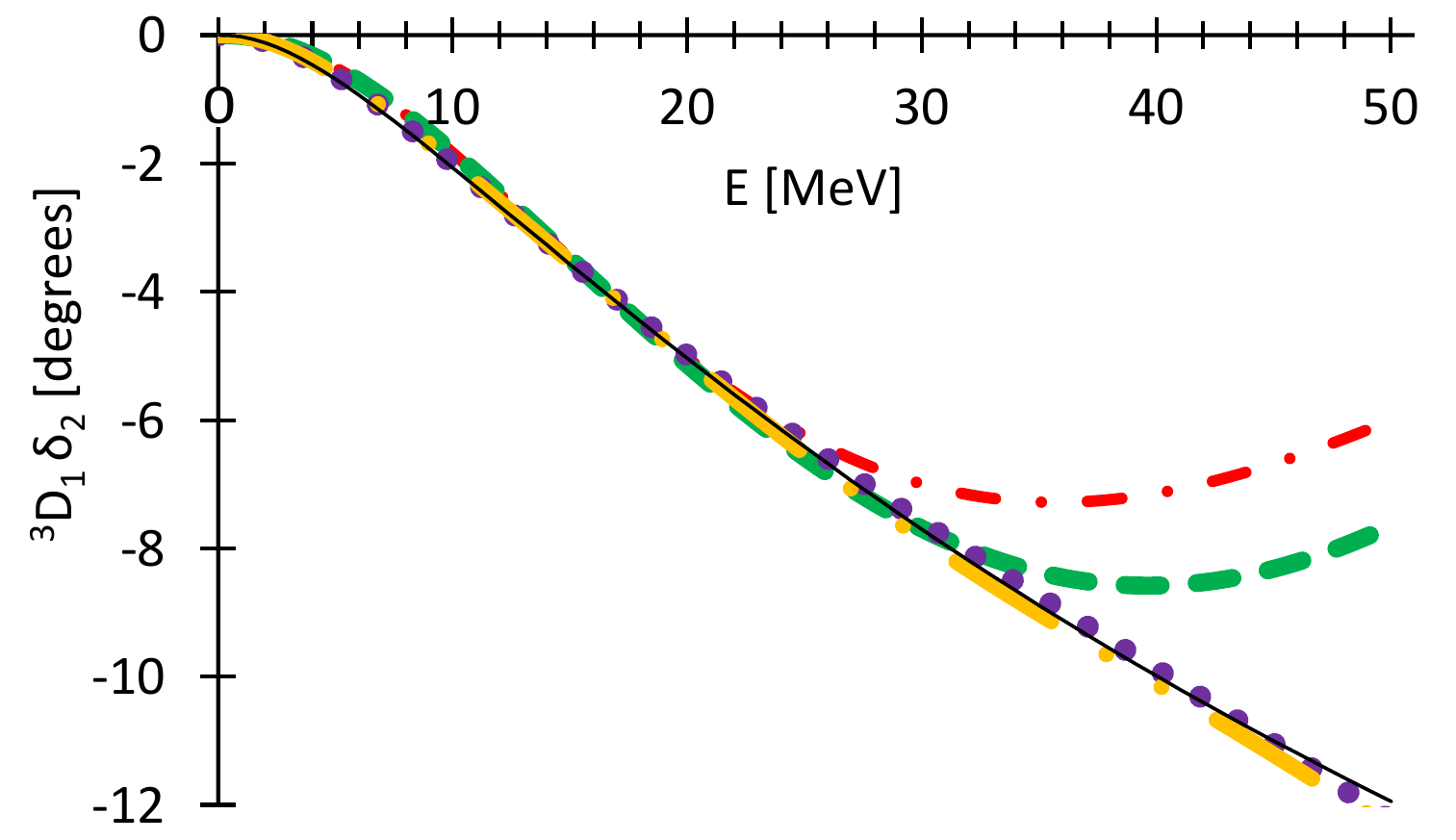}%
}\hfill
\subfloat[]{%
  \includegraphics[width=1\columnwidth]{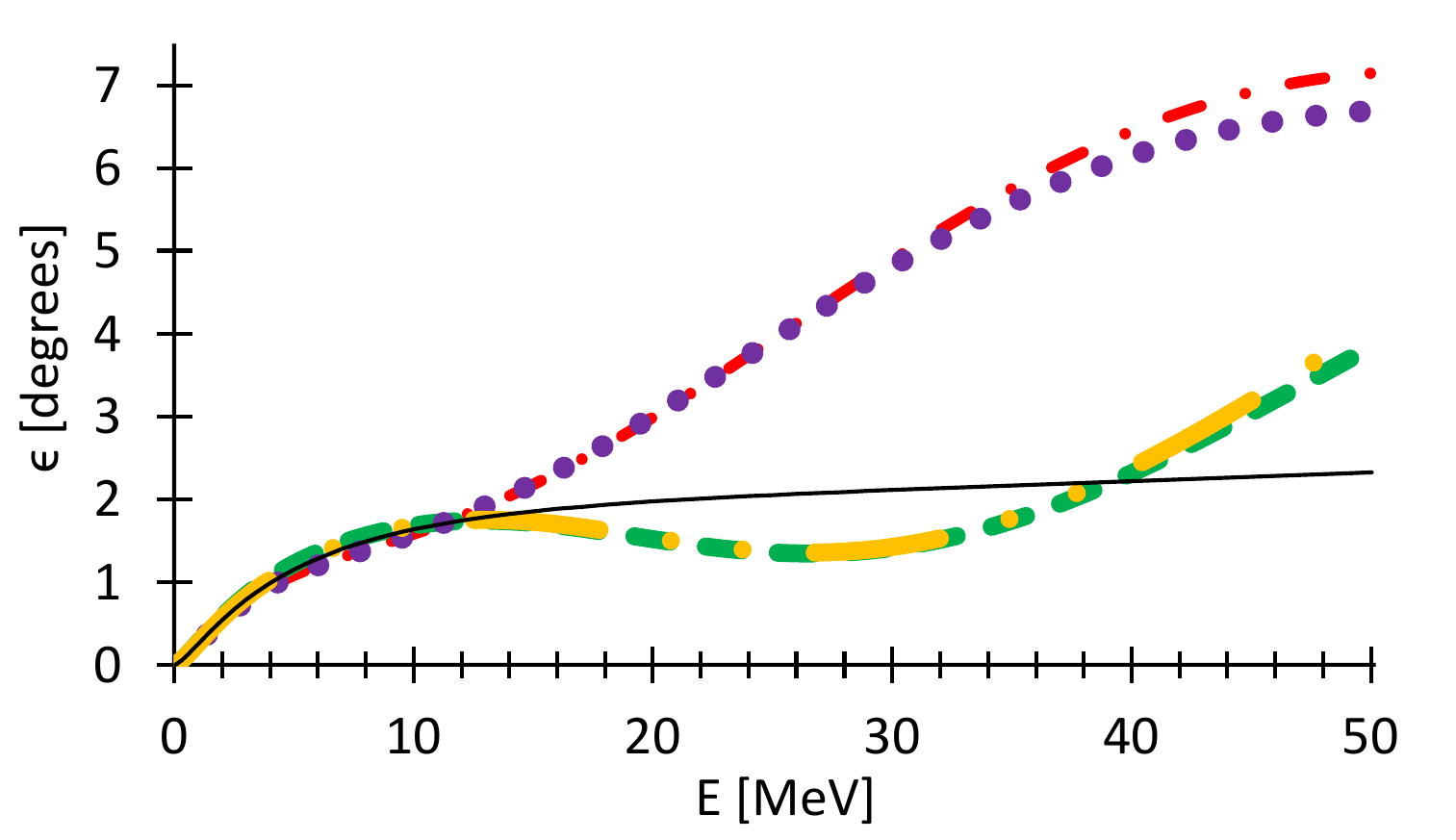}%
}\hfill
\caption[]{\clg a) $^3S_{1}$ 
{\clg phase shifts}~$\delta_{0}$,  b) $^3D_{1}$ 
{\clg phase shifts}~$\delta_{2}$, c)~$^3SD_{1}$ {\clg mixing parameter}~$\epsilon$
{\clg obtained with different}~$N$, $N_{\textrm{max}}=12$, 
$\hbar \Omega=30$~MeV, {\clg and}
with smoothing  
{\clg parameter} ${a=2.5}$ using the eigenfunction {\clg SRFs and} 
$^3D_{1}$ {\clg as the first  channel with~$i=1$} 
 and rearranging the set of SRFs to eliminate 
 {\clg the} vanishing {\clg point in the} determinant of~$A$.  This elimination is checked over a range of~$E$~from 0 to 50 MeV. 
}
\label{fig18}
\end{figure}

In the case of a limited set of SRFs, we use
the~$^3D_{1}$
{as the first channel corresponding to~$i=1$~and set~$l_{1}=2$ 
and~$l_{2}=0$}
[i.\,e., first solving { the complete set of the}~$^3D_{1}$ channel~equations, 
{obtaining~$\mathbb{S}_{12}(E)$,}~and~next 
solving the reduced set of the~$^3S_{1}$ {equations
imposing the symmetry by setting~$\mathbb{S}_{12}(E)$ 
 equal to~$\mathbb{S}_{21}(E)$}].  These results differ from those obtained using
 the~$^3S_{1}$~as the first channel corresponding to~$i=1$ and setting~$l_{1}=0$~and~$l_{2}=2$
(i.\,e., {solving first the complete set of equations in}  the~$^3S_{1}$ {partial wave}
and { next switching to} 
the reduced { set of equations in} the~$^3D_{1}$ {partial wave).  We present here the first choice since in this case the
convergence is more stable, though both choices converge to the HORSE method result when the sets of SRFs are ``complete''.  \color{black} In Ref.~\cite[Figs.~S14~and~S16]{Supplemental}), we present \color{black} the results using  $i=1$ as the $^3S_{1}$ channel.  \color{black}

\color{black} In Fig.~\ref{fig18}, we present the convergence of the results with $N$. \color{black}  The results for the phase shifts and the mixing parameter~$\epsilon$~presented in Fig.~\ref{fig18}~were obtained using $H^{tr}$ eigenfunctions as SRFs.  We prefer to use the lowest-lying eigenfunctions of~$H^{tr}$~in calculations.  Let us introduce the notation \color{black} for SRF selection \color{black} used in the label for Fig.~\ref{fig18} by way of an explicit example.  The $N=5\,\{0,1,2\}$ in the figure legends means that we use 5 unknowns in the~$i=1$ partial wave [3~SRF coefficients~$b^{1}_{q}(E)$, $q=0,1,2$
plus 2 $S$-matrix elements~{$\mathbb{S}_{11}(E)$ and~$\mathbb{S}_{21}(E)$]}, 
4 unknowns in the~$i=2$ partial wave {[3 SRF coefficients~$b^{2}_{q}(E)$, $q=0,1,2$}
plus 1 $S$-matrix element~{$\mathbb{S}_{22}(E)$], that is} 
a total of~$\mathbb{N}=9$ independent unknowns, and take as the SRFs the lowest (ground state) 
eigenfunction~$\beta_{0}^{\textrm{Efn}}(\vec r)$, $\beta_{1}^{\textrm{Efn}}(\vec r)$, and $\beta_{2}^{\textrm{Efn}}(\vec r)$.  We}
 omit~{${\beta}_{2}^{\textrm{Efn}}(\vec r)$} in the set~$\{\bar{\beta}_{q}^{2}\}$ used for the orthonormalization of 
the reduced set of equations { in the~$i=2$ channel}, but not in the  
set~{$\{\bar{\beta}_{q}^{1}\}$ used in the} 
equations
{ in the~$i=1$ channel}.~For $N=7, 8$, we use \color{black} different SRF selections from $N=7\,\{0,1,2,3,4\}$ and $N=8\,\{0,1,2,3,4,5\}$ \color{black} due to \color{black} jumps around $E=3$ MeV \color{black} for $\delta_{2}$ and $\epsilon$.  Instead, we use the SRF selections $N=7\,\{0,1,4,6,2\}$ and $N=8\,\{0,1,2,3,4,6\}$ to get  \color{black} rid of these jumps.   \color{black} See \color{black} Ref. \color{black} \cite[Figs.~S15--S17]{Supplemental} for more details on those numerical issues.  \color{black} Modified sets of SRFs
are indicated by the red colors in the legends.  It is seen that with~$N=8\,\{0,1,2,3,4,6\}$, we obtain an excellent description of both phase shifts~$\delta_{0}$
and~$\delta_{2}$ but some inaccuracy in the description of the mixing parameter~$\epsilon$.

{  \color{black} The description of the phase shifts with increasing~$N$
approaches that of the HORSE method.  The HORSE method 
description of the scattering parameters, in particular, of the mixing parameter~$\epsilon$,
is still not ideal at~$N_{\max}=12$ with~$\hbar\Omega=30$~MeV, and smoothing parameter~$a=2.5$.  Therefore, to}
 {\clg 
obtain a good description of all scattering observables of~$np$ scattering
 in the ${^{3}SD}_{1}$ coupled partial waves in a wide energy range, one needs to perform calculations with
 a larger~$N_{\max}$ value. \color{black} We present in \color{black} Ref. \cite[Fig.~S18]{Supplemental} \color{black} the results of calculations with~${N_{\max}=20}$, with the rest of the parameters the same as in \color{black} Fig.~\ref{fig18}, \color{black} where it is seen that we get an excellent description of all scattering observables
up to the energy of 50~MeV if we use a large enough set of SRFs~$N=11\,\{0,1,2,3,4,5,6,7,8\}$.  Convergence for the expansion coefficients of the scattering wave functions occurs in a similar manner \color{black} \cite[Figs.~S19~and~S20]{Supplemental}. \color{black} }
 \color{black}

\section{\protect\clg\boldmath$
{S}$-matrix poles
}
\label{V}
We present and discuss calculations of the~$S$-matrix poles within \color{black} the adapted \color{black} Efros approach
using three simple examples.  {\clg In particular, 
we explore a resonance in~$n \alpha$ scattering using}
 the Woods--Saxon interaction with parameters fitted in Ref.~\cite{WoodsSaxon1979}.
Next, we consider 
 {\clg an $S$-matrix pole associated with a bound state using as an example}
 the deuteron bound state  {\clg supported by} 
 the Daejeon16 {\clg$NN$} interaction.  \color{black} In \color{black} Ref. \color{black} \cite[Section IV]{Supplemental}, we also present convergence of two-channel resonances using a model originally presented in Ref.~\cite{Norro1980} as well as in Refs.~\cite{Moiseyev1990,Rakityansky2006} (see \color{black} Ref. \color{black} \cite[Figs.~S24~and~S25]{Supplemental}).  \color{black}
  We first obtain the results by  the HORSE method presented in \color{black} SM \color{black} \cite[Section IV]{Supplemental}.  \color{black} We discuss in the main text the results obtained within \color{black} the adapted \color{black} Efros method for a given~$N_{\textrm{max}}$~with smoothing.  
\subsection{\clg Single-channel resonance \boldmath$S$-matrix 
pole}

{\clb To  describe the $n \alpha$~scattering with}
the Woods--Saxon interaction  
fitted in Ref.~\cite{WoodsSaxon1979},
we use {\clg the} 
reduced mass of~$m=\frac{4}{5}m_{N}=751.136~
{\textrm{MeV}}/{c^{2}}$, 
${\hbar \Omega=30}$~MeV, and 
the smoothing parameter~$a=5$.  We investigate  {\clg a resonance in}
the~$^2P_{\frac{3}{2}}$ {\clg partial wave by locating  numerically  the
corresponding $S$-matrix pole by solving Eq.~\eqref{eq44} in the complex momentum plane.\color{black}} 
From the HORSE method results shown in \color{black} Ref. \color{black} \cite[Fig.~S21]{Supplemental}, \color{black} we find that at~$N_{\textrm{max}}=10$, {we obtain results very close to the exact values 
of~$E_{r}=0.837$~MeV and~$\Gamma=0.780$~MeV  (see  
Refs.~\cite{AMShirokovHORSE2016, BlokhintsevSSHorse}).  For the case of the Woods-Saxon interaction, the ``exact" method refers to the numerical integration result of the Schrödinger equation such as with the Numerov method.

{\clg In the search for the $S$-matrix pole by \color{black} the adapted \color{black} Efros method, which can use  a smaller 
number of eigenstates, we perform the calculations
with the \color{black} SRF selection \color{black} with increasing eigenvalues~$N\,\{0,1,2,...\}$
at~${ N_{\textrm{max}}=12}$  and compare with the result obtained within the HORSE method. \color{black}
The results are presented in Fig.~\ref{fig22}. Both~$E_{r}$
and~$\Gamma$ exhibit oscillations, but are still reasonably accurate.  In particular, 
the \color{black} energy and width are seen to converge with~$N$ reasonably fast to the exact values,
and starting from~$N=5$ \color{black} differ from the exact values by less than 5\%.\color{black}}

\begin{figure}[t!]
\subfloat[]{%
  \includegraphics[width=1\columnwidth]{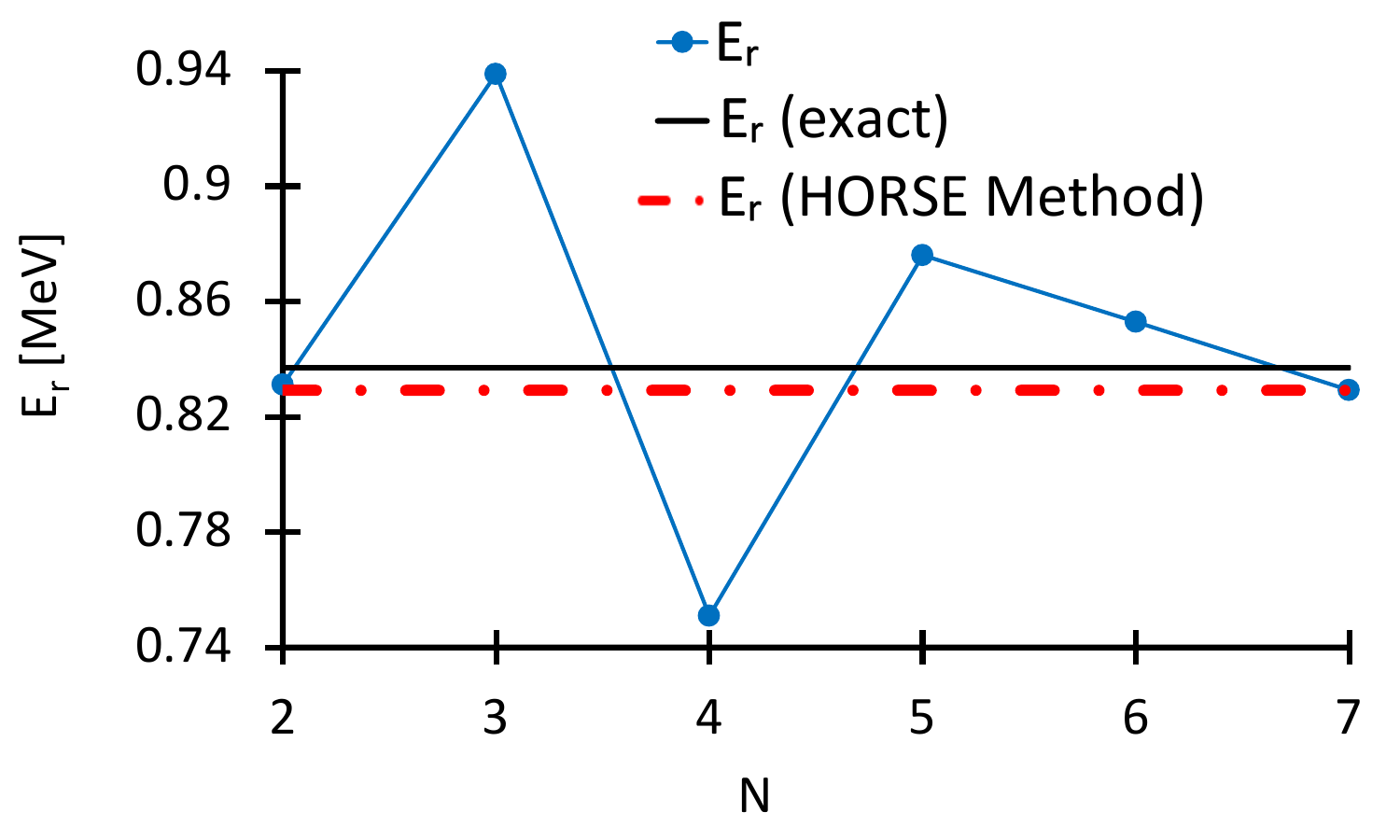}%
}\hfill
\subfloat[]{%
  \includegraphics[width=1\columnwidth]{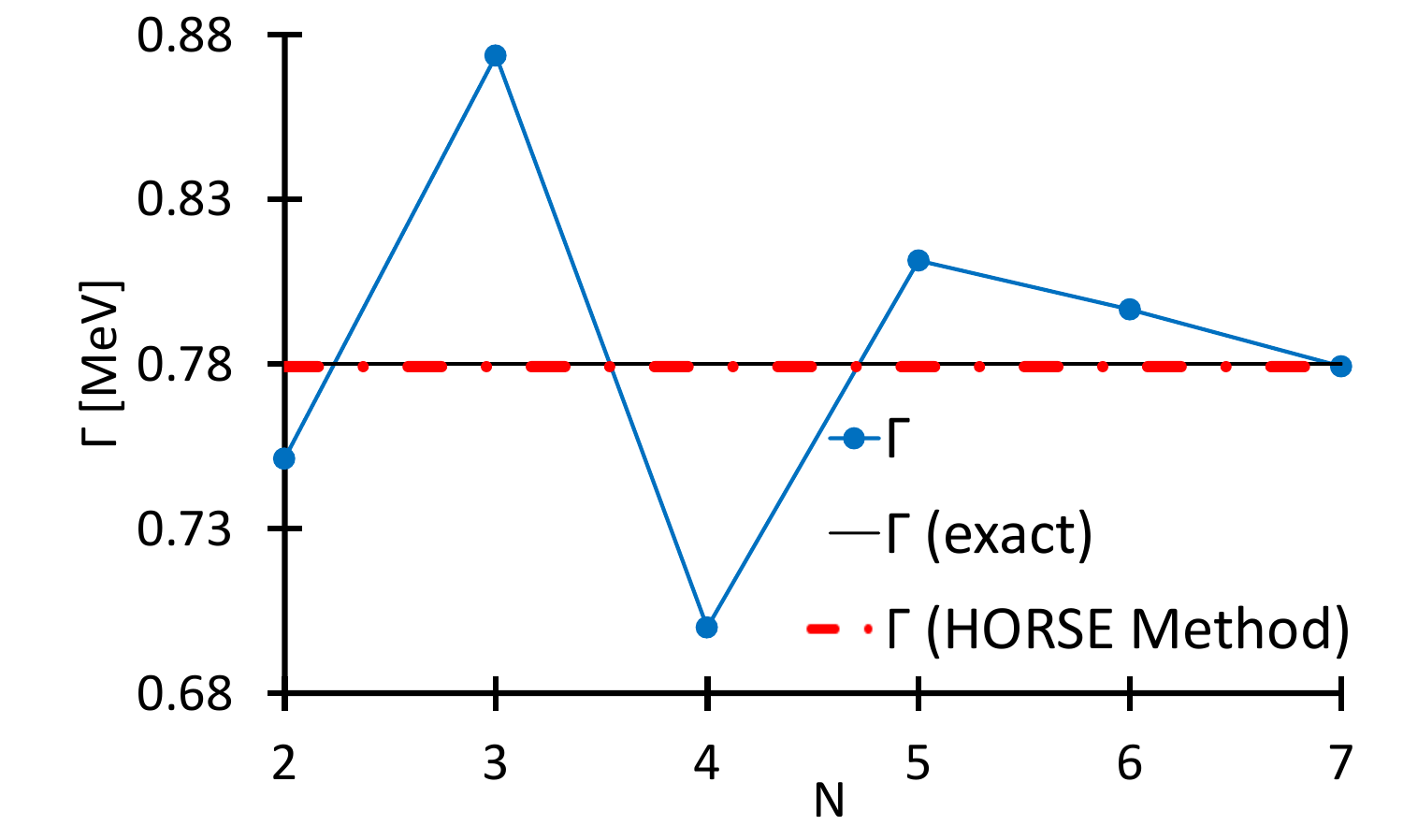}%
}\hfill
\caption[]{\clg $\frac32^{-}$ resonance in~$n\alpha$ scattering supported by the 
Woods--Saxon potential: a) resonance energy~$E_{r}$ and b)~resonance width~$\Gamma$
 obtained by locating the $S$-matrix pole within \color{black} the adapted \color{black} Efros
method with eigenfunction SRFs using various~$N$, {$N_{\max}=12$}, 
 $\hbar \Omega=30$~MeV, {\clg and} 
 smoothing 
 parameter~$a=5$ 
 {\clb in comparison} with the exact {\clb values} 
 and with the HORSE method results. \color{black}
}
\label{fig22}
\end{figure}

\begin{figure}[t!]
  \includegraphics[width=1\columnwidth]{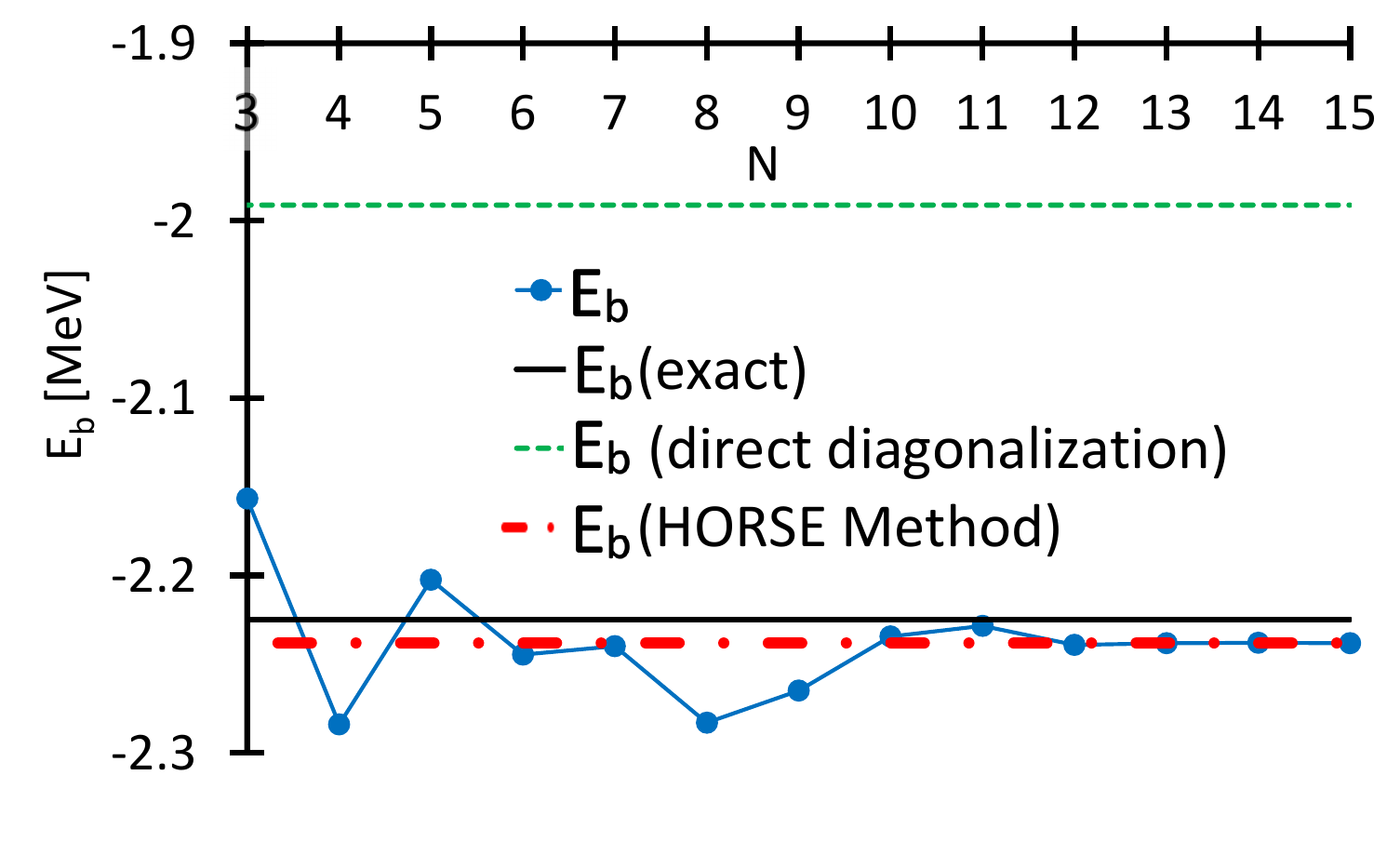}%
\caption[]{\clg
Deuteron bound state energy~$E_{b}$  obtained with Daejeon16 $NN$ interaction 
by locating the $S$-matrix pole via \color{black} the adapted \color{black} Efros method using the eigenfunction SRFs,
various~$N$, $N_{\textrm{max}}=12$, \color{black} 
$\hbar \Omega=30$~MeV, and 
smoothing parameter~
$a=5$ 
 {\clb in comparison}  with the exact 
 {\clb value} and with the HORSE method result obtained with the same~$N_{\textrm{max}}$~and~$\hbar \Omega$.
}
\label{fig24}
\end{figure}

\color{black}
\subsection{\clg Bound state \boldmath$S$-matrix 
pole 
}

We next {\clg search for the $S$-matrix pole in the coupled $^3SD_{1}$ partial waves
associated with the deuteron bound state 
energy~$E_{b}$} 
using the Daejeon16 {\clg$NN$} interaction {\clg smoothed with the parameter~$a=5$.  From the HORSE method results shown in \color{black} Ref. \color{black} \cite[Fig.~S22]{Supplemental}, we find that~$N_{\textrm{max}}=12$}  and \color{black}
$\hbar \Omega=30$~MeV give highly accurate results. \color{black}

{\clg We obtain also a fast convergence by locating the $S$-matrix pole using \color{black} the adapted \color{black}
Efros method which utilizes}
a limited number of {\clg  SRF eigenfunctions with increasing eigenvalues $N\:\{0, 1, 2, ...\}$.
The results for the ground state energy~$E_{b}$ obtained with}~%
  $N_{\textrm{max}}=12$ \color{black}
 {\clg are presented}  in Fig.~\ref{fig24}.  {\clg$E_{b}$ exhibits jumps and instability as~$N$
 increases up to  $N=12$; \color{black} note, however, an extended scale in  Fig.~\ref{fig24}.} 
 In particular, 
 even at small~$N$~starting from~$N=3$, the fluctuations {\clg of~$E_{b}$} never deviate {\clg by} more than  {4\%} \color{black} 
 from the exact value.  The results are much more accurate than those obtained in the variational approach via direct diagonalization.  \color{black}

 {\clg In the case of bound states, there are no entrance (exit) channels and the solution is unique 
and independent from the selection of a channel as the 
first channel.
The expansion of the bound
state wave function in the series of oscillator functions takes the form
\begin{equation}
\label{eqbound}
u(E_{b};\vec r)=\sum_{i=1}^{w}\sum_{n=0}^{\infty}\breve{d}_{n,i}(E_{b})\,\phi_{n{i}}(\vec r). 
\end{equation}
 \color{black} In \color{black} Ref.~\cite[Fig.~S23]{Supplemental}, \color{black} we present~$\breve{d}_{n,i}(E_{b})$~of the deuteron wave function obtained by \color{black} the adapted \color{black} Efros method with different~$N$~using eigenfunction SRFs, 
 { $N_{\max}=12$}, $\hbar \Omega=30$~MeV, and smoothing parameter~$a=5$.  We see a high accuracy of expansion coefficients  calculated even with small~$N$.  \color{black}   \color{black}

We note that by the diagonalization of the truncated Hamiltonian matrix~$H^{tr}$, we obtain
the expansion coefficients~$\breve{d}_{n,i}(E_{b})$ only with~$n\leq \mtc{N}_{i}-1$.  That is, in
our case of~$N_{\max}=12$, we have~$n\leq6$ in the $^3S_{1}$ and~$n\leq5$ \color{black} in the $^3D_{1}$
partial waves. Within the HORSE and \color{black} the adapted \color{black} Efros methods, we obtain 
coefficients~$\breve{d}_{n,i}(E_{b})$ with~$n$ ranging up to infinity in each channel.
The coefficients~$\breve{d}_{n,i}(E_{b})$ with~$n>\mtc{N}_{i}-1$ decrease monotonically and are
responsible for the description of asymptotic tails of the bound state wave 
functions~$A_{i}e^{-\varkappa r\!}$, where~$A_{i}$ is the so-called asymptotic 
normalization constant in the channel~$i$ and~$\varkappa =\frac{\sqrt{2mE_{b}}}{\hbar}$.
The description of the asymptotic tails of
bound state wave functions is important especially in weakly bound systems such as the
deuteron since they contribute to the rms radius of the system, to its 
quadrupole moment, to electromagnetic transitions, etc.
The asymptotic normalization constants~$A_{i}$ are important observables related to the
residue of the $S$-matrix pole~\cite{Baz} which are widely used in calculations of various
reactions.
}

\section{Conclusions}
\label{VI}

{\clg  \color{black} We adapted the Efros method \cite{VDEfros} for calculating various observables in the continuum and discrete spectrum using two-body interactions in single- and coupled-channel systems. \color{black} Additional features are introduced to the \color{black} adapted method with the goal of utilizing it in future many-body applications. \color{black} These features include using achievements developed
within the HORSE formalism~\cite{BangHORSE}, constructing basis wave functions by using different choices of linearly independent superpositions of oscillator functions as SRFs, and utilizing a smoothing scheme suggested in Refs.~\mbox{\cite{HungOne,HungTwo}}. 

\color{black} We demonstrated the utility of these features by investigating \color{black} in detail possibilities of \color{black} the adapted \color{black} Efros method in describing scattering parameters, wave functions, and in locating $S$-matrix poles to obtain resonant energies and widths or accurate bound state energies and wave functions.  To study the convergence and accuracy of \color{black} the adapted \color{black} Efros method, we utilized various two-body systems where the exact results may be obtained by other 
approaches.~In particular, we studied the description of~$np$ scattering in the ${^{1}S}_{0}$
partial wave in the single-channel version of the Efros approach and in the ${^{3}SD}_{1}$
coupled partial waves in its two-channel version using the realistic Daejeon16 $NN$
interaction.~To investigate the location of $S$-matrix poles, we explored the~$\frac32^{-}$
resonance in~$n\alpha$ scattering described by the Woods--Saxon potential. We 
{\clb located also} the
$S$-matrix poles in the two-channel systems by calculating the
deuteron bound state using the Daejeon16 $NN$ interaction.~\color{black} In~\color{black} SM \color{black}\cite{Supplemental}, we also calculated the \color{black}  resonance parameters
in the Noro--Taylor two-channel model~\cite{Norro1980}.  \color{black} In each of those cases, we have demonstrated that the adapted Efros method \color{black} combined with the smoothing of potential energy matrix elements suggested in Refs.~\mbox{\cite{HungOne,HungTwo}} significantly improve the convergence of all calculations and lead to accurate results using a limited number of basis functions \color{black} even when restricted to oscillator quanta accessible for modern NCSM computations. \color{black}

\color{black} We note that obtaining accurate results with limited oscillator quanta is essential for making our method applicable for many-body problems.~\color{black} Having in mind future applications in many-body systems in combination
with the NCSM~\cite{NCSMProg}, we exploited the harmonic oscillator basis and introduced the 
Hamiltonian~$H^{tr}$ truncated in the oscillator basis at some maximal oscillator
quanta~$N_{\max}$ as is done in the NCSM. We demonstrated how this truncated
Hamiltonian~$H^{tr}$ can be utilized within the Efros method. We suggested some
modifications of the formalism of the Efros method using achievements developed
within the HORSE formalism~\cite{BangHORSE} in scattering theory. The Efros method
is equivalent to the HORSE approach if the set of SRFs includes all oscillator functions
with oscillator quanta up to~$N_{\max}$.  \color{black} Calculations within the HORSE method \color{black} however,  \color{black} are \color{black} impractical in many-body applications since \color{black} it requires the calculation of all eigenstates with definite $J$ and $0s$ center-of-mass motion which \color{black} is extremely large \color{black} and is unknown.  \color{black} The great advantage of \color{black} the adapted \color{black} Efros method is that it provides a reasonable convergence with significantly reduced SRF sets.  One can use as SRFs a limited number of
the oscillator functions describing
the relative motion in each channel, a number of the eigenfunctions of~$H^{tr\!}$, or
their mixture. The use of the~$H^{tr}$ eigenfunctions seems to be most convenient for future
many-body applications of the Efros method in combination with the NCSM. Therefore, most of our studies were performed with the eigenfunction SRF sets.  \\

Generally, \color{black} the adapted  \color{black} Efros method provides accelerated convergence and good accuracy in the description of various observables with a number of eigenfunction SRFs that are
accessible in modern NCSM \color{black} calculations. 
\color{black} \hspace{-0.158cm}~We did not find problems in calculations of 
single-channel scattering but in calculations of two-channel scattering, we obtained
spikes and discontinuities of some $S$-matrix elements discussed in the SM Ref.~\cite{Supplemental}.~These shortcomings,
however, can be eliminated by \color{black} choosing a suitable SRF selection. \color{black}\\

We conclude that \color{black} the adapted \color{black}
Efros method in combination with the NCSM is promising for {\it ab initio} studies  of
elastic scattering and reactions with light nuclei. By locating the $S$-matrix poles
within this approach, one can investigate resonant states in light nuclei and derive
resonant energies and widths. The $S$-matrix poles associated with bound states which
can be calculated within the Efros extension of the NCSM, will improve the results of calculations of bound state energies and wave functions. In particular, it will be possible
to describe the asymptotically decreasing tails of the wave functions.
These improvements will result in converged and accurate predictions of various nuclear
observables like the rms radii, quadrupole moments, asymptotic normalization constants,
electromagnetic transition rates, etc. The most important improvements are expected in 
 weakly bound systems, e.\,g., in drip-line nuclei which are a focus of modern
research in  nuclear physics.
} \\

\section*{Acknowledgments}

 We are thankful to V. D. Efros and P. Maris for valuable discussions.  A. M. Shirokov is thankful for the hospitality of colleagues from Pacific National University (Khabarovsk, Russia) where part of this work was done.  This work is supported by new faculty startup funding by the Institute of Modern Physics, Chinese Academy of Sciences, grant No. E539951SBH, by the Gansu International Collaboration and Talents Recruitment Base of Particle Physics (2023–2027), by the Senior Scientist Program funded by Gansu Province grant No. 25RCKA008, by the Ministry of Science and Higher Education of the Russian Federation, project No. FEME-2024-0005), and by the U. S. Department of Energy Grants No. DE-SC0023692 and No. DE-SC0023707 under the Office of Nuclear Physics Quantum Horizons program for the ``Nuclei and Hadrons with Quantum computers (NuHaQ)" project.

\clearpage
\onecolumngrid
\begin{center}
  \textbf{\large Supplemental Material: Applications of the Modified Hulthén--Kohn Method for Bound and Scattering States}\\[0.5em]
Mamoon A. Sharaf, Andrey M. Shirokov, Weijie Du, and James P. Vary
\end{center}
\twocolumngrid

\renewcommand\thesection{\Roman{section}} 
\setcounter{section}{0} 

\renewcommand{\thefigure}{S\arabic{figure}}
\setcounter{figure}{0} 

\section{Naive application} 
\label{SectionI}

\color{black} In this section, we demonstrate the necessity of employing the Hamiltonian that includes coupling to the continuum [see Eq.~\eqref{eq65}] in order to perform correct scattering calculations via the adapted Efros method.  We do this by first showing \color{black} that if we naively apply the Hamiltonian structure conventionally employed in \textit{ab initio} NCSM calculations to scattering, we get incorrect results.  In particular, we incorrectly set $H=H^{tr}$, where $H^{tr}$ is given in Eq.~\eqref{eq48}, because we have \color{black} in mind future many-body applications combining the NCSM and \color{black} the adapted \color{black} Efros method with the overly naive assumption that solutions are obtainable within \color{black} this \color{black} approach.

In the case of single-channel two-body scattering, the relative motion oscillator functions~$\phi_{n}(\vec{r})$ are given by Eq.~\eqref{eq3} and $2n+l \leq N_{\textrm{max}}$, where $n=0,1,...,\mtc{N}-1$. \color{black} The wave function is expanded as in Eq.~\eqref{eq49}, where the expansion coefficients are expected to be of the form \color{black} given by Eqs.~\eqref{eq50} and \eqref{eq51} \color{black} and $\mtc{N}$ energy-dependent coefficients $g_{n}(E)$ should be found within the approach.  \color{black} We consider \color{black} $NN$ scattering in the $^1S_{0}$ partial wave, where the fixed quantum numbers are $l=0$, $S=0$, $J=0$, and $M_{J}=0$; therefore, $\mtc{N}=\frac{1}{2}N_{\textrm{max}}+1$.

We need to specify the sets of SRFs~$\{\beta_{q}\}$ and~${\{\bar{\beta}_{q}\}\equiv\{\bar{\beta}^{1}_{q}\}}$~entering Eqs.~\eqref{eq32}.  A natural option is to use~$N\le \mtc{N}$ eigenfunctions~$\beta_{q}^{\textrm{Efn}}(\vec r)$ of the truncated Hamiltonian \eqref{eq48},
\begin{subequations}
\renewcommand{\theequation}{S1\alph{equation}}
\begin{align}
\tag{S1a}\label{eq52a}
\bar{\beta}_{q}(\vec r) &= \beta_{q}^{\textrm{Efn}}(\vec r), \qquad q = 0, \ldots, N-1, \\
\tag{S1b}\label{eq52b}
\beta_{q}(\vec r) &= \beta_{q}^{\textrm{Efn}}(\vec r), \qquad q = 0, \ldots, N-2
\end{align}
\begingroup
\renewcommand{\theequation}{S1}
\refstepcounter{equation}
\label{eq52}
\endgroup
\end{subequations} 
\hspace{-0.3cm} which solve the Schr\"odinger equation \color{black} [see Eq.~\eqref{eq53}] \color{black} and can be expressed as superpositions of oscillator functions \color{black} [see Eq.~\eqref{eq54}].~\color{black}  In this case, $\mtc{N}$~coefficients~$g_{n}(E)=\sum_{q=0}^{N-2}b_{q}(E)\,a^{\textrm{Efn}}_{qn}$ are expressed through~${v=N-1}$~unknown energy-dependent coefficients~$b_{q}(E)$~in the wave function expansion~\eqref{eq1} which should be found within the approach \color{black} [see Eq.~\eqref{eq50}]. \color{black}
 Setting~$\bar{\beta}_{q}=\beta_{q}^{\textrm{Efn}^{\vphantom{P}}}(\vec r)$ in Eq.~\eqref{eq21} and using the standing-wave functions~${u}(E;\vec r)$ given by Eq.~\eqref{eq18}, with the help of { Eq.~\eqref{eq53}}
we obtain:
\begin{equation}
\label{eq56}
(E_{q}-E) \langle \beta^{\textrm{Efn}}_{q}|
u(E;\vec r) \rangle=0 , \quad q=0,...\,,N-1 .
\tag{S2}
\end{equation}
Therefore, at any energy~$E \neq E_{q}$, 
\begin{gather}
\label{eq57}
\langle \beta_{q}^{\textrm{Efn}}|{u}(E;\vec r) \rangle=0, \quad q=0,...\,,N-1  ,
\tag{S3}
\end{gather}
and the last of these equations with~$q=N-1$~results in
\begin{gather}
\label{eq58}
\tan\delta_{l}(E)=\frac{B_{N-1}}{A_{N-1,N-1}}.
\tag{S4}
\end{gather}
\color{black}

 \begin{figure}[t!]
{%
  \includegraphics[width=1\columnwidth]{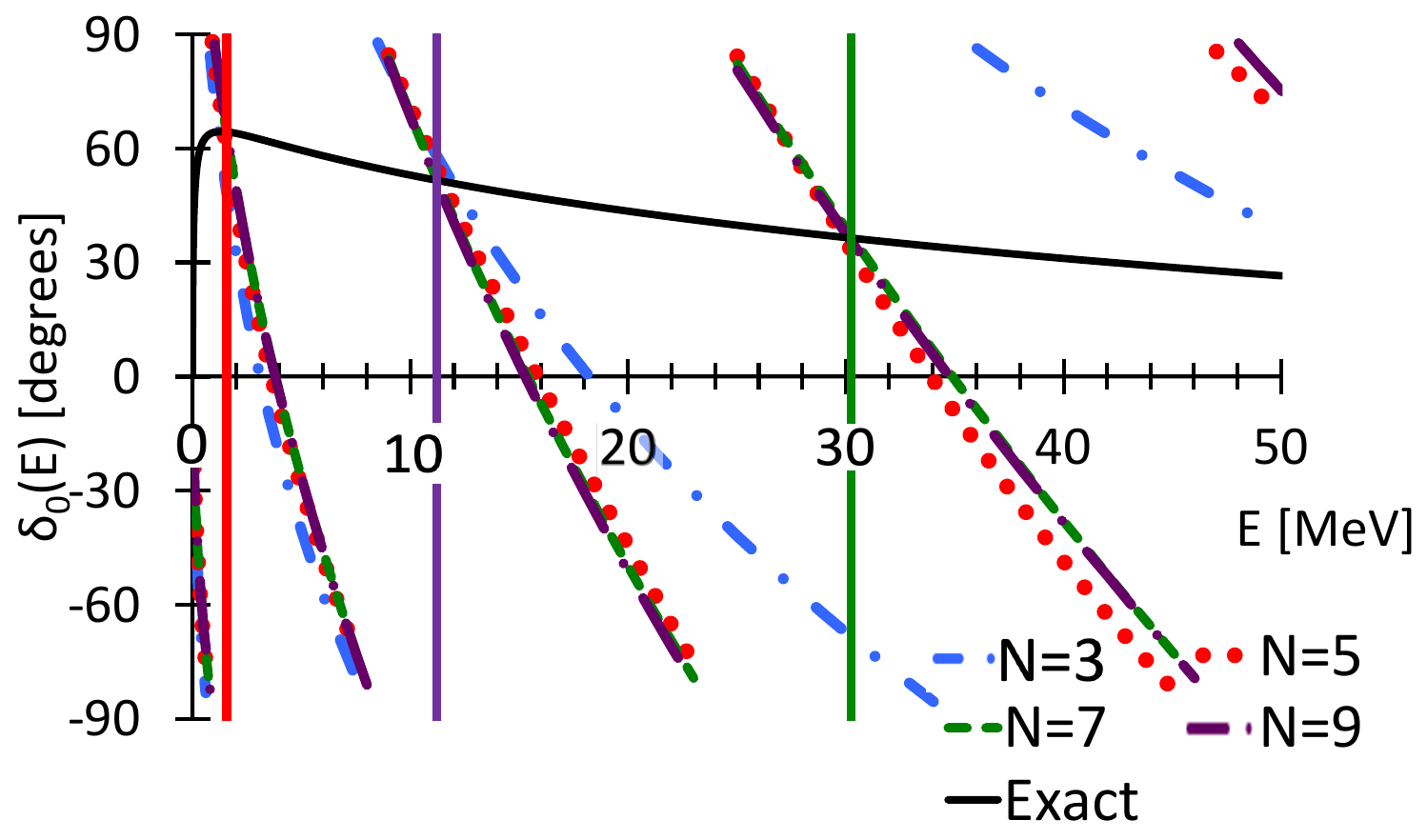}%
}
\caption{$^1S_{0}$ phase shifts~$\delta_{0}(E)$ as  functions  of energy~$E$ obtained for various $N$  with $N_{\textrm{max}}=16$ and~$\hbar \Omega=30$~MeV  using eigenfunctions~$\beta_{q}^{\textrm{Efn}} $ as 
SRFs. Vertical lines indicate the eigenvalues of~$H^{tr}$.\color{black}  }
 \label{fig:S1}
\end{figure}

The resulting phase shifts~$\delta_{0}(E)$~in the~$^1S_{0}$~partial wave of~$np$~scattering obtained with the Daejeon16~$NN$~interaction \color{black} (see Ref.~\cite{Daejeon16}) \color{black} using the oscillator basis with~${\hbar \Omega=30}$~MeV truncated at~$N_{{\max}}=16$~are presented in Fig.~\ref{fig:S1}.~The results are compared with the exact results.~It is seen that the obtained phase shifts at any~$N$~have nothing to do with the exact ones. This issue stems from the structure of the Hamiltonian~$H^{tr}$~in Eq.~\eqref{eq48} and the use of its eigenfunctions~$\beta_{q}^{\textrm{Efn}}(\vec r)$~as SRFs~$\bar{\beta}_{q}(\vec r)$~which result in producing the wave function~${u}(E;\vec r)$~as an expansion in a truncated set of oscillator functions~$\phi_{n}(\vec r)$~with~${n\le \mtc{N}-1}$ with no coupling to the continuum.~The wave function components~$\phi_{n}(\vec r)$~with~$n=\mtc{N},\mtc{N}+1,...\,,\,\infty$~are also needed
since only the infinite sum of oscillator functions in Eq.~\eqref{eq49} can describe the continuum spectrum wave function. 
Therefore, to calculate the contribution of the wave function components~$\phi_{n}(\vec r)$~with~$n\geq \mtc{N}$, we obtain an oversimplified \color{black} Eq.~\eqref{eq58} \color{black} for~$\tan\delta_{l}(E)$ which enters the equation for~$a^{as}_{n}(E)$ \color{black} [see Eq.~\eqref{eq51}]. \color{black} Equation~\eqref{eq58} \color{black} contains information only about a single 
eigenstate~$\beta_{N-1}^{\textrm{Efn}}(\vec r)$ of the Hamiltonian~$H^{tr}$ and thus cannot describe
its complete structure and the interaction~$V$ in the system. So, we cannot expect to obtain
adequate phase shifts using \color{black} Eq.~\eqref{eq58} \color{black} in a range of energies~$E$.

\begin{figure}[t!]
{%
  \includegraphics[width=1\columnwidth]{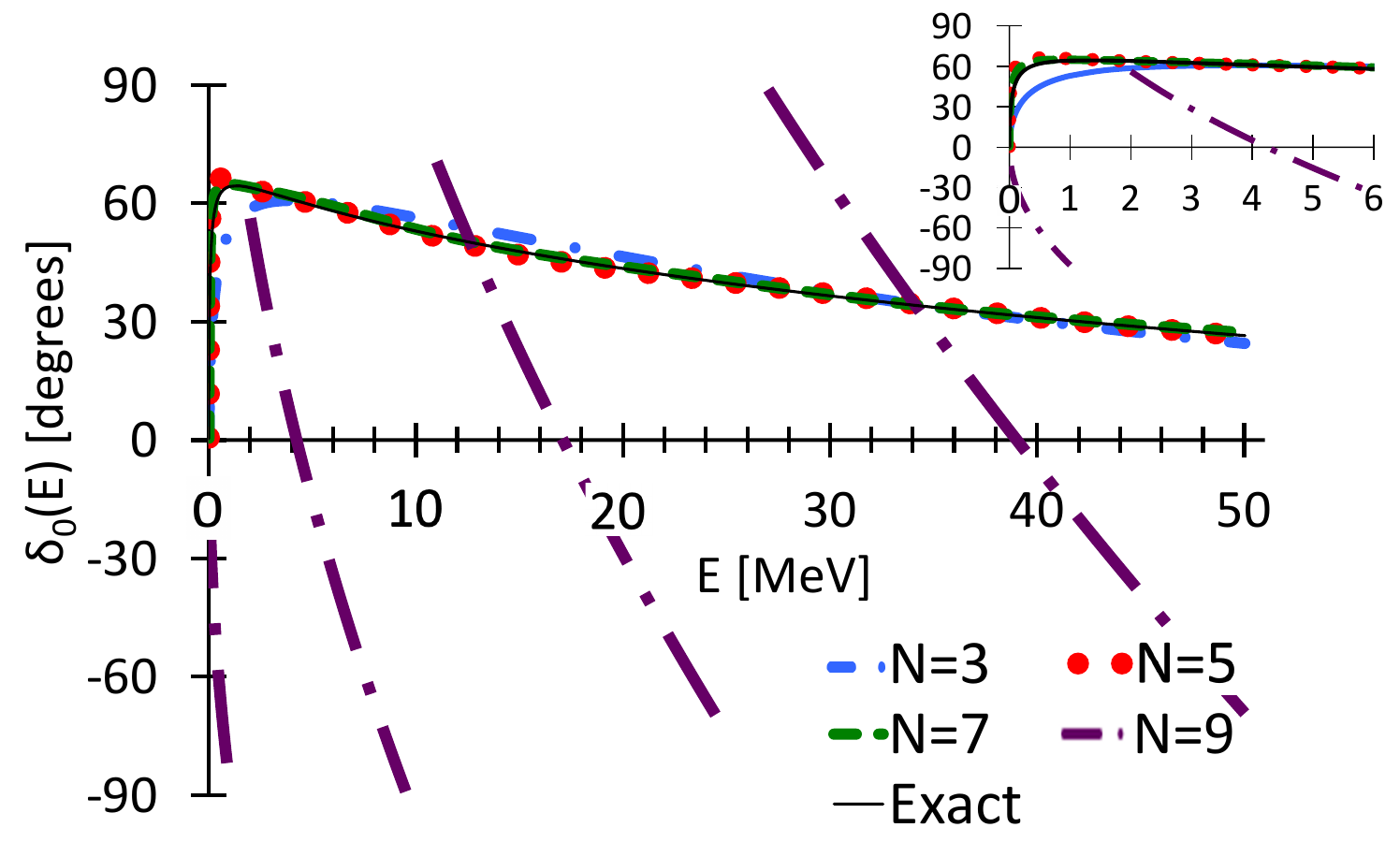}%
}
\caption[]{$^1S_{0}$ phase shifts obtained for
various
~$N$ with $N_{\textrm{max}}=16$~and~$\hbar \Omega=30$ MeV using the harmonic oscillator SRF.   The inset in this and in some of the following figures shows either low-energy phase shifts or other observables in different scales. }
 \label{fig:S2}
\end{figure}

One can{, however,} 
expect that in this case we obtain  at least  good phase shifts~$\delta_l(E_q)$ at energies~$E=E_{q}$ 
\color{black} [see Eq.~\eqref{eq56}] \color{black}
and can use them for evaluating phase shifts in some energy interval as is done within the SS-HORSE approach \color{black} (see Refs. \cite{AMShirokovHORSE2016,BlokhintsevSSHorseC,AMShirokovHORSE2018}). \color{black}  We see in Fig.~\ref{fig:S1} that the curves describing phase shifts obtained with different~$N$ cross the
curve  representing the exact phase shifts at energies close to the lowest~$E_{q}$ values,
however, contrary to the SS-HORSE, calculations of~$\tan\delta_{l}(E_{q})$~using \color{black} Eq.~\eqref{eq58} \color{black} are 
not accurate and this inaccuracy increases for higher eigenenergies.

Another reasonable option for SRFs~$\beta_{q}$ and~$\bar{\beta}_{q}$ are the oscillator functions:
\begin{subequations}
\begin{align}
\tag{S5a}\label{eq60a}
\bar{\beta}^{\rm HO}_{q}(\vec r) &= \phi_{q}(\vec r), \qquad q \leq v = N - 1, \\
\tag{S5b}\label{eq60b}
{\beta}^{\rm HO}_{q}(\vec r) &= \phi_{q}(\vec r), \qquad q \leq v - 1 = N - 2.
\end{align}
\end{subequations}
\begingroup
\renewcommand{\theequation}{S5}
\refstepcounter{equation}
\label{eq60}
\endgroup
\hspace{-0.4cm} This choice is supported by our intention to express the wave function~$u(E;\vec r)$ as an infinite series of oscillator functions. However, in many-body applications it will be difficult to select from a huge number of Slater determinants comprising the NCSM basis a reasonable set of a  relatively small  number of antisymmetrized SRFs with a given~$J$ value. One, of course, can use some assumptions, e.g., the ideas of the RGM and define SRFs as a set of few relative motion functions times an antisymmetrized product of cluster wave functions supposing that the structure of colliding nuclei is unchanged during the collision.  This 
 is, of course, an approximation which works well in some cases but would be better to avoid for more general applications including the NCSM.

 With the harmonic oscillator SRFs \color{black} \eqref{eq60}, \color{black} we expect the expansion coefficients of the wave function Eq.~\eqref{eq49}, to be 
\begin{gather}
\label{eq61}
\tag{S6}
d_{n}(E)= 
\begin{cases} 
    b_{n}(E)+a^{as}_{n}(E), &  n =0,...\,,N-2 , 
    \\ 
      a^{as}_{n}(E), &  n = N-1,...\,,\infty ,
   \end{cases}
\end{gather}
with~$a^{as}_{n}(E)$ given by Eq.~\eqref{eq51}.~The results of the $^1S_{0}$ phase shift calculations using the oscillator basis with~${\hbar \Omega=30}$~MeV 
truncated at~$N_{{\max}}=16$~are presented in Fig.~\ref{fig:S2} in comparison with the exact phase shifts for this~$NN$~interaction.

In the case of $N=\mtc{N}=9$, the set of SRFs~\eqref{eq60} results in the unphysical phase shifts {of the
same type as the set of eigenfunction SRFs~\eqref{eq52} presented in Fig.~\ref{fig:S1}.~This is not surprising since both the functions \color{black} \eqref{eq60a} \color{black} and \color{black} \eqref{eq52a} \color{black} form a complete basis in the Hilbert space spanned by the oscillator functions~$\phi_{n}(\vec r)$ with~$n\leq \mtc{N}-1$. Hence, any of the functions~$\bar{\beta}^{\rm HO}_{q}(\vec r)$~can be expressed as a superposition of the functions~$\bar{\beta}^{\rm Efn}_{q}(\vec r)$} and the set of equations \color{black} \eqref{eq57} \color{black} guarantees at any energy~$E\ne E_{q}$ that
\begin{gather}
\label{eq62}
\tag{S7}
\langle { \bar{\beta}_{q}^{\textrm{HO}}}|{u}(E;\vec r) \rangle=0, \quad q=0,...\,,N-1  
\end{gather}
with no coupling to the continuum and results in obtaining expansion 
coefficients~$d_{n}(E)=0$~for all~$n=0,$ ...\,, $\mtc{N}-1$.

The situation improves dramatically when~$N<\mtc{N}$.  The phase shifts~$\delta_{0}(E)$~demonstrate clear convergence patterns with increasing~$N$~and a good description of exact phase shifts is achieved even at relatively small values of~$N=5$.

The success of the phase shift calculations with harmonic oscillator SRFs with~$N<\mtc{N}$~seems surprising.  The problem is that we were using an inconsistent formulation of the problem which disappears in the case of~$N<\mtc{N}$ for the oscillator SRFs. In particular, our Hamiltonian \eqref{eq48} was confined to the Hilbert space spanned by the oscillator functions~$\phi_{n}(\vec r)$~with~$n \leq \mtc{N}-1$.  At the same time, we supposed that our wave function~$u(E;\vec r)$~has an infinite expansion in oscillator functions~\eqref{eq49} where, according to Eq.~\eqref{eq50}, the expansion coefficients~$d_{n}(E)=a^{as}_{n}(E)$~at~$n\geq \mtc{N}$.  The asymptotic expansion coefficients~$a^{as}_{n}(E)$ [see Eq.~\eqref{eq51}] describing the free motion at energy~$E$ and within the HORSE formalism fit the free Schr\"odinger equation \cite{Zaitsev1998,BangHORSE} which, due to the tridiagonal structure of the kinetic energy matrix elements, Eqs.~\eqref{eq39}, reduces to the three-term recurrent relation  \color{black} \eqref{eq64}.  Therefore, the Hamiltonian \eqref{eq48} should be extended to infinity [see Eq.~\eqref{eq65}] \color{black} in the oscillator space by the kinetic energy matrix elements.

  However, in the case of eigenfunction SRFs~$\beta_{q}^{\textrm{Efn}}(\vec r)$ and~$\bar\beta_{q}^{\textrm{Efn}}(\vec r)$,~these coupling terms \color{black} in~Eq.~\eqref{eq65} \color{black} were completely suppressed since all matrix elements~$\langle\bar\beta_{q}^{\textrm{Efn}}|H^{tr}|\phi_{\mtc{N}}\rangle=\langle\phi_{\mtc{N}}|H^{tr}|\beta_{q}^{\textrm{Efn}}\rangle=0$.  As a result, we arrived at an unphysical {\color{black} Eq.~\eqref{eq58} \color{black} for the phase shifts.}
The case of a ``complete'' set of harmonic oscillator SRFs~$\bar{\beta}^{\rm HO}_{q}(\vec r)$, \color{black} Eq.~\eqref{eq60}, \color{black} with~$N=\mtc{N}$ is equivalent to that of the eigenfunction SRFs~$\beta_{q}^{\textrm{Efn}}(\vec r)$.   In the expansion of the wave function Eq.~\eqref{eq49}, all coefficients~$d_{n}(E)=0$~with~$n\leq \mtc{ N}-1$~and we have an analog of the scattering by an impenetrable sphere in the coordinate space with the unphysical phase shifts of the same type. 
\color{black}

 \begin{figure}[t!]
{%
  \includegraphics[width=1\columnwidth]{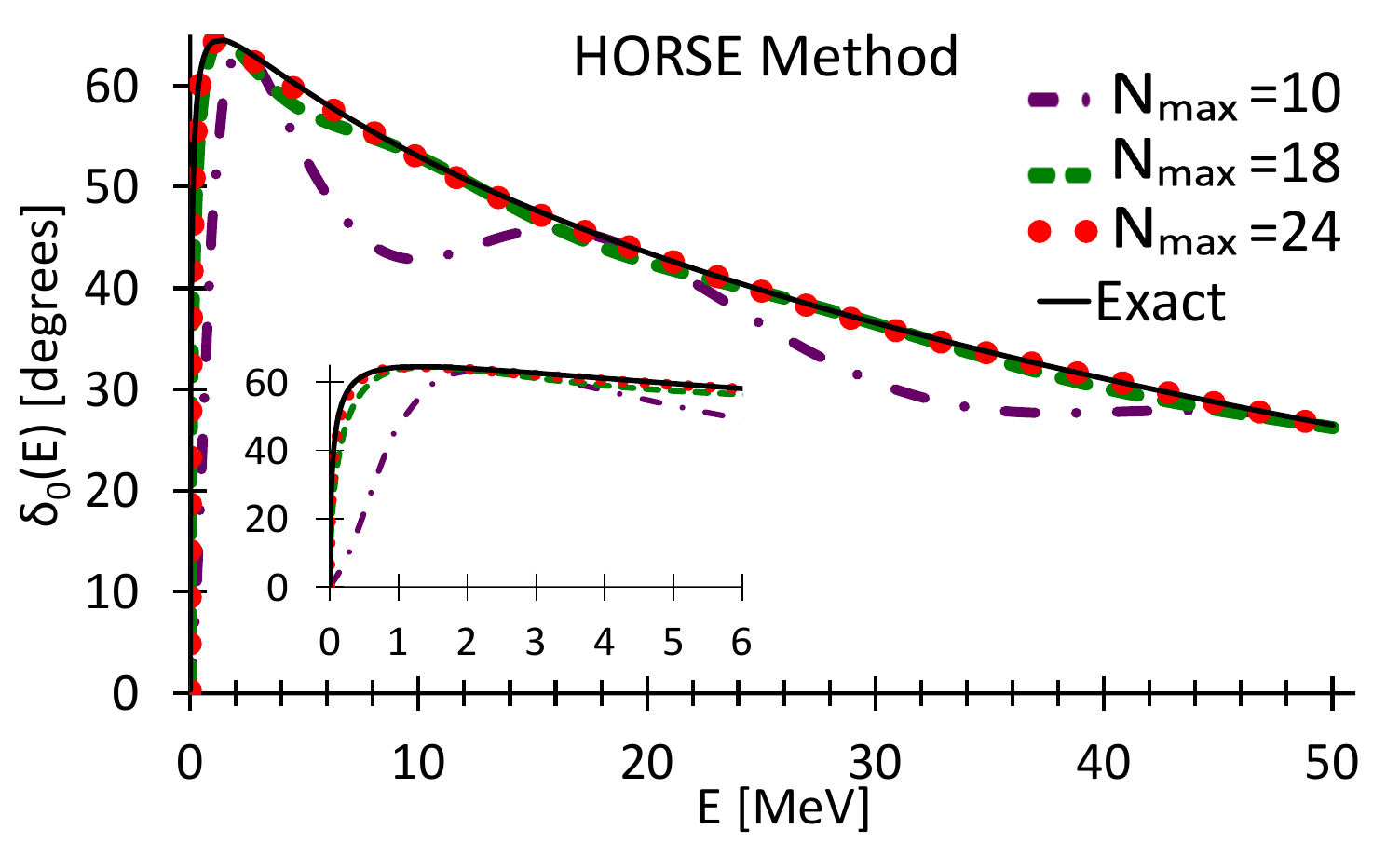}%
}
\caption{$^1S_{0}$ phase shifts 
obtained by the HORSE method for various~$N_{\textrm{max}}$ with~$\hbar \Omega=30$~MeV.  }
 \label{fig:S3}
\end{figure}

\begin{figure}[t!]
{%
  \includegraphics[width=1\columnwidth]{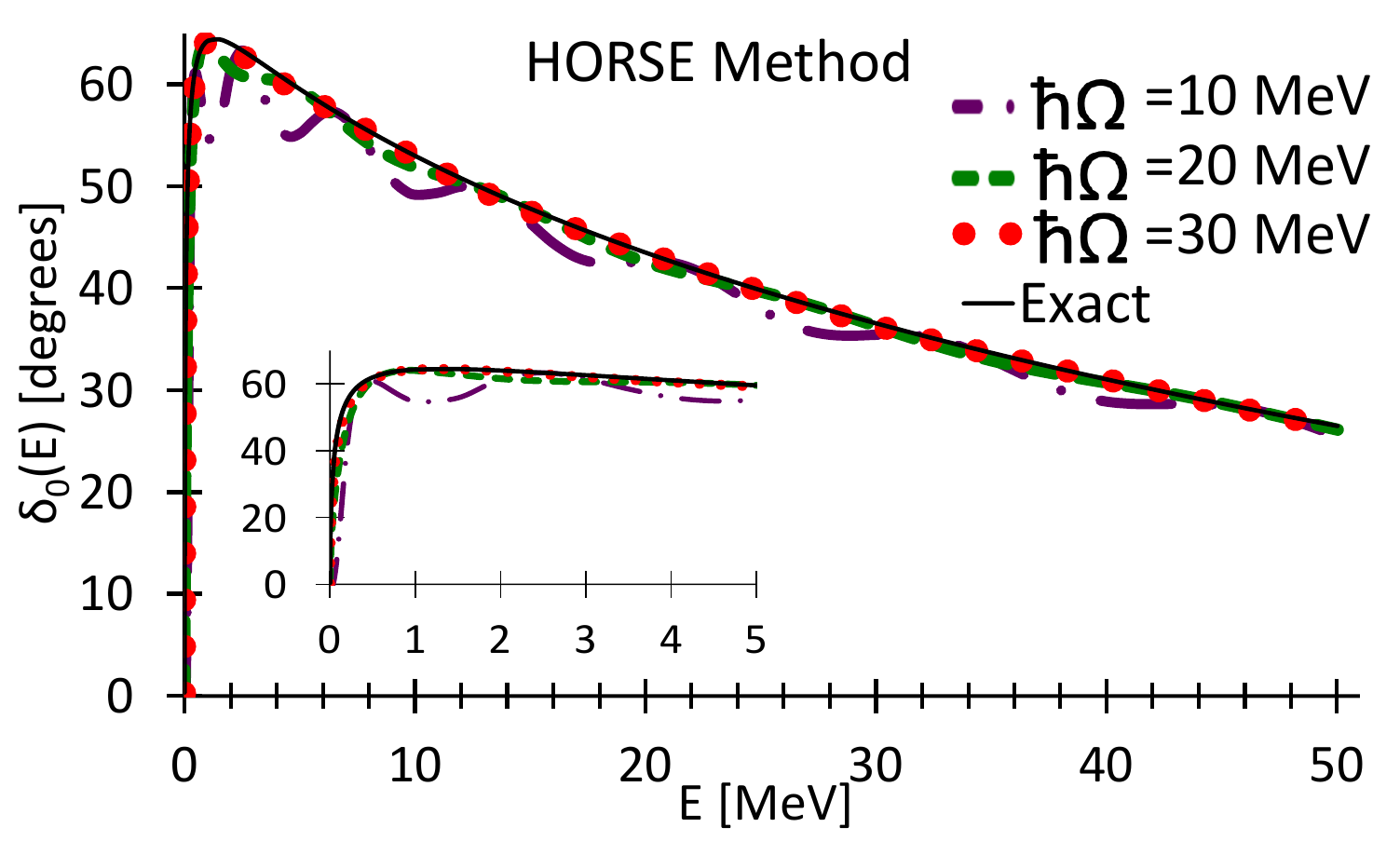}%
}
\caption{$^1S_{0}$ phase shifts 
obtained by the HORSE method for various~$\hbar \Omega$ at~$N_{\textrm{max}}=24$.}
 \label{fig:S4}
\end{figure}

\begin{figure}[t!]
{%
  \includegraphics[width=1\columnwidth]{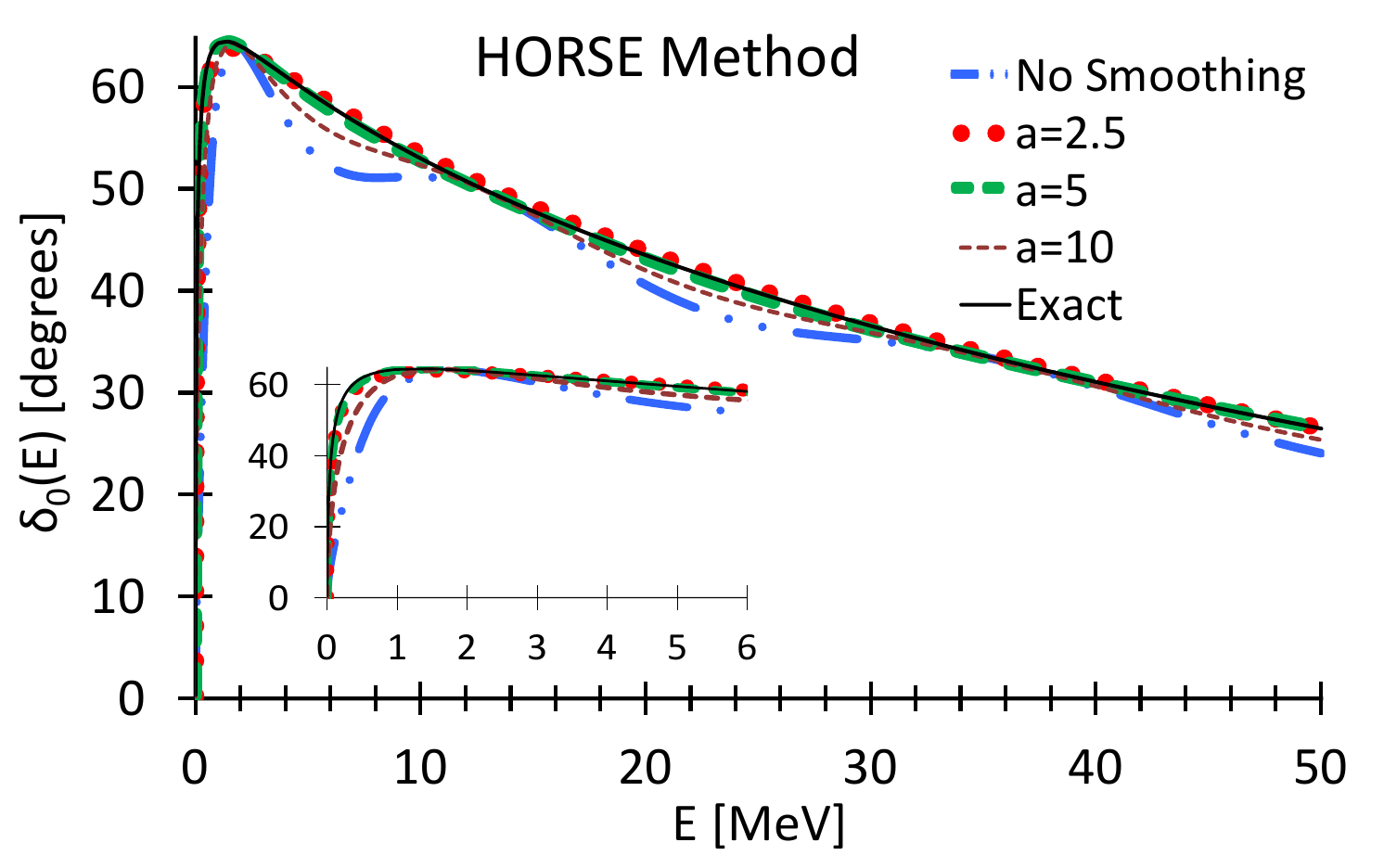}%
}
\caption[]{The effect of smoothing with various parameters~$a$ on calculations of the
$^1S_{0}$ phase shifts 
by the HORSE method with $N_{\textrm{max}}=14$ and~$\hbar \Omega=30$~MeV. 
}
 \label{fig:S5}
\end{figure}

\begin{figure}[t!]
{%
  \includegraphics[width=1\columnwidth]{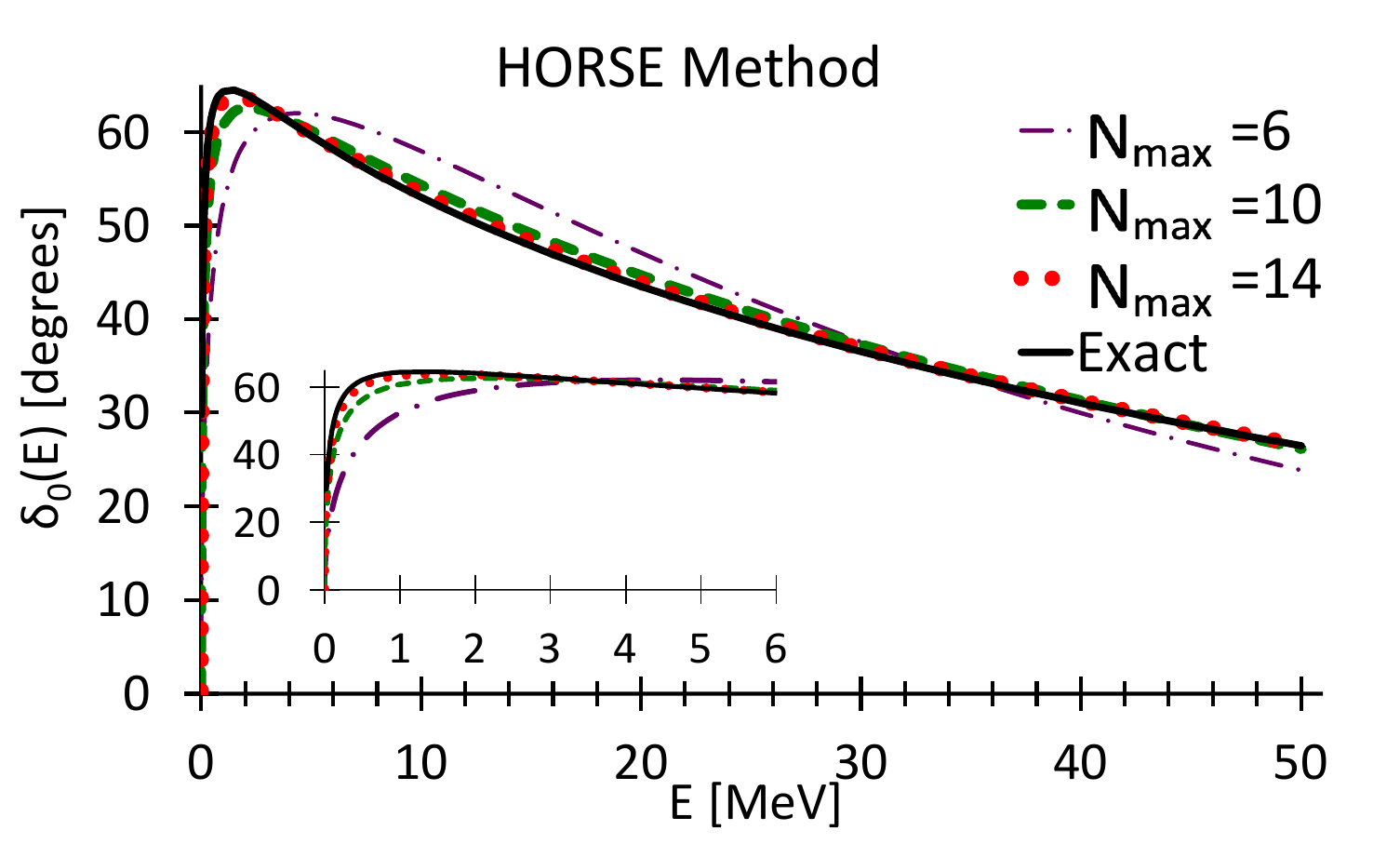}%
}
\caption[]{$^1S_{0}$  phase shifts 
obtained by the HORSE method 
with various~$N_{\textrm{max}}$ 
at~$\hbar \Omega=30$~MeV with smoothing with parameter~$a=2.5$. }
 \label{fig:S6}
\end{figure}

However in the case of an ``incomplete'' set of harmonic oscillator SRFs~$\bar{\beta}^{\rm HO}_{q}(\vec r)$ with~$N<\mtc{N}$, by solving Eq.~\eqref{eq31}, we obtain a set of functions~$g_{n}(E)$ with~$n=0$, ...\,, $N-2$~in Eq.~\eqref{eq50} and get nonzero expansion coefficients~${d_{n}(E)=a^{as}_{n}(E)}$~with~$n\geq N-1$~which mimic the role of the last three terms in Eq.~\eqref{eq65} though these terms were not explicitly included in $H^{tr}$.
\color{black}

\section{Single-channel scattering }
\label{SectionII}

\begin{figure}[t!]
\subfloat[]{
  \includegraphics[width=1\columnwidth]{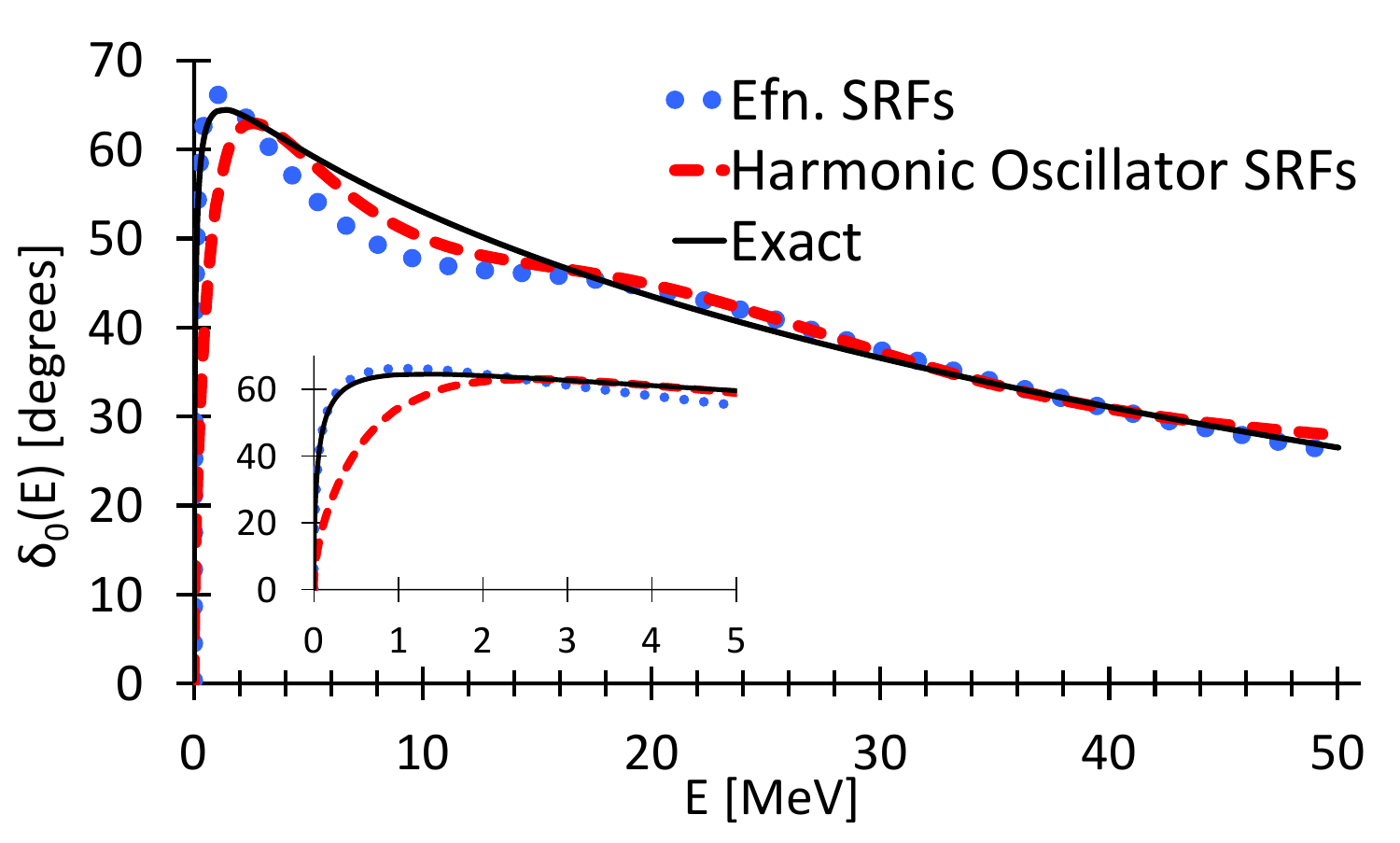}%
}\hfill
\subfloat[]{%
  \includegraphics[width=1\columnwidth]{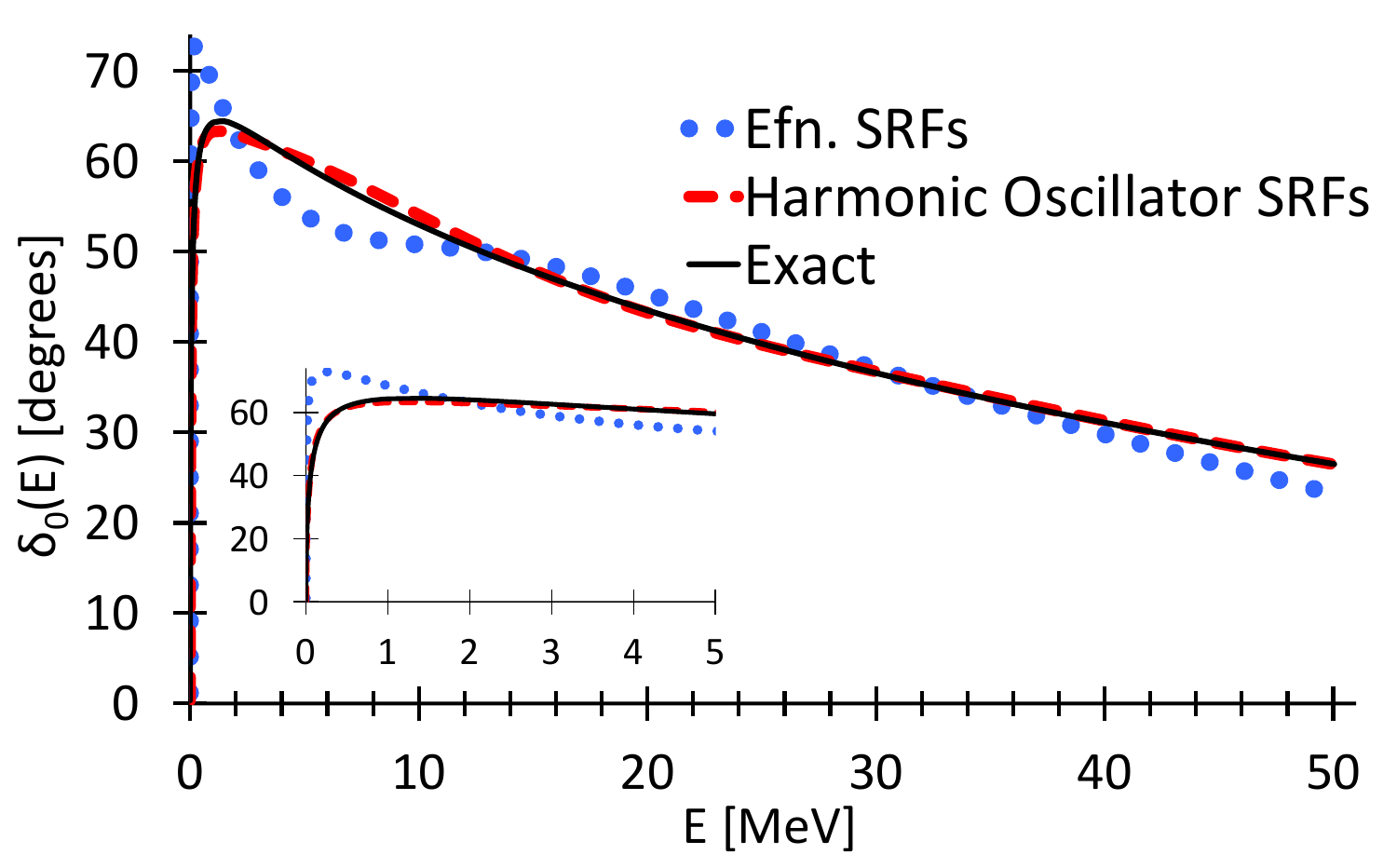}%
}\hfill
\subfloat[]{%
  \includegraphics[width=1\columnwidth]{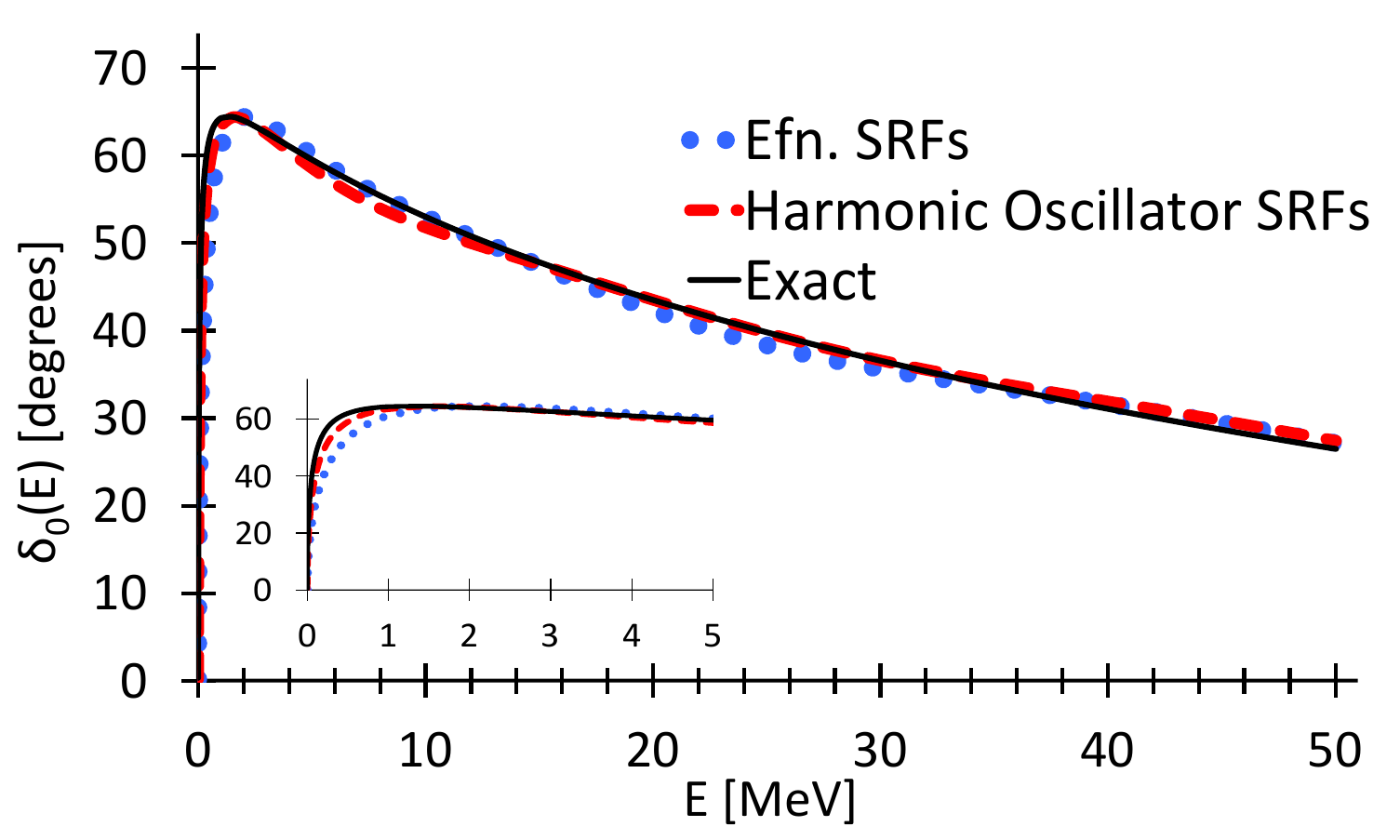}%
}
\caption[]{ $^1S_{0}$  phase shifts 
obtained by the \color{black} adapted \color{black} Efros method  using a)~$N=4$ and~$N_{\textrm{max}}=10$, 
b)~$N=4$ and~${N_{\textrm{max}}=14}$, and  
c)~$N=5$ and~${N_{\textrm{max}}=14}$
with  both the harmonic oscillator and eigenfunction SRFs, ${\hbar \Omega=30}$~MeV.}
 \label{fig:S7}
\vspace{-3ex}
\end{figure}

\begin{figure}[t!]
{%
  \includegraphics[width=1\columnwidth]{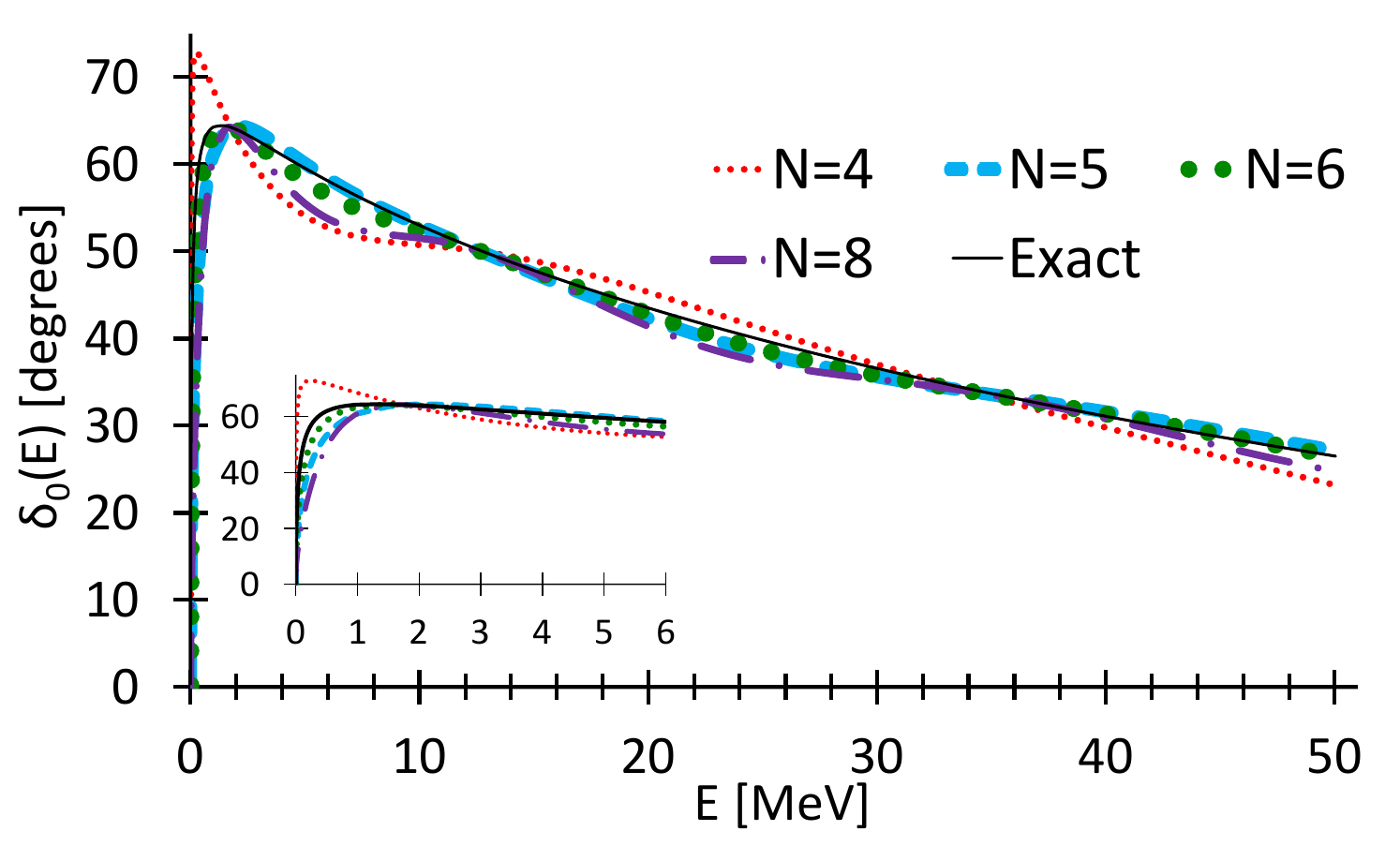}%
}
\caption[]{$^1S_{0}$  phase shifts obtained by \color{black} the adapted \color{black} Efros method with 
various~$N$ at $N_{\textrm{max}}=14$ and~$\hbar \Omega=30$~MeV using the eigenfunction SRFs.}
 \label{fig:S8}
\end{figure}

\begin{figure}[t!]
{%
  \includegraphics[width=1\columnwidth]{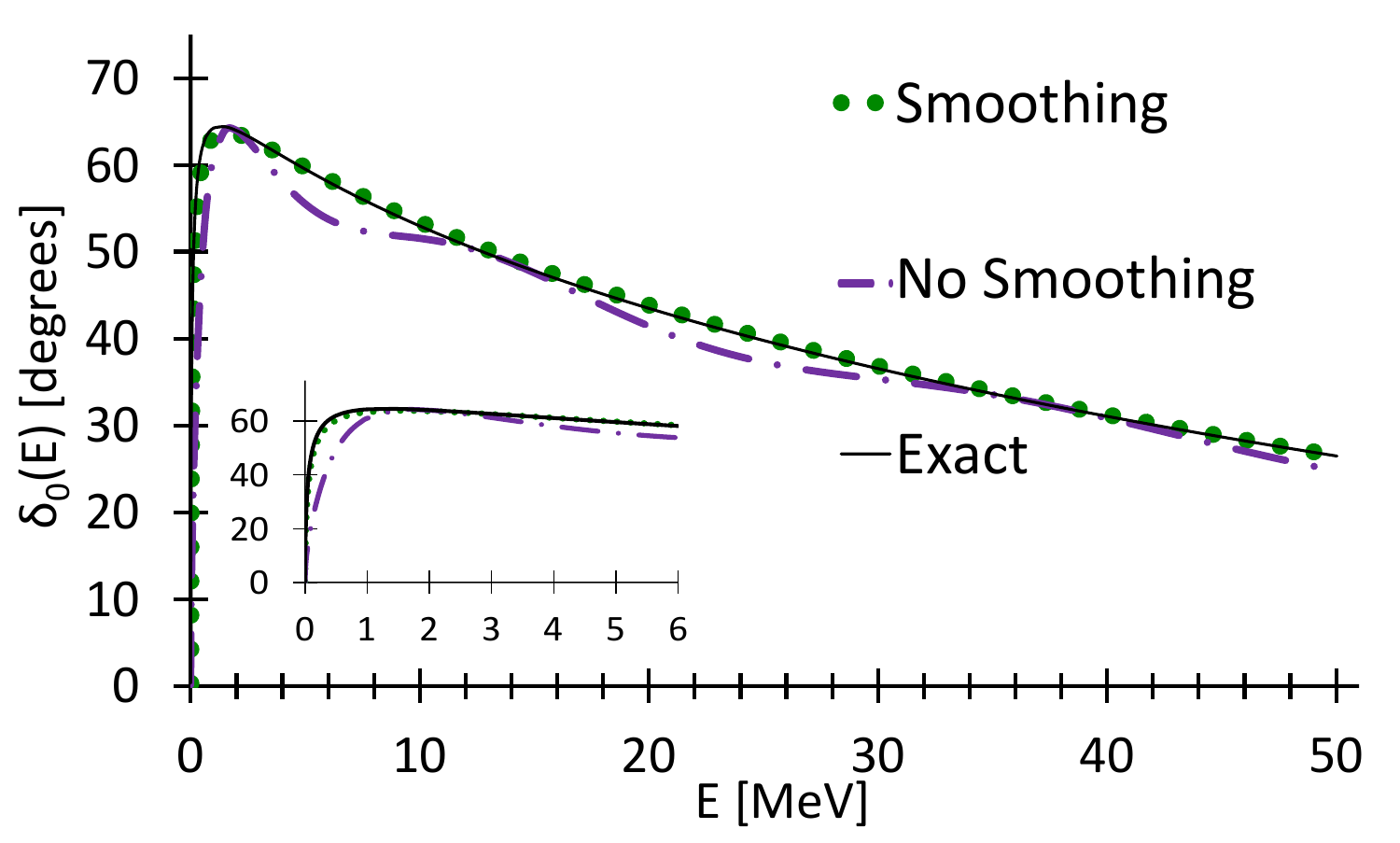}%
}
\caption[]{$^1S_{0}$  phase shifts
obtained by \color{black} the adapted \color{black} Efros method  using $N=8$, $N_{\textrm{max}}=14$,  and~${\hbar \Omega=30}$~MeV using the eigenfunction SRFs without smoothing and with smoothing at~$a=2.5$.
}
 \label{fig:S9}
\end{figure}

\begin{figure}[t!]
  \includegraphics[width=1\columnwidth]{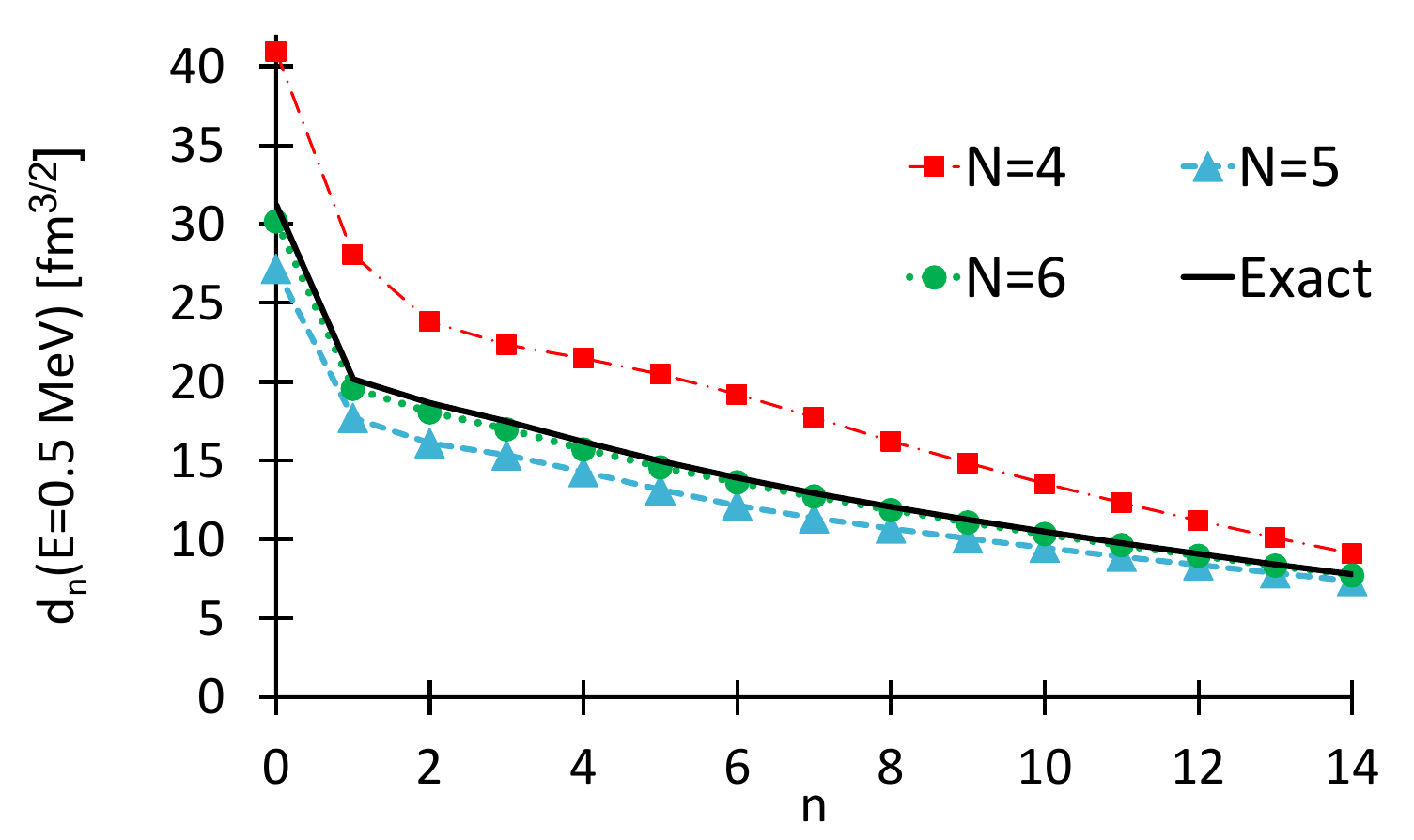}%
\caption[]{{Coefficients $d_{n}(E)$  of the wave function expansion~\eqref{eq49}
at energy~${E=0.5}$~MeV in the
 $^1S_{0}$ partial wave obtained with various~$N$ 
 at~${N_{\textrm{max}}=14}$ and~$\hbar \Omega=30$~MeV with smoothing parameter~$a=2.5$
using the eigenfunction SRFs.
} }
 \label{fig:S10}
\end{figure}

\begin{figure}[t!]
{%
  \includegraphics[width=1\columnwidth]{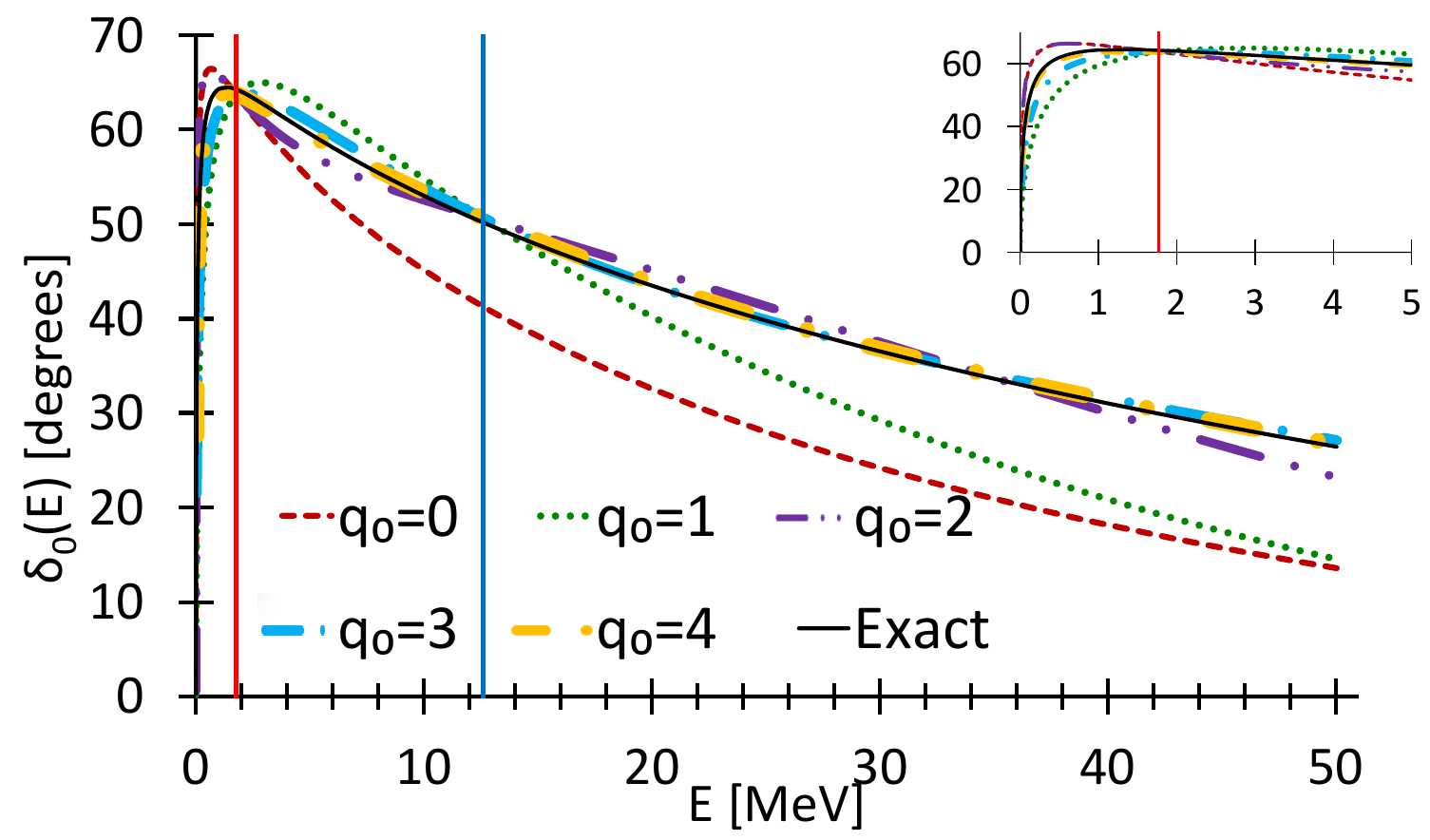}%
}
\caption[]{$^1S_{0}$ phase shifts obtained by \color{black} the adapted \color{black} Efros method with various~$q_{0}$
using the hybrid SRF set given by Eq.~\eqref{eq69}, 
$N_{\textrm{max}}=14$, $N=6$, $\hbar \Omega=30$~MeV, and the smoothing 
parameter~$a=2.5$.  Vertical lines show the energies of eigenstates of the Hamiltonian~$H^{tr}$. }
 \label{fig:S11}
\vspace{4ex}
{%
  \includegraphics[width=1\columnwidth]{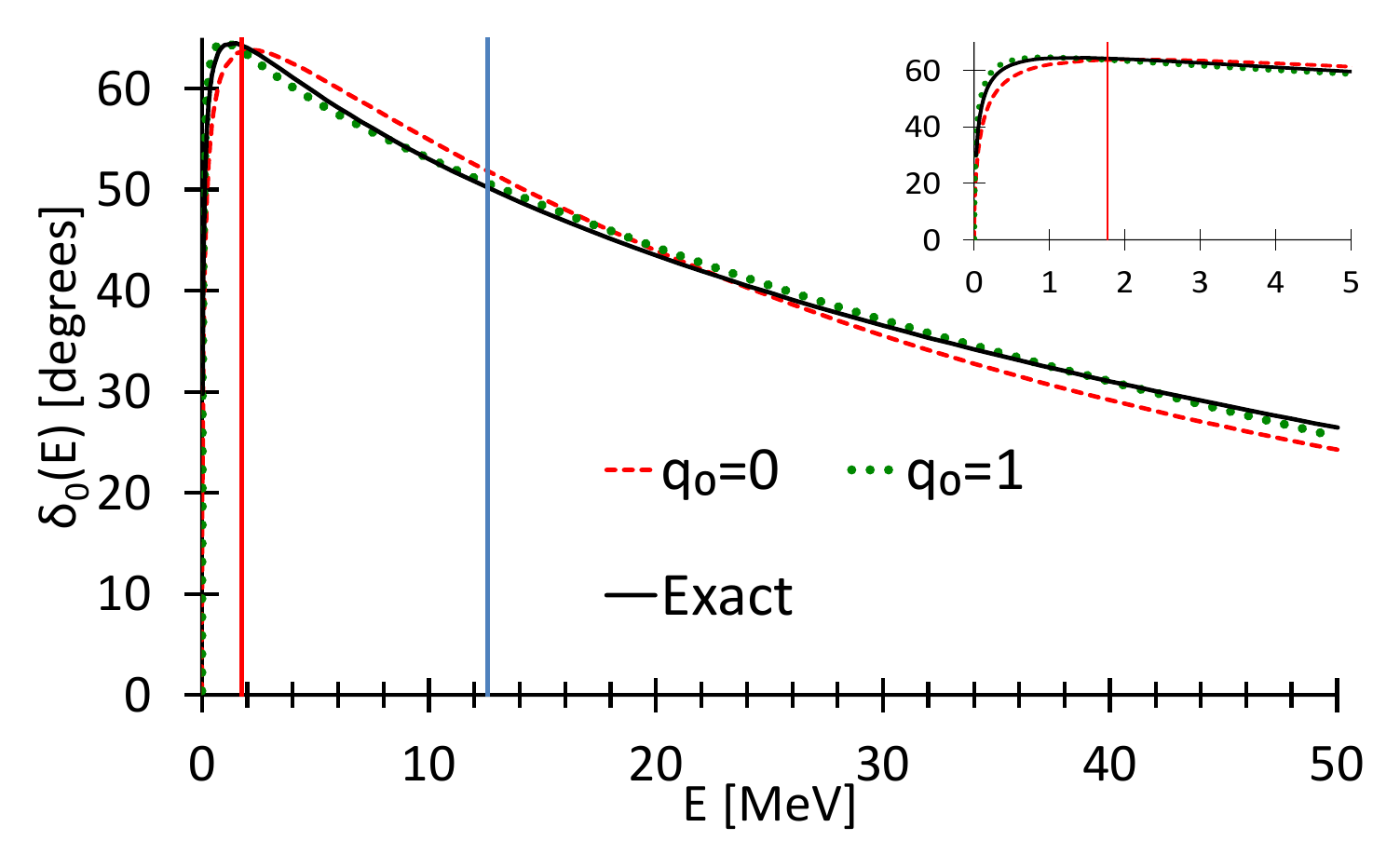}%
}
\caption[]{ The same as Fig.~\ref{fig:S11} but
using the hybrid SRF set given by Eq.~\eqref{eq70} and $N_{\max}=12$.  }
 \label{fig:S12}
\end{figure}

In this section, we start by discussing the convergence of single-channel scattering calculations using the HORSE method.  Figure~\ref{fig:S3} shows the convergence of the ${^{1}S_{0}}$ $np$ scattering phase shifts supported by the Daejeon16~$NN$~interaction
with~$N_{\textrm{max}}$ at~$\hbar \Omega=30$~MeV.  
It is seen that at~$N_{\textrm{max}}=18$, we obtain a very good description of the phase shifts while
at~$N_{\textrm{max}}=24$, a complete 
convergence is achieved in a wide range of energies.

Figure~\ref{fig:S4}~shows the change of the phase shift description 
with~$\hbar \Omega$~at~$N_{\textrm{max}}=24$.  We see that~$\hbar \Omega=30$~MeV is optimal from the point of view of convergence
of calculations.

Figure~\ref{fig:S5} demonstrates the effect of smoothing on the phase shift calculations within the HORSE method with~$N_{\textrm{max}}=14$.  We see that our choice of using the parameter values of~$a=2.5$~or~$a=5$~in our studies of~$NN$~scattering is justified.

In Fig.~\ref{fig:S6}, we find that excellent convergence of the phase shifts within 
the HORSE method is achieved with smoothing at~$N_{\textrm{max}}=14$ 
as opposed to the~$N_{\textrm{max}}=24$ needed for convergence without smoothing. \color{black}

Figure~\ref{fig:S7} compares the $^1S_{0}$ phase shifts~$\delta_{0}(E)$ obtained within \color{black} the adapted \color{black} Efros approach with the eigenfunction and harmonic oscillator SRFs for different truncations, namely~($N$,$N_{\textrm{max}}$)$=(4,10), (4,14)$, and $(5,14)$.  We find that it is not obvious whether the harmonic oscillator SRFs or the eigenfunction SRFs perform better for the phase shifts.  We see that the two sets of SRFs 
are comparable in terms of convergence.

  {\clb 
Figure~\ref{fig:S8}~shows the convergence of~$\delta_{0}(E)$~obtained by \color{black} the adapted \color{black} Efros method
with~$N$ for~$N_{\textrm{max}}=14$.  We note a reasonable convergence at~$N=5$. However, at larger~$N$, namely at~$N=8$, the results 
start deviating significantly from the exact result.  The irregularity of convergence is a consequence of the short-range behavior 
of the $NN$ interaction and a sudden drop-off of its matrix elements~$V_{nn'}$ at $n,n'=\cal N$ due to 
the sharp truncation at relatively small~$N_{\textrm{max}}$. }  

\begin{figure}[t!]
\subfloat[]{%
  \includegraphics[width=1\columnwidth]{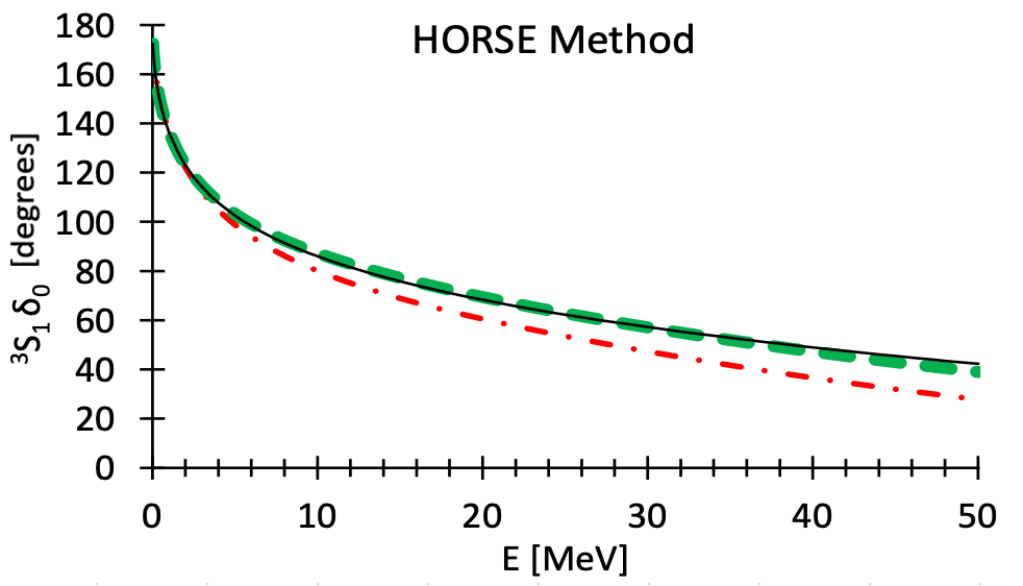}%
}\hfill
\subfloat[]{%
  \includegraphics[width=1\columnwidth]{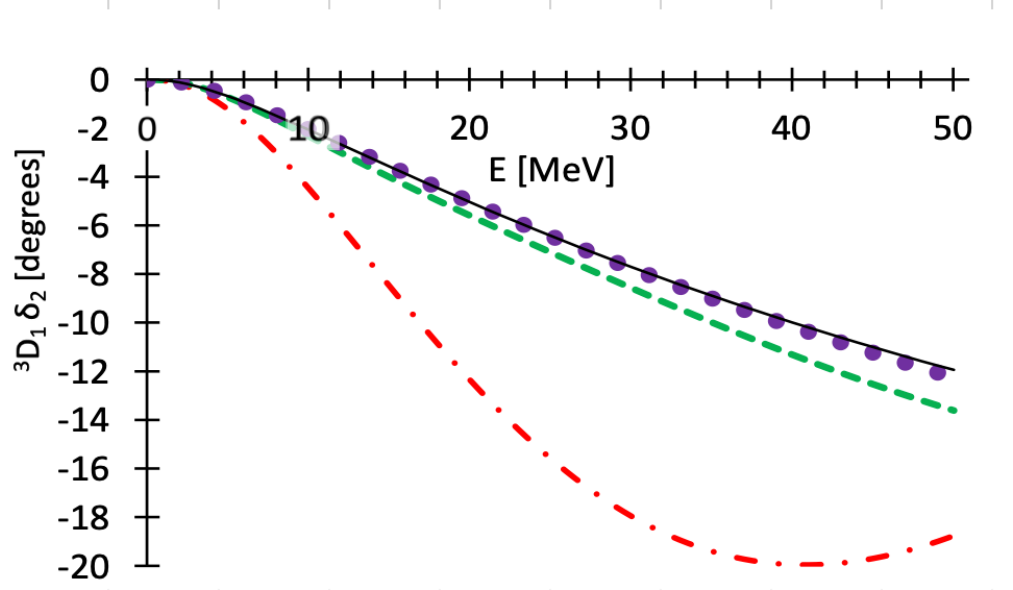}%
}\hfill
\subfloat[]{%
  \includegraphics[width=1\columnwidth]{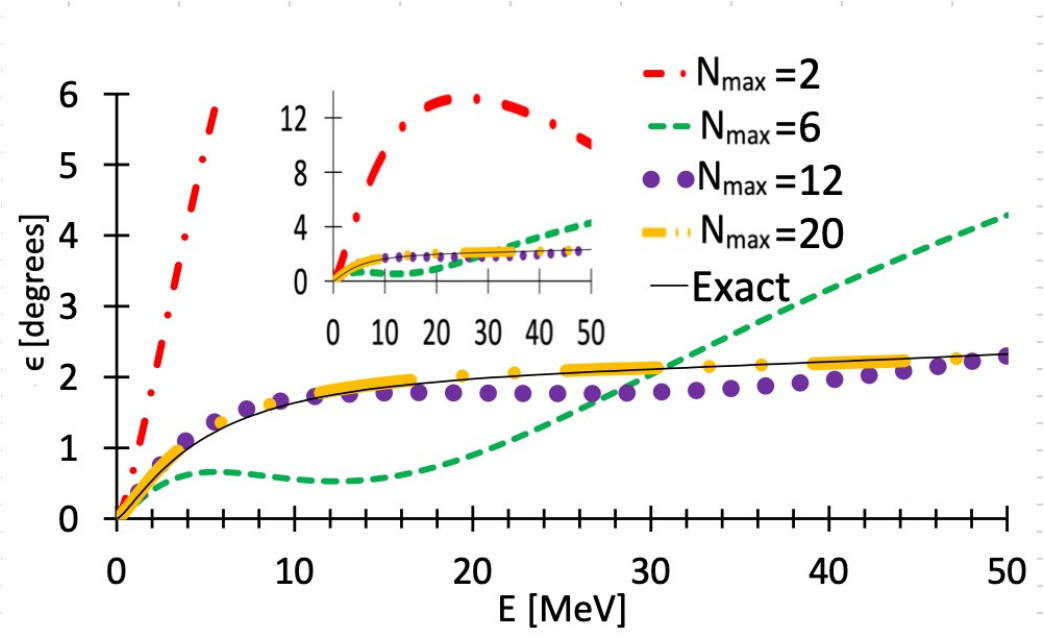}%
}\hfill
\caption[]{a) $^3S_{1}$ 
{\clg phase shifts}~$\delta_{0}$, b) $^3D_{1}$ 
{\clg phase shifts}~$\delta_{2}$, c)~$^3SD_{1}$ {\clg mixing parameter}~$\epsilon$
obtained by the HORSE method 
{\clg with~$\hbar \Omega=30$~MeV and various}~$N_{\textrm{max}}$;
{\clg the smoothing parameter~$a=2.5$}.  The inset in (c) displays~$\epsilon$~in a larger scale on the y-axis.}
 \label{fig:S13}
\end{figure}

\begin{figure}[t!]
\subfloat[]{%
  \includegraphics[width=1\columnwidth]{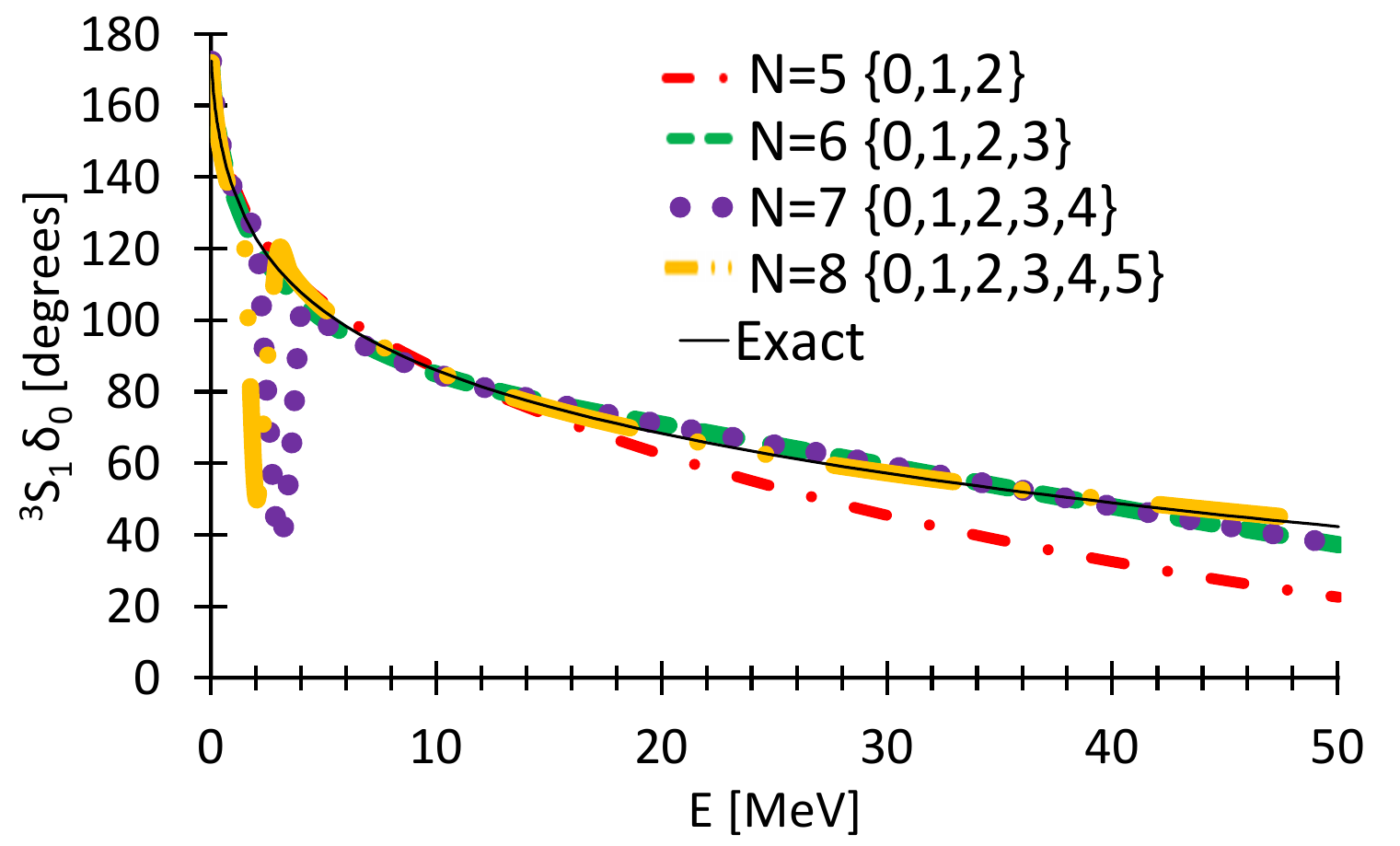}%
}\hfill
\subfloat[]{%
  \includegraphics[width=1\columnwidth]{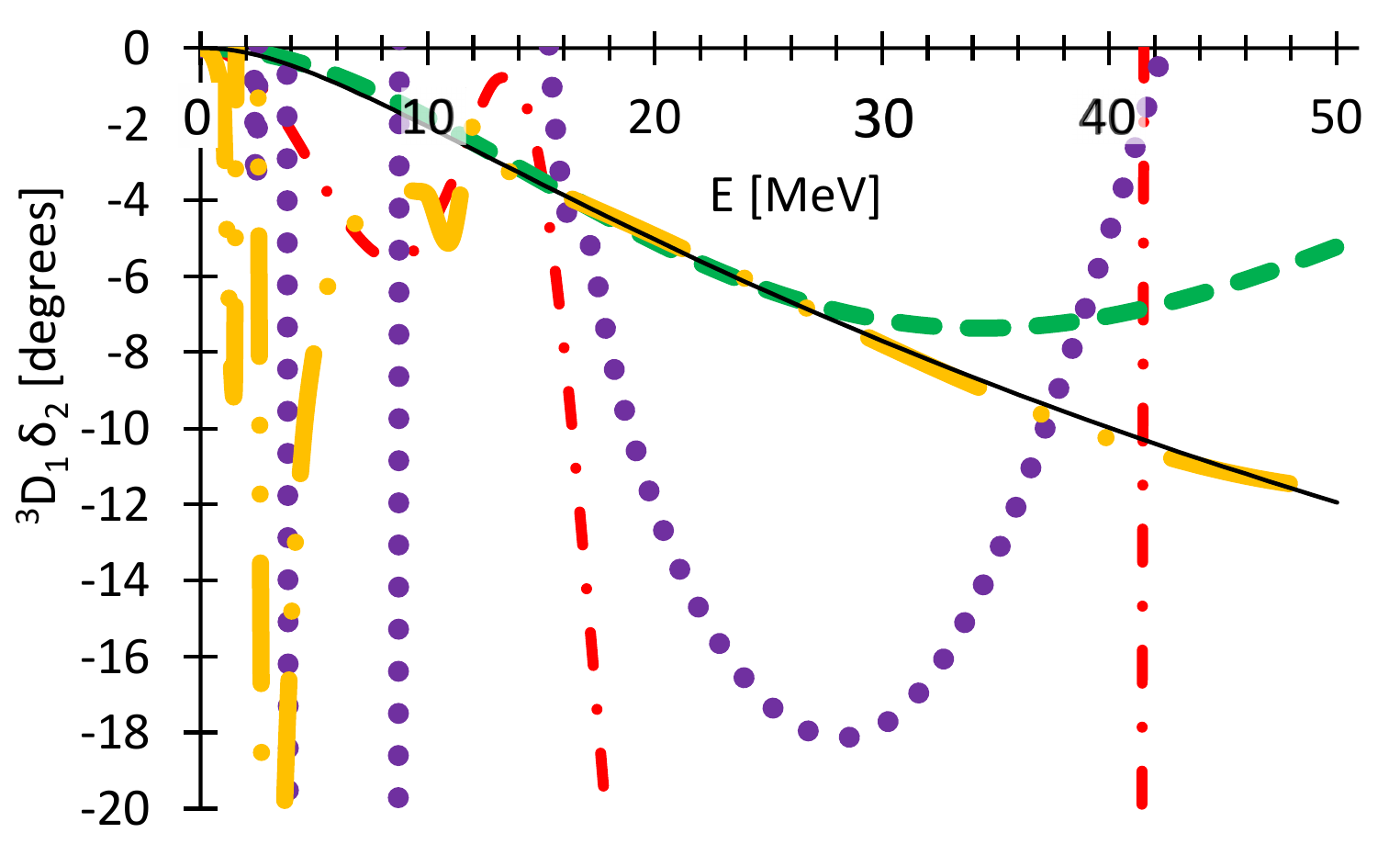}%
}\hfill
\subfloat[]{%
  \includegraphics[width=1\columnwidth]{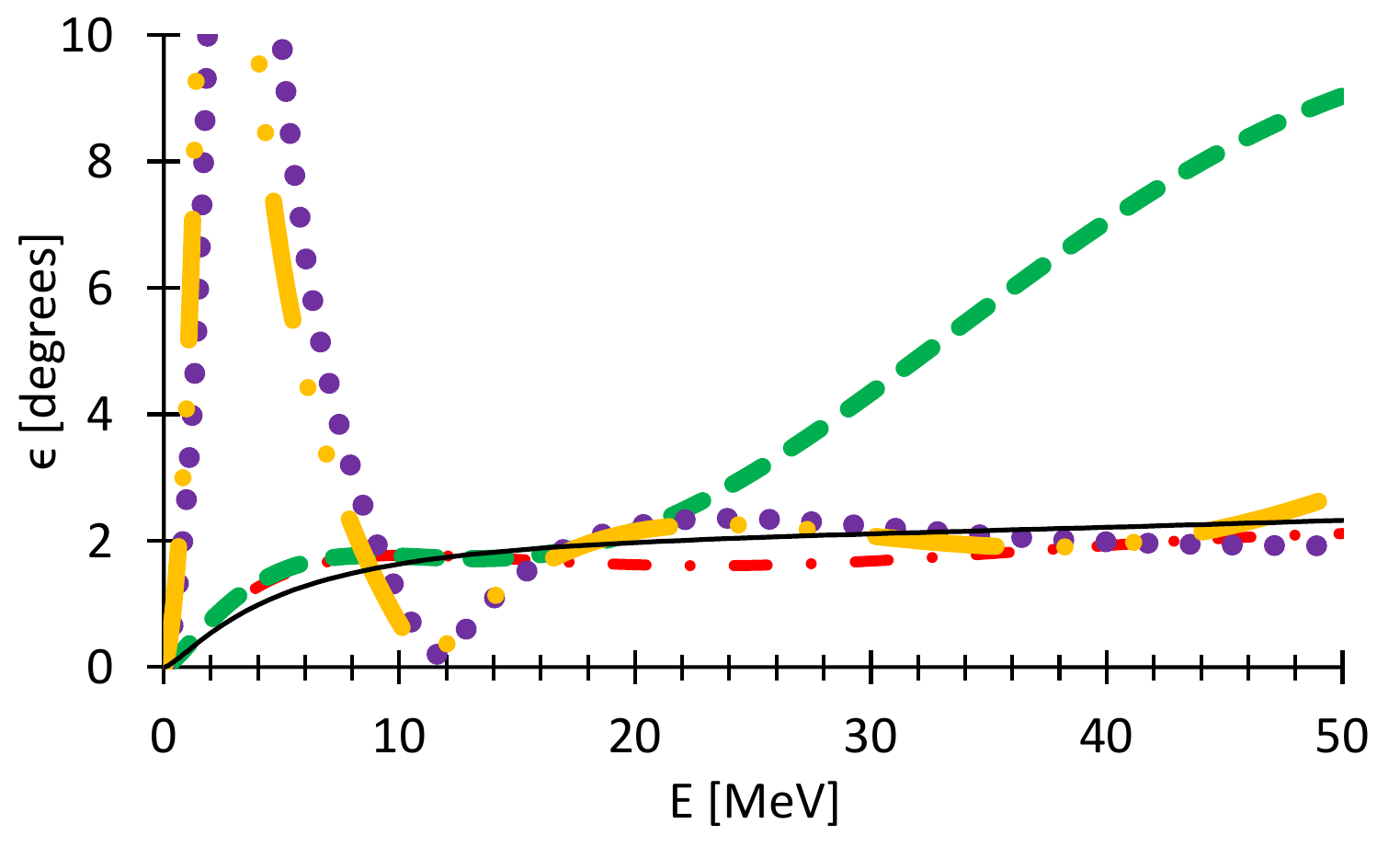}%
}\hfill
\caption[]{a) $^3S_{1}$ 
{\clg phase shifts}~$\delta_{0}$,  b) $^3D_{1}$ 
{\clg phase shifts}~$\delta_{2}$, c)~$^3SD_{1}$ {\clg mixing parameter}~$\epsilon$
{\clg obtained by \color{black} the adapted \color{black} Efros method with different}~$N$, $N_{\textrm{max}}=12$, 
$\hbar \Omega=30$~MeV, {\clg and} with smoothing 
{\clg parameter~${a=2.5}$} using the eigenfunction SRFs  
 {\clg and}~$^3S_{1}$ {\clg as the first  channel with~$i=1$, i.\,e.,} 
 imposing {\clg the} symmetry by setting~\clg$\mathbb{S}_{12}(E)$ 
 equal to~$\mathbb{S}_{21}(E)$.}
 \label{fig:S14}
\end{figure}

 Figure~\ref{fig:S9} demonstrates the effect of smoothing with  parameter~$a=2.5$
 on the phase shift calculations within \color{black} the adapted \color{black} Efros method with~$N_{\textrm{max}}=14$, $N=8$.  We see complete convergence when smoothing is used in contrast to when it is not used.

Figure~\ref{fig:S10} shows the convergence with~$N$ of the wave function expansion
coefficients~$d_{n}(E)$~[see Eq.~\eqref{eq49}] at the energy~${E=0.5}$~MeV where the ${^1S_{0}}$ phase shift obtained with smoothing at~$N_{\max}=14$ 
and~$\hbar \Omega=30$~MeV is close to the exact.  The oscillator amplitudes~$d_{n}(E)$ are seen to converge reasonably well.

We consider {also a couple of sets} 
of nonorthonormal { linearly independent}
SRFs { combining} 
both the harmonic oscillator and the eigenfunctions, 
\begin{gather}
\label{eq69}
\tag{S8}
 \bar{\beta}_{q}(\vec r)= 
\ \begin{cases} 
    \beta_{q}^{\textrm{Efn}}(\vec r), & q=0,...\, , q_{0},  \\ 
     \phi_{\mtc{N}-q+q_{0}}(\vec r), & q= q_{0}+1,...\, ,N-2,    \\ 
     \phi_{\mtc{N}}(\vec r), &  q=N-1, 
   \end{cases}
\end{gather}
and
\begin{gather}
\label{eq70}
\tag{S9}
 \bar{\beta}_{q}(\vec r)= 
\ \begin{cases} 
    \beta_{q}^{\textrm{Efn}}(\vec r), & q=0,...\,,q_{0},  \\ 
     \phi_{q-q_{0}-1}(\vec r),   & q=q_{0}+1,...\,,N-2,  \\ 
      \phi_{\mtc{N}}(\vec r), &  q=N-1. \hspace{1mm} 
   \end{cases}
\end{gather}
Here, $q_{0}+1$~is the number of low-lying eigenfunctions used.  For both 
Eqs.~\eqref{eq69} and \eqref{eq70}, the set of SRFs~$\{\beta_{q'}\},$
$q'=0,...\,,N-2$ appearing explicitly in Eq.~\eqref{eq18}~is the same as
{ the set of SRFs~$\{\bar{\beta}_q\}$} 
for $q=0,...\,, N-2$~{but} it excludes {$\bar{\beta}_{N-1}= \phi_{\mtc{N}}(\vec r)$.} 

{In Fig.~\ref{fig:S11} we present the $^1S_{0}$ phase shifts obtained with the SRF set
of Eq.~\eqref{eq69} with $N_{\textrm{max}}=14$,  $\hbar \Omega=30$~MeV, and~$N=6$. Vertical lines correspond to the energies of eigenstates of the Hamiltonian~$H^{tr}$
at which the phase shifts are well-described.}   
We see that at least~{$q_{0}=3$ is needed to get  reasonable phase shifts in a wide
range of energies while with}
$q_{0}=4$ {the results for the phase shifts are excellent.}

{Figure~\ref{fig:S12} presents the same results but obtained with the SRF set  of 
Eq.~\eqref{eq70} and smaller~${N_{\textrm{max}}=12}$.}  We see that in this case, we get reasonable results even with~$q_{0}=0$ and 
much more accurate 
with~$q_{0}=1$. Thus it is preferable to adopt the SRF set of Eq.~\eqref{eq70}.

\begin{figure}[t!]
\subfloat[]{%
  \includegraphics[width=1\columnwidth]{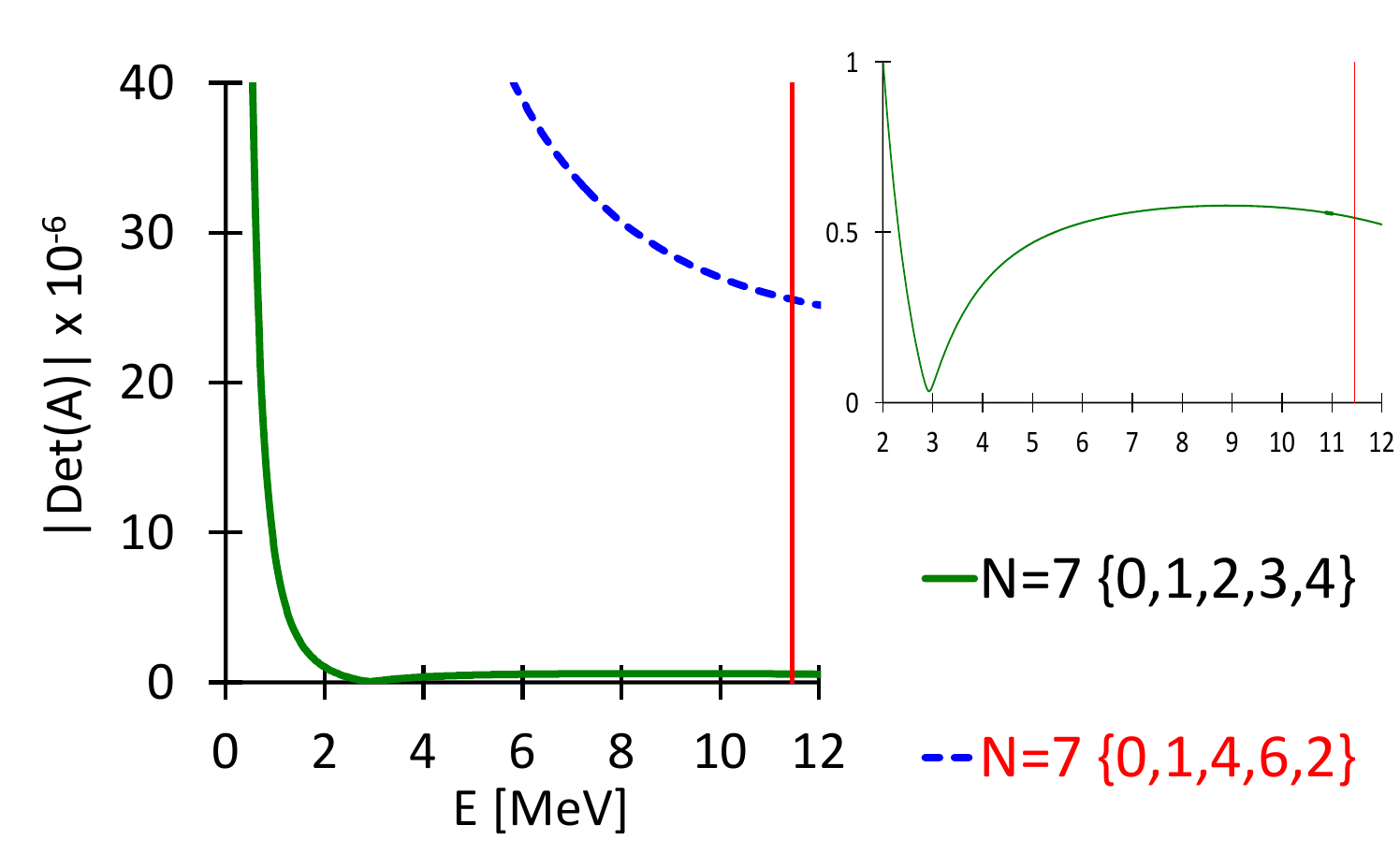}%
}\hfill
\subfloat[]{%
  \includegraphics[width=1\columnwidth]{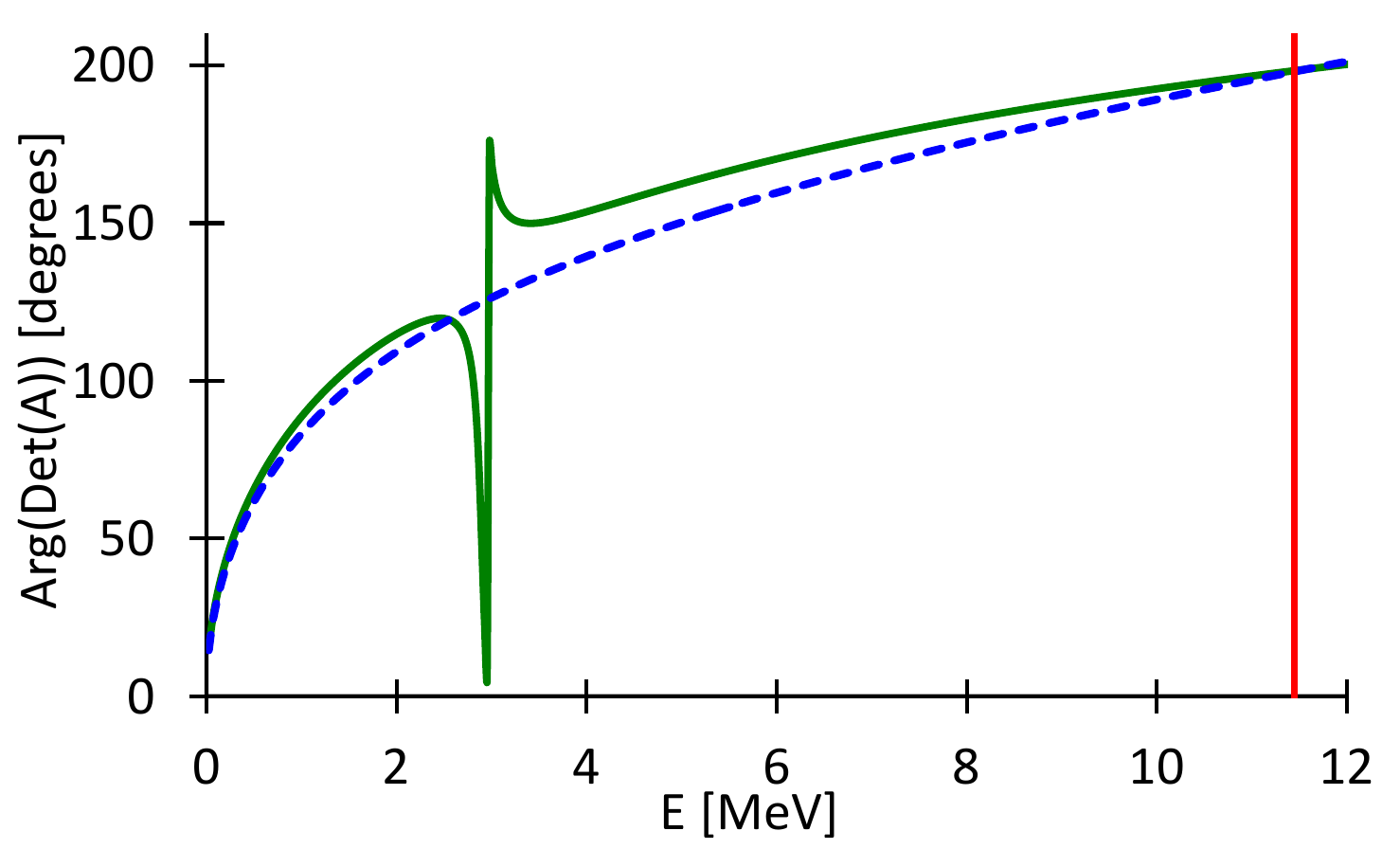}%
}
\caption{$^3SD_{1}$~channel: a)~$|\textrm{det}(A)|$, b)~$\textrm{Arg}(\textrm{det}(A))$ 
{\clg obtained in calculations with}
$N_{\textrm{max}}=12$, $\hbar \Omega=30$~MeV, 
{\clg and} smoothing 
{\clg parameter}~{\clg$a=2.5$} using the eigenfunction SRFs and the ${^{3}D}_{1}$ partial
wave as the first channel
 with and without 
{\clg the} basis rearrangement.  The vertical line corresponds to the first excited state 
{\clg eigenenergy} 
of~$\clg H^{tr}\!$.
}
 \label{fig:S15}
\end{figure}

\begin{figure}[t!]
\subfloat[]{%
  \includegraphics[width=1\columnwidth]{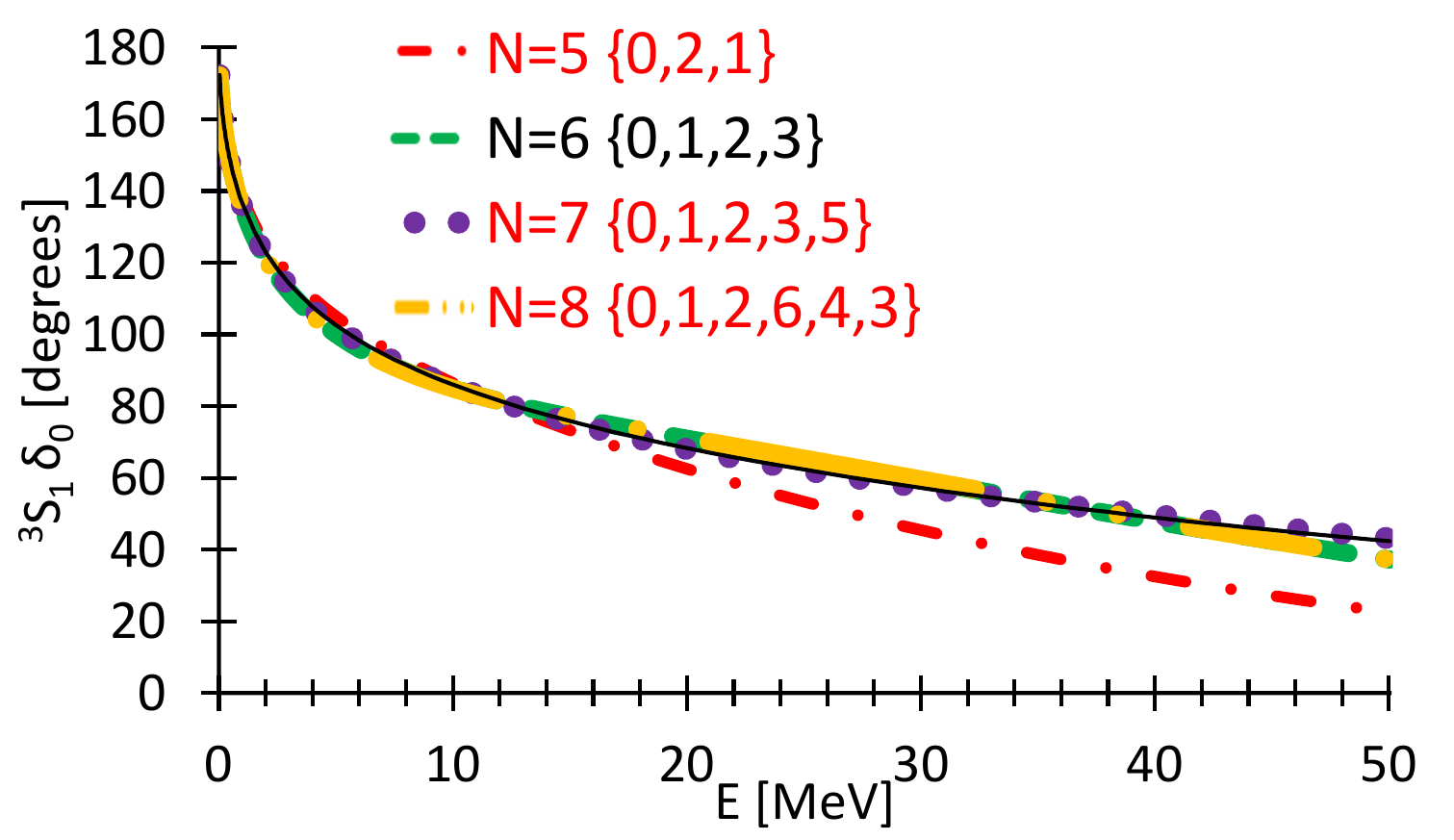}%
}\hfill
\subfloat[]{%
  \includegraphics[width=1\columnwidth]{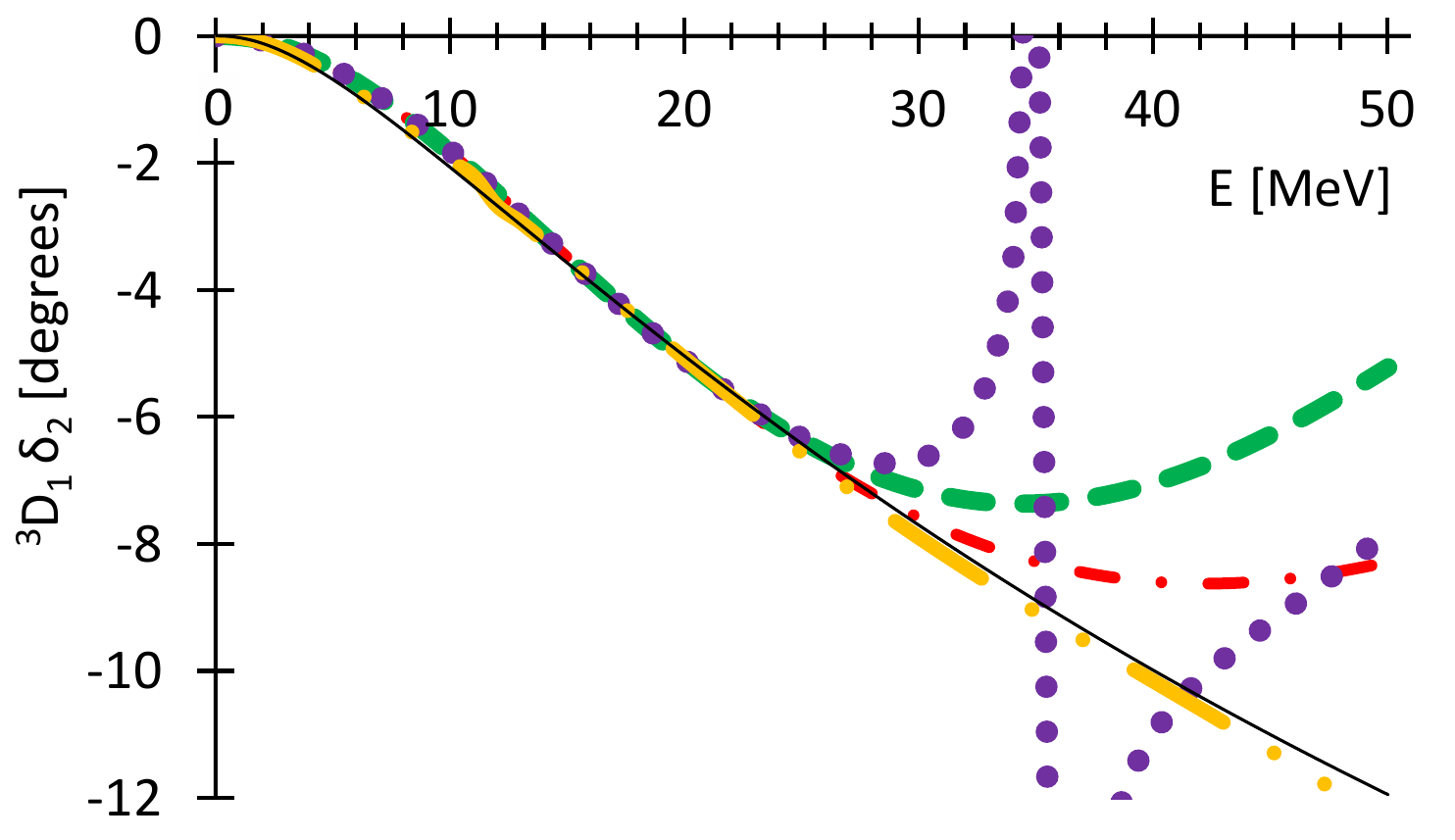}%
}\hfill
\subfloat[]{%
  \includegraphics[width=1\columnwidth]{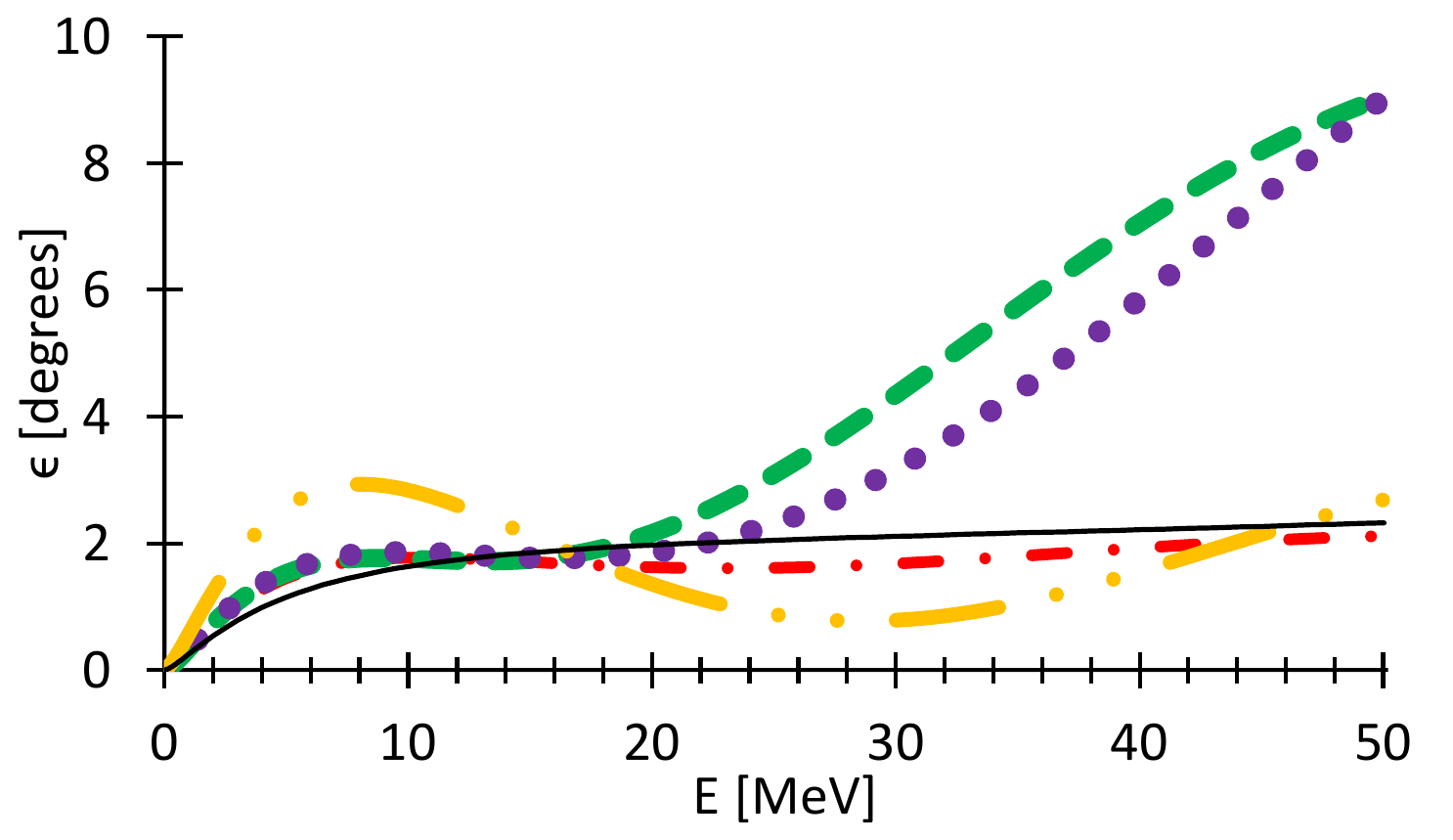}%
}\hfill
\caption[]{a) $^3S_{1}$ 
{\clg phase shifts}~$\delta_{0}$,  b) $^3D_{1}$ 
{\clg phase shifts}~$\delta_{2}$, c)~$^3SD_{1}$ {\clg mixing parameter}~$\epsilon$
{\clg obtained with different}~$N$, $N_{\textrm{max}}=12$, 
$\hbar \Omega=30$~MeV, {\clg and}
with smoothing 
{\clg parameter} ${a=2.5}$ using the eigenfunction {\clg SRFs and} 
$^3S_{1}$ {\clg as the first  channel with~$i=1$} 
 and rearranging the set of SRFs to eliminate 
 {\clg the} vanishing {\clg point in the} determinant of~$A$.  This elimination is checked over a range of~$E$~from 0 to 50 MeV. 
}
 \label{fig:S16}
\end{figure}

\section{Two-channel scattering}
\label{SectionIII}
{\clg We present in Fig.~\ref{fig:S13} convergence with~$N_{\textrm{max}}$~of the phase shifts and mixing parameter~$\epsilon$
in the ${^{3}SD}_{1}$ coupled partial waves
of $np$ scattering using the HORSE method and smoothing 
with the parameter~$a=2.5$.  We see that~$N_{\textrm{max}}=12$~gives reasonable results for both phase shifts but still some small deviations from the
exact results for the mixing parameter~$\epsilon$ while the complete convergence is
in a wide energy range is
achieved at~$N_{\max}=20$.}

If we use a ``complete'' set of SRFs~$\{\beta_{q}\}$ in Eq.~\eqref{eq1} which is enough
to form a complete basis in the subspace of the oscillator functions~$\phi_{ni}(\vec r)$ 
with~$n=0,1,...\,,{\cal N}_{i}$, $i=1,2$,  \color{black} the adapted Efros method \color{black}  \color{black} is equivalent to the HORSE method.
This set of SRFs~$\{\beta_{q}\}$ may be arbitrary: it is only important that the 
functions~$\beta_{q}$ are linearly independent. In this case, one can use the
over-complete Eq.~\eqref{eq25} which suggest additional relations for independent
calculations of the unknowns that should be equal 
such as~$\mathbb{S}_{12}(E)=\mathbb{S}_{21}(E)$ in the case of two-channel scattering. As an
example we consider the $S$-matrix
calculations at the energy of~$E=0.5$~MeV in the coupled~$^3SD_{1}$ partial waves of $np$ scattering.~We employ
the set of~$H^{tr}$~eigenfunction SRFs
for~$N_{\textrm{max}}=12$~and~$N=15$, i.\,e., all of the 13 {$H^{tr}$} eigenfunctions are used in the
set of the SRFs~$\{\beta_{q}\}$~and two additional oscillator functions  
used exclusively for orthogonalization are included in the set~$\{\bar{\beta}_{q}\}$ as in Eq.~\eqref{eq74}. 
 We obtain the $S$ matrix
$$
\mathbb{S}(E)=\begin{pmatrix}
   0.4193 - 0.9078 i & -0.0022 - 0.0035 i  \\[0.3em]
   -0.0022 - 0.0035 i & 1 - 0.0002 i  \\[0.3em]
     \end{pmatrix}\!, 
     $$
which is 
unitary and~$\mathbb{S}_{12}(E)=\mathbb{S}_{21}(E)$.  Since
we can switch from the ``complete'' 
set of eigenfunction SRFs to an {equivalent} 
set of harmonic oscillator SRFs via a unitary transformation, both sets of SRFs
yield the same solutions. 
{ Thus the ``complete'' set of SRFs~$\{\beta_{q}\}$ guarantees the consistency of relations
between unknowns following from Eq.~\eqref{eq25}.}

{ As it was already noted, in the above case of the ``complete'' set of SRFs~$\{\beta_{q}\}$,
\color{black} the adapted \color{black} Efros and HORSE methods give the same results.  Since the ``complete'' set of SRFs in
current {\it ab initio} nuclear structure applications is extremely large, it}
is more interesting to investigate the { essential simplification suggested within \color{black} the adapted \color{black}
Efros method as compared to the HORSE method which makes it possible to use a reduced
set of SRFs. Let us}
consider the same case as above, but for~$N=12$.  In this case, Eq.~\eqref{eq74}
results in the $S$ matrix
     $$
     \mathbb{S}(E)=\begin{pmatrix}
   0.4169 - 0.909 i & -0.0021 - 0.0033 i  \\[0.3em]
   -0.0016 - 0.0025 i & 1 - 0.0002 i  \\[0.3em]
     \end{pmatrix} 
     $$
with broken unitarity and
$\mathbb{S}_{12}(E) \neq \mathbb{S}_{21}(E) $.~In the case of a reduced set of
SRFs~$\{\beta_{q}\}$, one is pushed to use Eq.~\eqref{eq31} and impose the $S$-matrix
symmetry either by calculating~$\mathbb{S}_{21}(E)$ and 
setting~$\mathbb{S}_{12}(E)$ equal to {\clb it} 
or vice versa.
\color{black}

\begin{figure}[t!]
\subfloat[]{%
  \includegraphics[width=1\columnwidth]{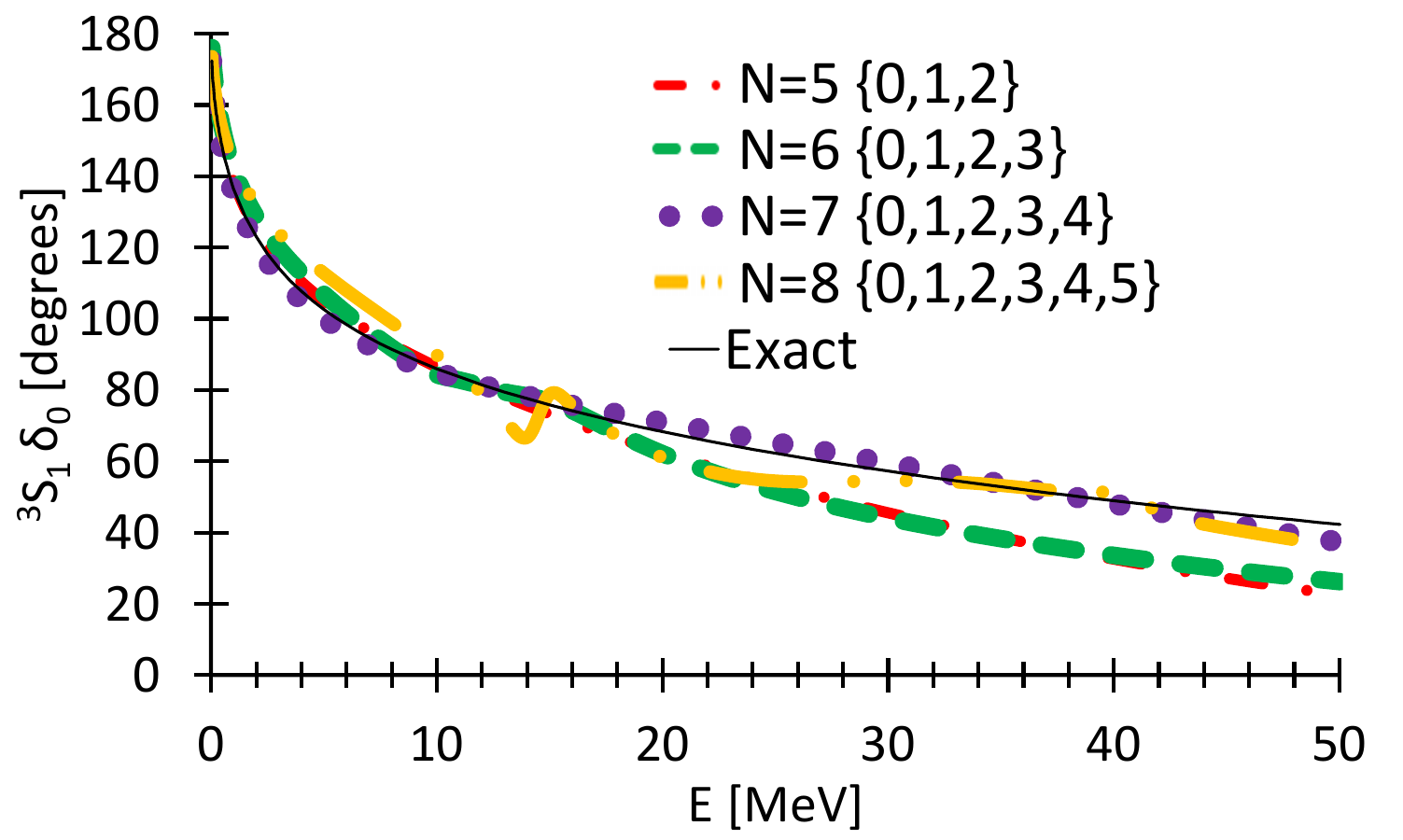}%
}\hfill
\subfloat[]{%
  \includegraphics[width=1\columnwidth]{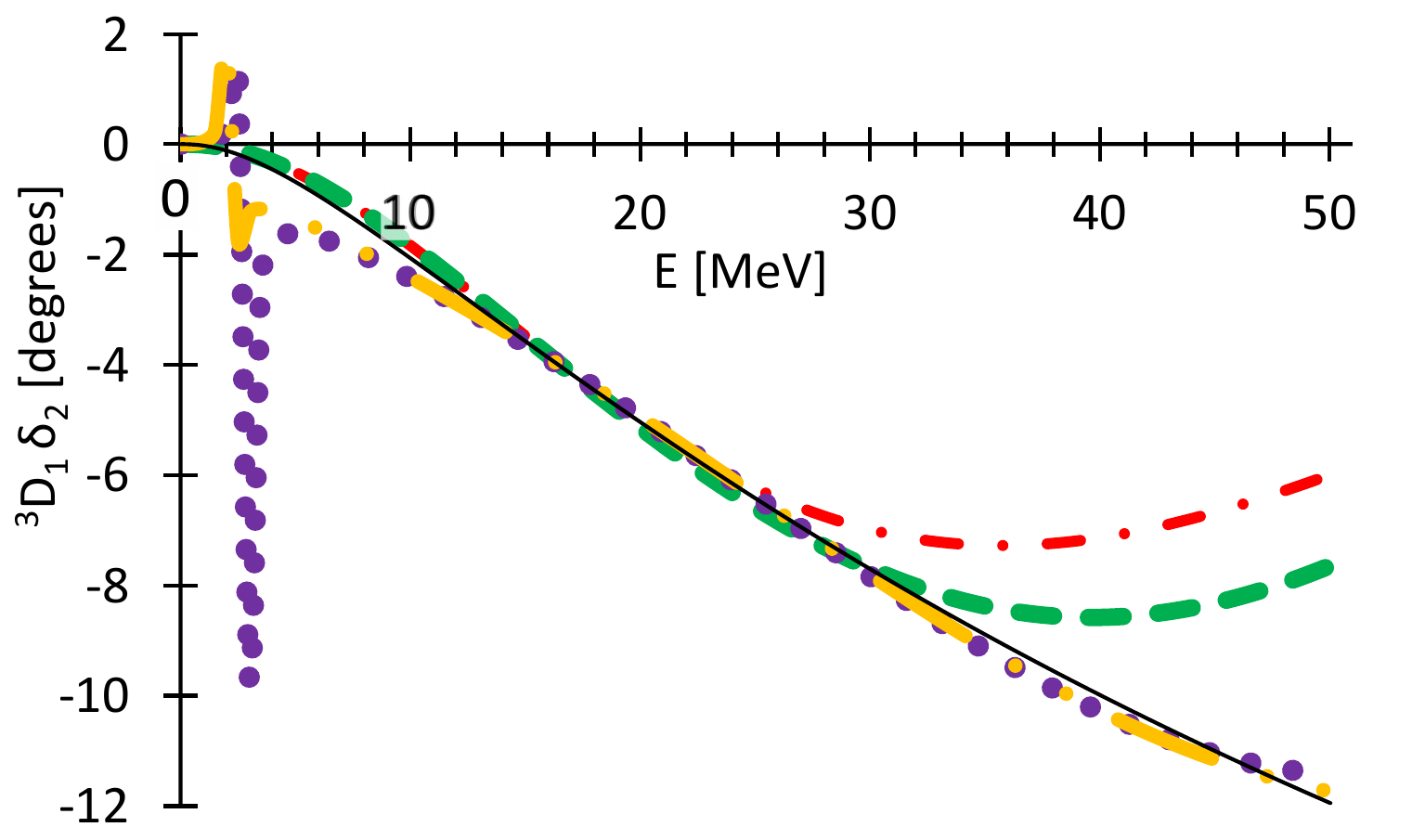}%
}\hfill
\subfloat[]{%
  \includegraphics[width=1\columnwidth]{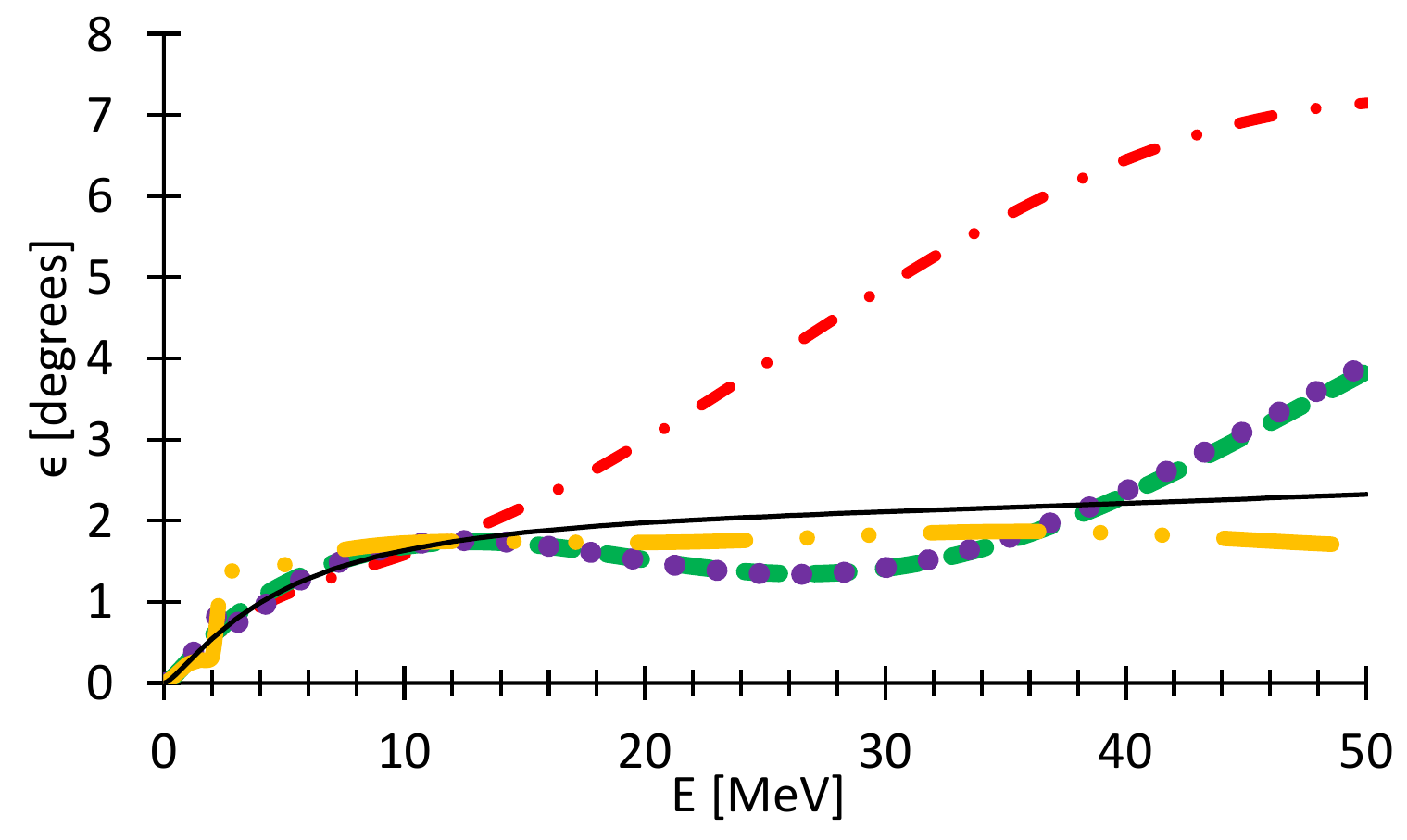}%
}\hfill
\caption{ a) $^3S_{1}$ 
{\clg phase shifts}~$\delta_{0}$,  b) $^3D_{1}$ 
{\clg phase shifts}~$\delta_{2}$, c)~$^3SD_{1}$ {\clg mixing parameter}~$\epsilon$~obtained by \color{black} the adapted \color{black} Efros method with different~$N$, $N_{\textrm{max}}=12$,
$\hbar \Omega=30$~MeV, {\clg and}~$^3D_{1}$~as the first  channel with~$i=1$, i.\,e., imposing {\clg the} symmetry by setting~\clg$\mathbb{S}_{21}(E)$ 
 equal to~$\mathbb{S}_{12}(E)$. 
}
 \label{fig:S17}
\end{figure}

In the case of \color{black} the adapted \color{black} Efros method with
a limited set of SRFs, we present in~Fig.~\ref{fig:S14}~the phase shift convergence \color{black} in which we {\clg use}
the~$^3S_{1}$
{\clg as the first channel corresponding to~$i=1$ and set~$l_{1}=0$
and~$l_{2}=2$}.  \color{black} We use the same notation in Fig.~\ref{fig:S14} as that used in the label for Fig. \ref{fig18} in the main text. \color{black}  In general, we see reasonable results for~$N=5$~and~$N=6$~at certain energy intervals but \color{black} emerging numerical problems for~${N=7}$ and~$8$~in the form of jumps.

\begin{figure}[t!]
\subfloat[]{%
  \includegraphics[width=1\columnwidth]{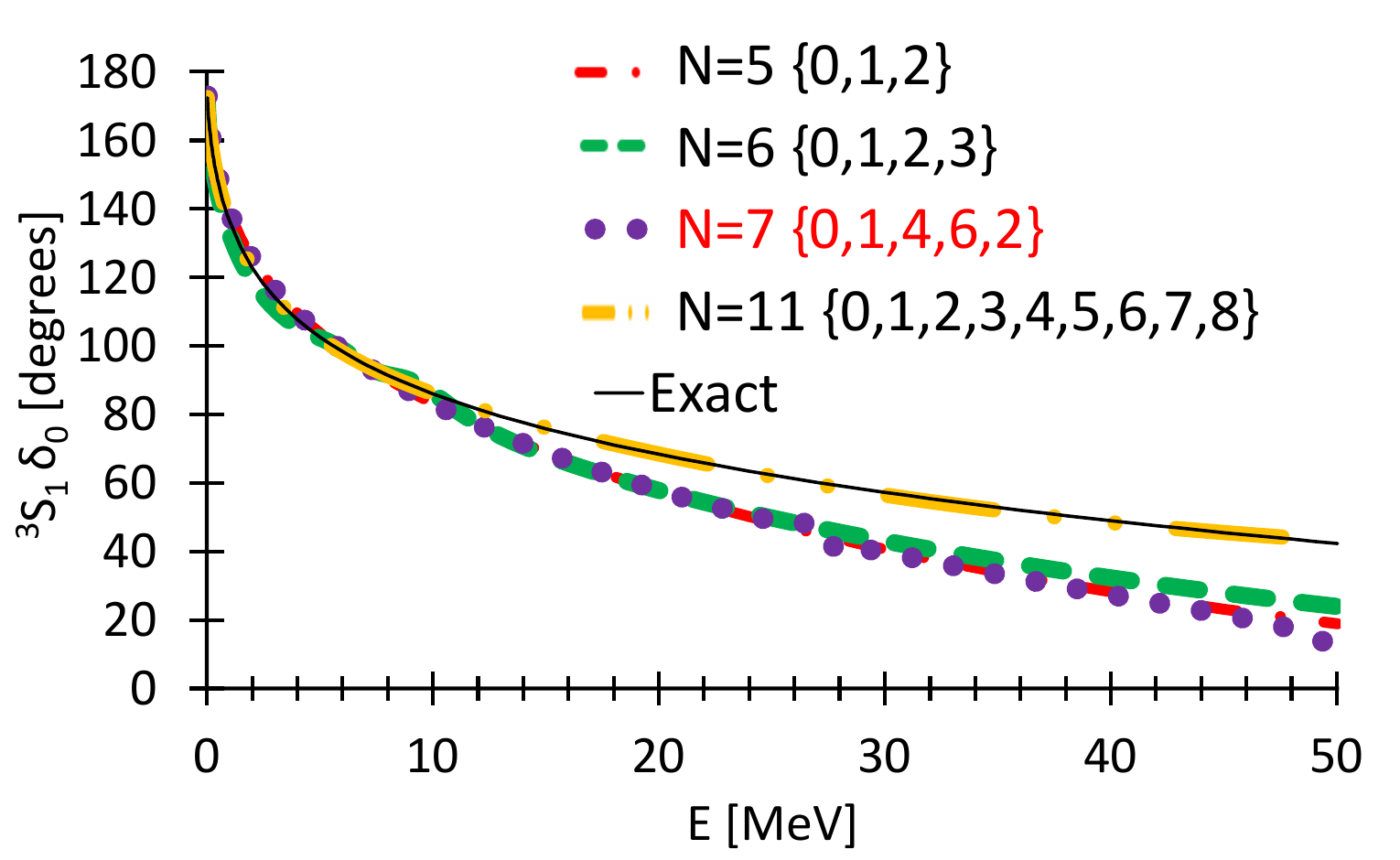}%
}\hfill
\subfloat[]{%
  \includegraphics[width=1\columnwidth]{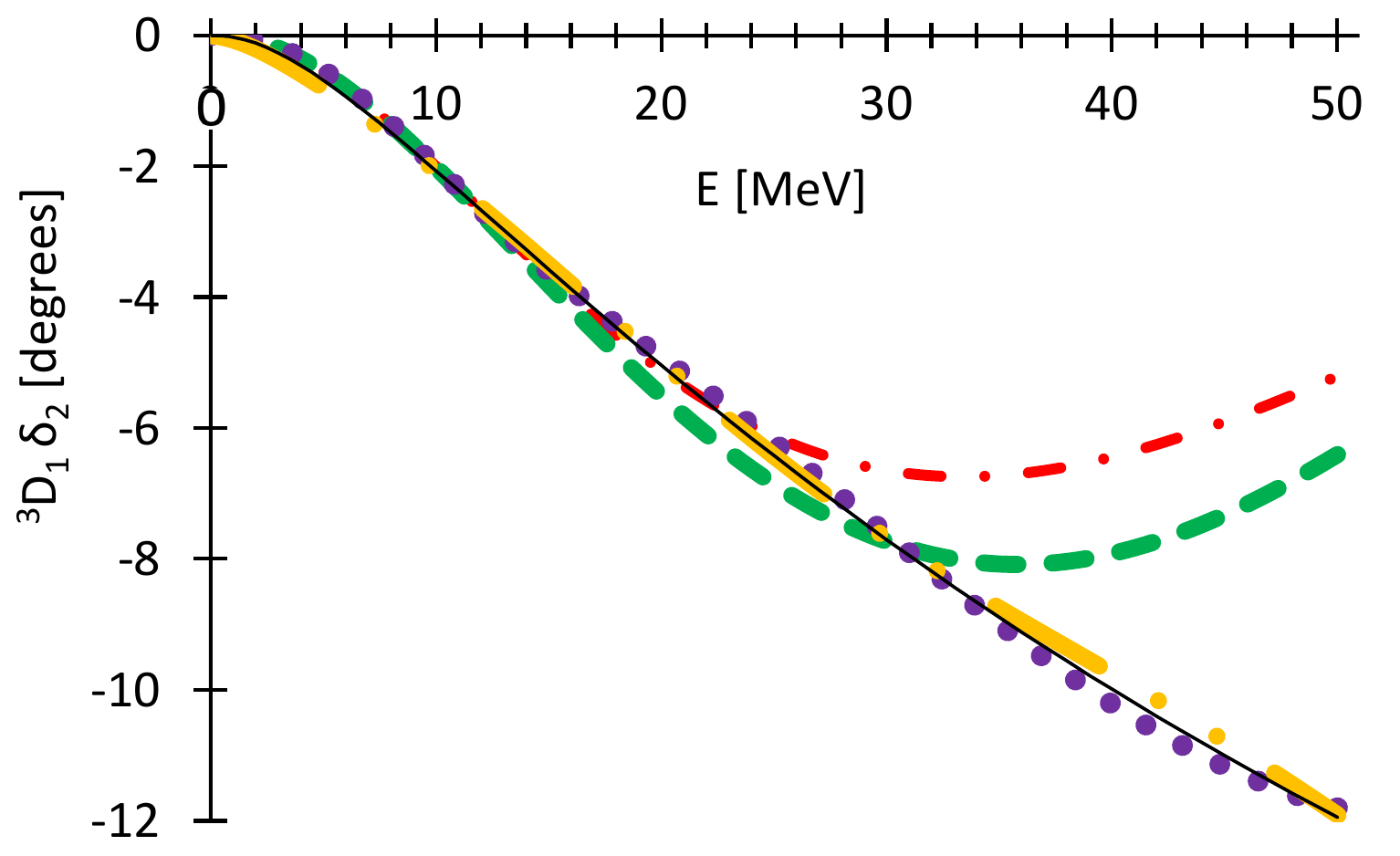}%
}\hfill
\subfloat[]{%
  \includegraphics[width=1\columnwidth]{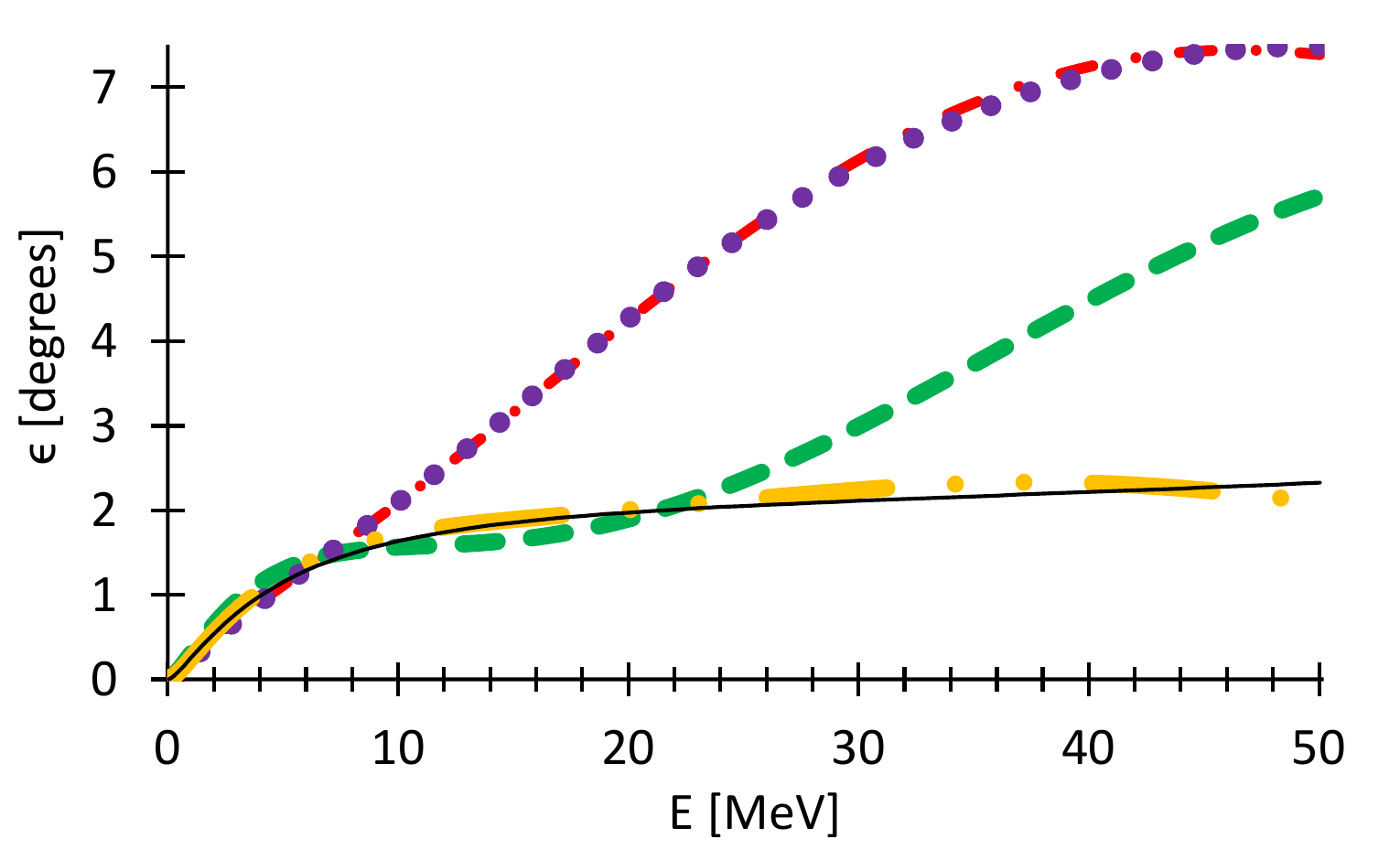}%
}\hfill
\caption[]{\clg The same as \color{black} Fig. \ref{fig18} \color{black} in the main text, but calculations are performed with
$N_{\max}=20$.
}
 \label{fig:S18}
\end{figure}

As illustrated in Fig.~\ref{fig:S15}, the origin 
of these spikes is due to the vanishing determinant of~$A$~given by Eq.~\eqref{eq33} at the energy~$E$ around 3~MeV.  Here, we plot \color{black} {\clg $|\det(A)|$,  the absolute magnitude} 
of the {\clg matrix~$A$} determinant,  
and 
{\clg $\textrm{Arg}\big(\textrm{det}(A)\big)$, the determinant} argument, are plotted as functions of~$E$ 
{\clg in the case of}~$N=7$ 
{\clg and using the $^3D_{1}$ as the first} channel.~If we use the \color{black} selection \color{black} of {\clg SRF} eigenfunctions with increasing eigenvalues, 
{\clg$N=7\,\{0,1,2,3,4\}$, there is a clearly seen}
discontinuity {\clg of~$\textrm{Arg}\big(\det(A)\big)$} 
around~$E=3$~MeV {\clg which} 
coincides with 
the spikes in Fig.~\ref{fig:S14} for $\clg\delta_{2}$ and~$\clg\epsilon$.  \color{black} The SRF selection~$\{0,1,4,6,2\}$ \color{black} 
fixes the discontinuities in~$\textrm{Arg}(\textrm{det}(A))$~and the spikes that they cause.    In particular, we simply 
replace one of the SRFs~$\beta_{q}^{\textrm{Efn}}(\vec r)$ and change their
 selection which results in replacing the omitted function~$\bar{\beta}_{q}^{\textrm{Efn}}(\vec r)$  
in the orthonormalization of the~$i=2$ equations.  We have not yet found a systematic way to change the SRF \color{black} selection \color{black}
to eliminate the vanishing {\clg of the~$A$} determinant.
Generally, {\clg our} 
preference is {\clg to retain the lowest-lying} 
eigenfunctions {\clg since they are  important in the formation of scattering
observables at low energies.} {\clg This adjustment} of  the {\clg set of SRFs}
improves the results as {\clg demonstrated} 
in Fig.~\ref{fig:S16}. It is seen that the
convergence trends with an increasing number of SRFs~$N$~can be complicated and
non-monotonic in the two-channel case.

\begin{figure*}[t!]
\subfloat[]{%
  \includegraphics[width=0.36\textwidth]{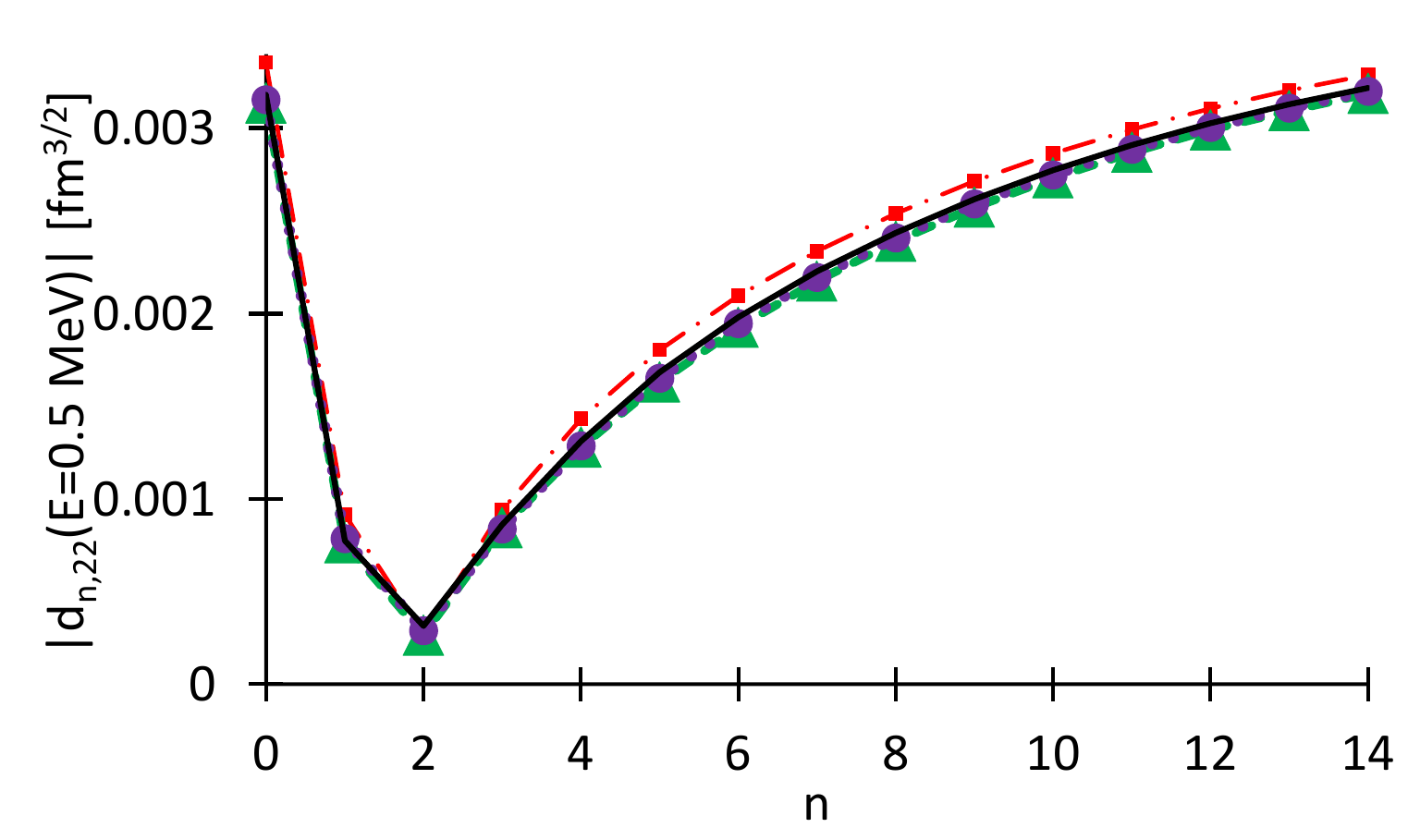}%
}\hspace{0.6cm}
\subfloat[]{%
  \includegraphics[width=0.36\textwidth]{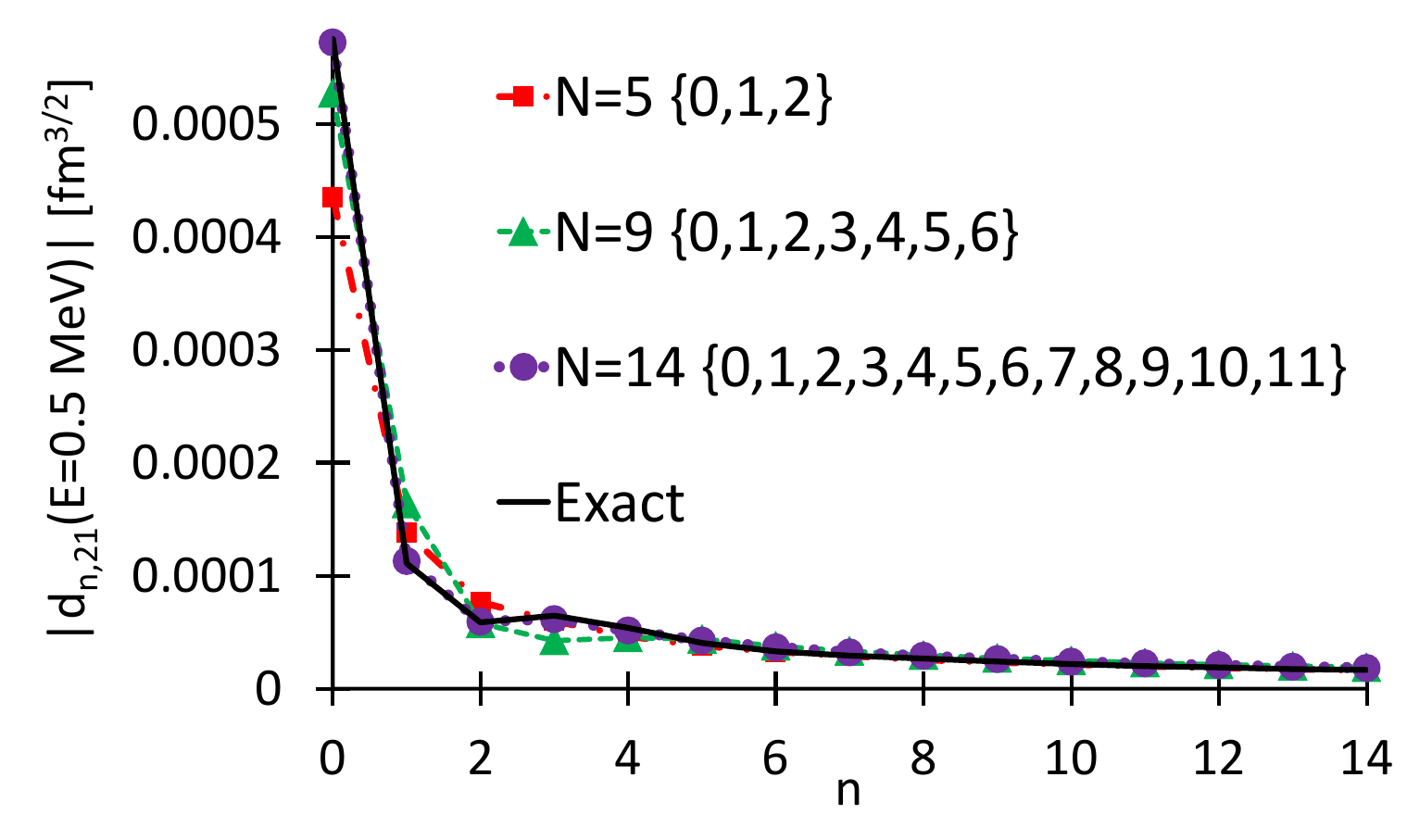}%
}\hspace{1cm}
\subfloat[]{%
  \includegraphics[width=0.4\textwidth]{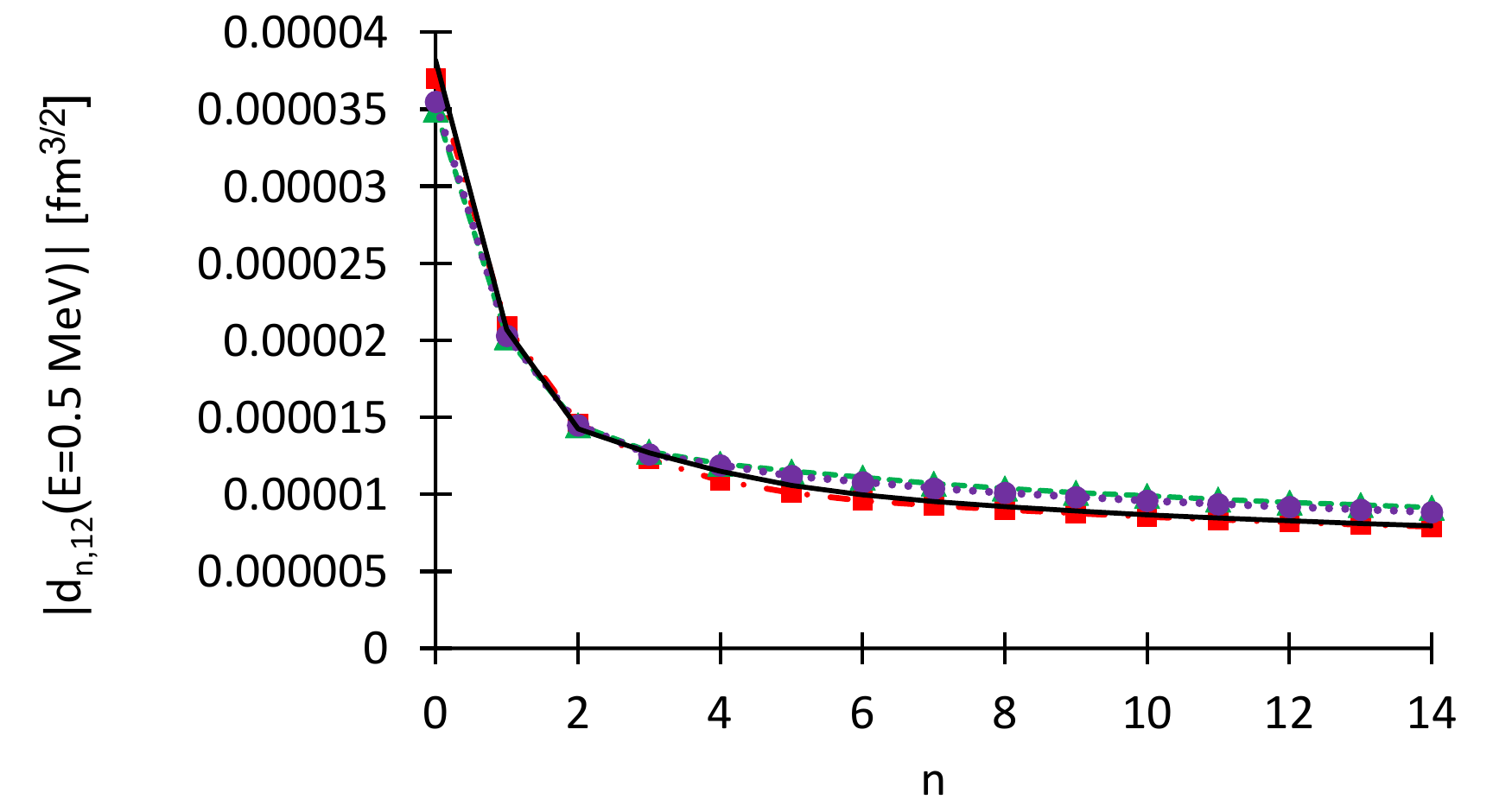}%
}\hspace{0.2cm}
\subfloat[]{%
  \includegraphics[width=0.38\textwidth]{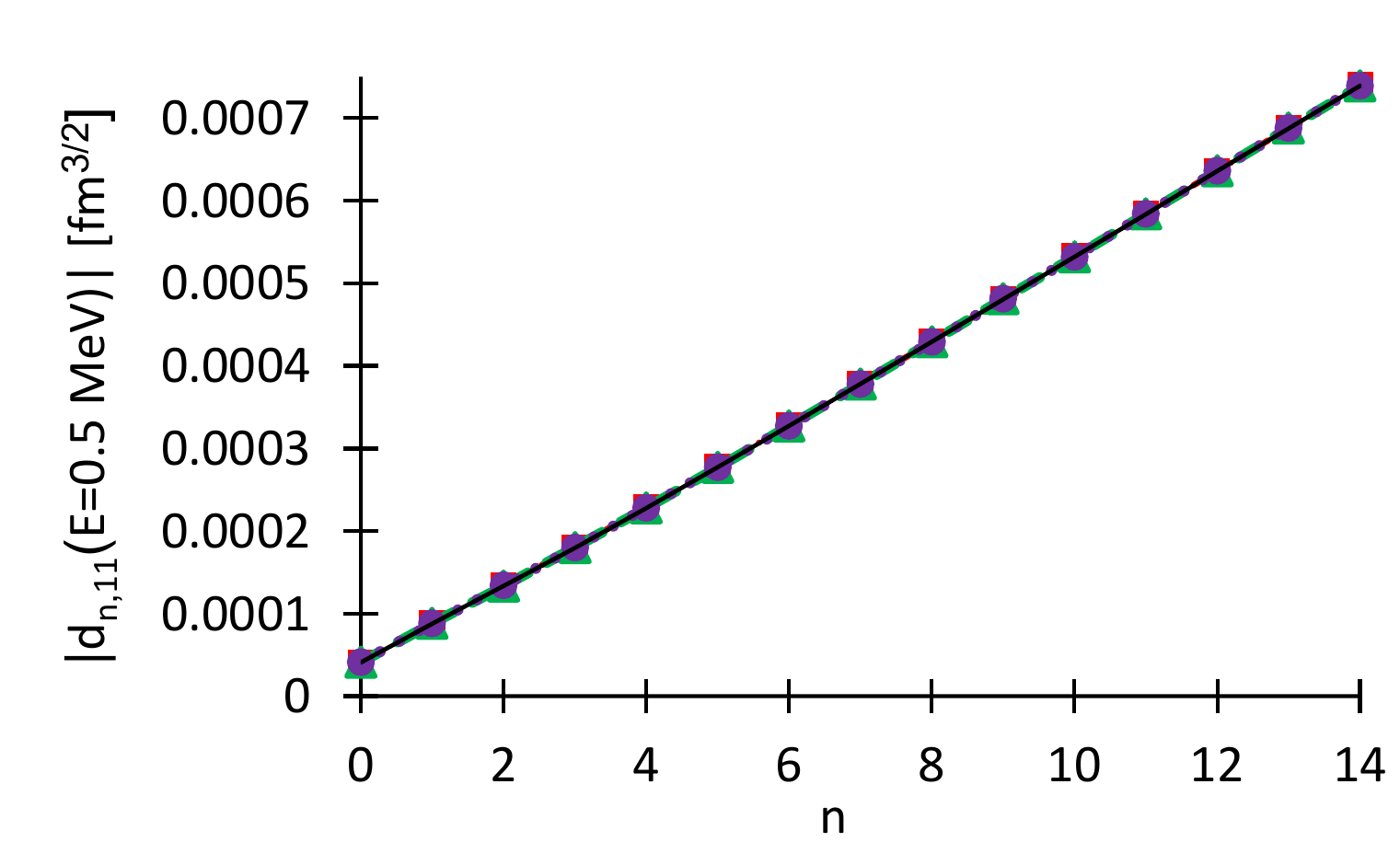}%
}\hspace{0.9cm}
\caption[]{\clg Absolute magnitudes of ${^{3}SD_{1}}$ wave function amplitudes 
at the energy of~$E=0.5$~MeV obtained  with 
different~$N$,  $N_{\textrm{max}}=12$, $\hbar \Omega=30$~MeV, and smoothing 
parameter~$a=2.5$  using the eigenfunction SRFs and  $^3D_{1}$ as the first  channel 
with~$i=1$:
a)~$|d_{n,22}(E)|$ (``strong'' $^{3}S_{1}$ component), 
b)~$|d_{n,21}(E)|$ (``weak'' $^{3}S_{1}$ component), 
c)~$|d_{n,12}(E)|$ (``weak''  $^{3}D_{1}$ component), 
d)~$|d_{n,11}(E)|$ (``strong'' $^{3}D_{1}$ component). 
}
 \label{fig:S19}
\end{figure*}

\begin{figure*}[t!]
\subfloat[]{%
  \includegraphics[width=0.36\textwidth]{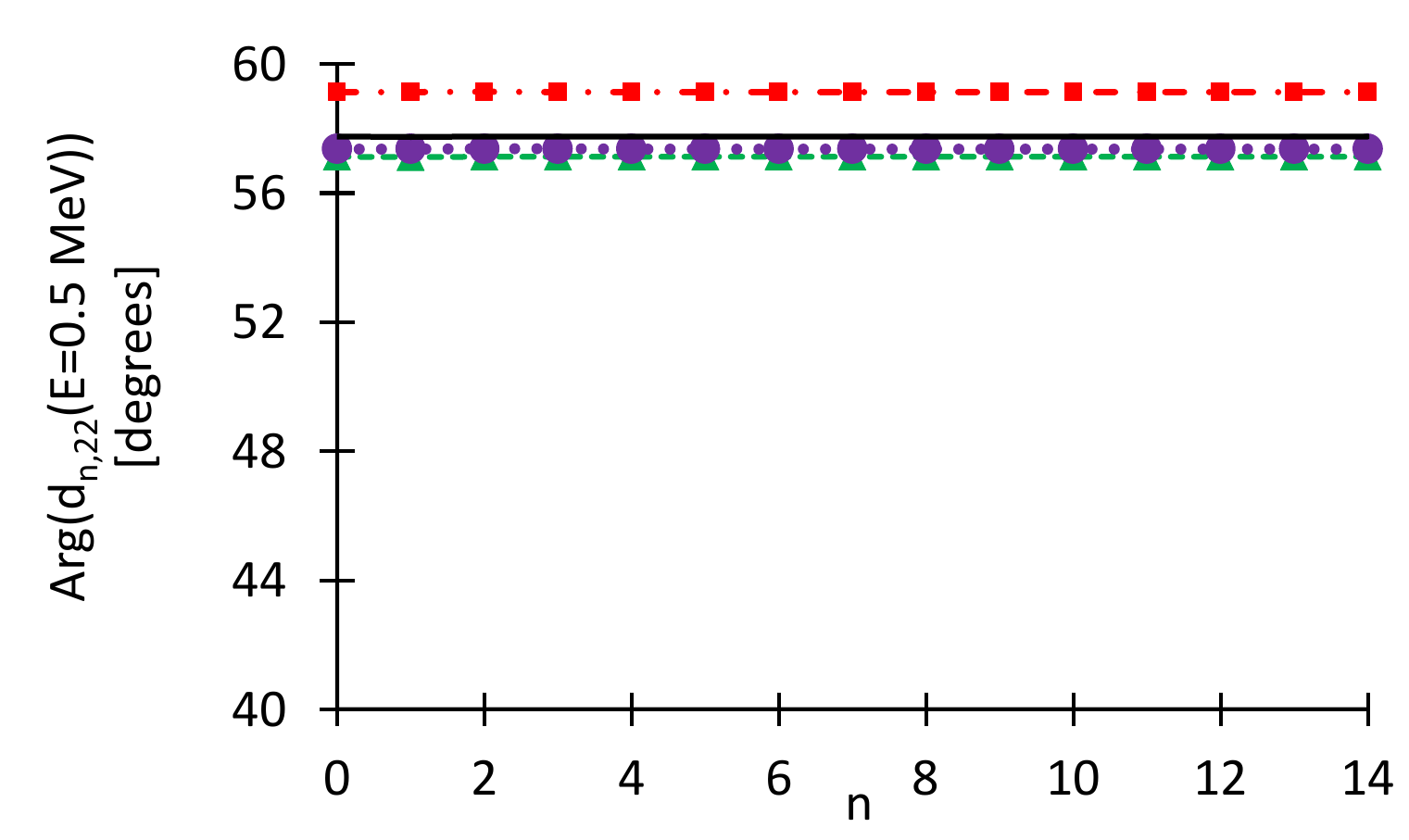}%
}\hspace{0.6cm}
\subfloat[]{%
  \includegraphics[width=0.36\textwidth]{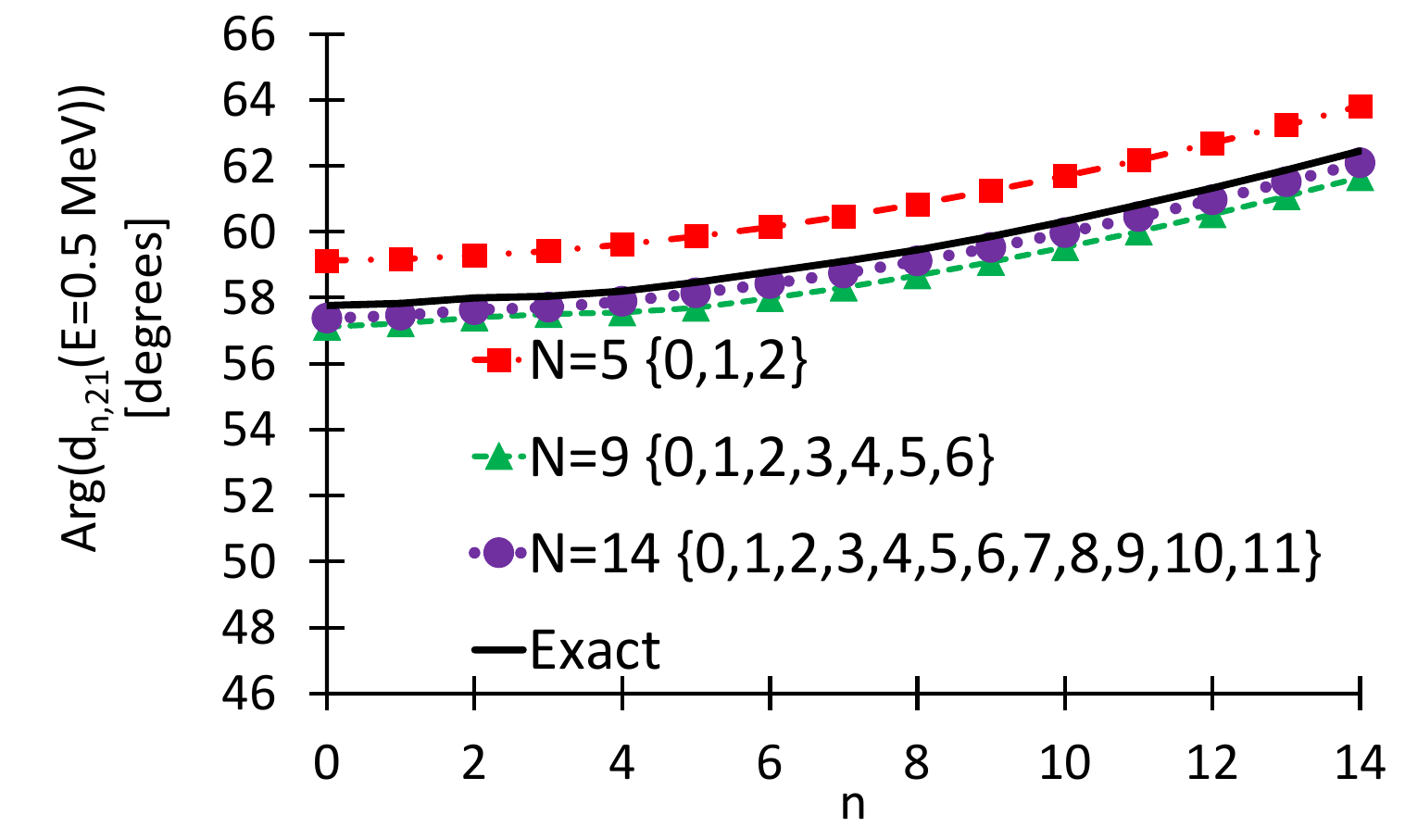}%
}\hspace{1cm}
\subfloat[]{%
  \includegraphics[width=0.36\textwidth]{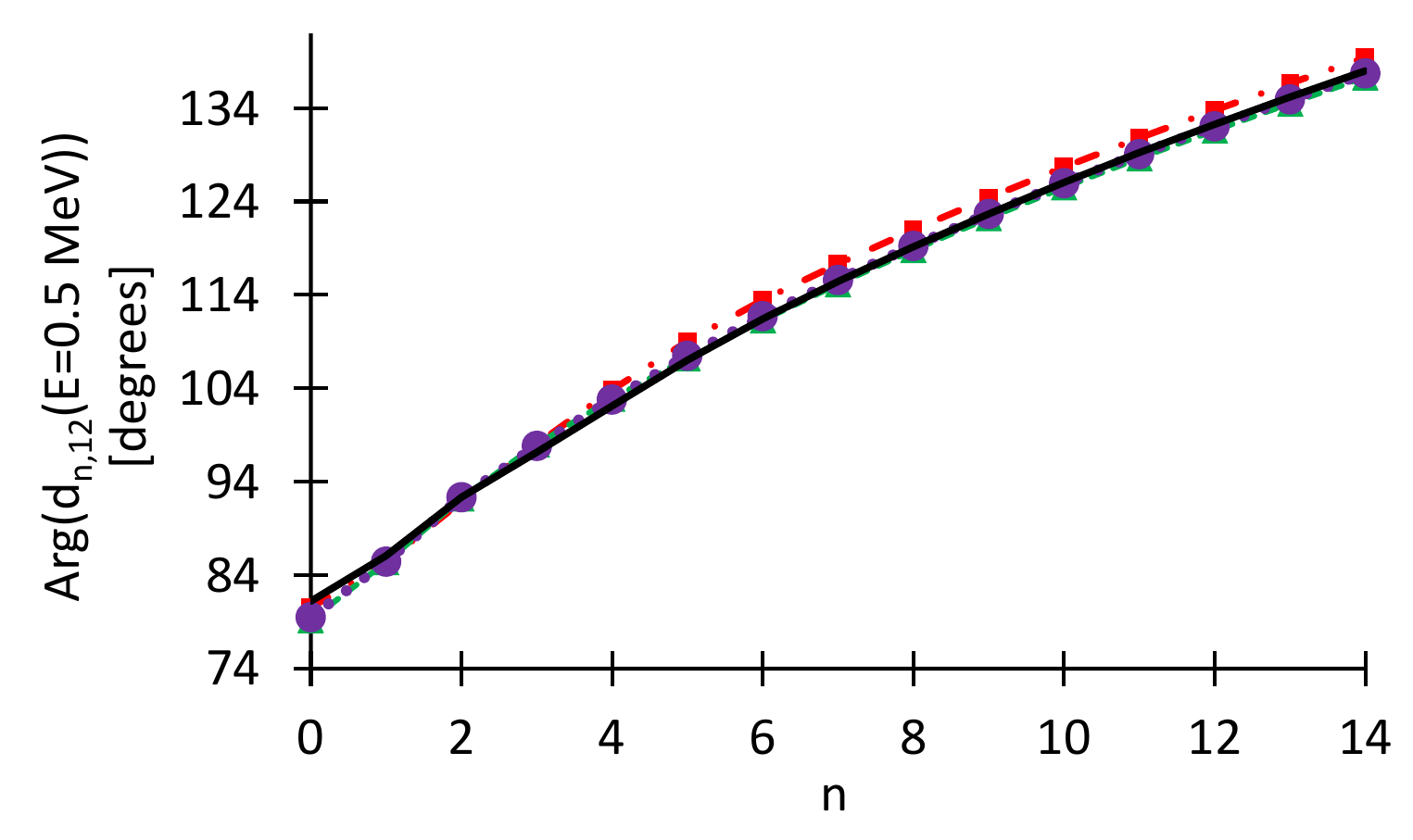}%
}\hspace{1cm}
\subfloat[]{%
  \includegraphics[width=0.36\textwidth]{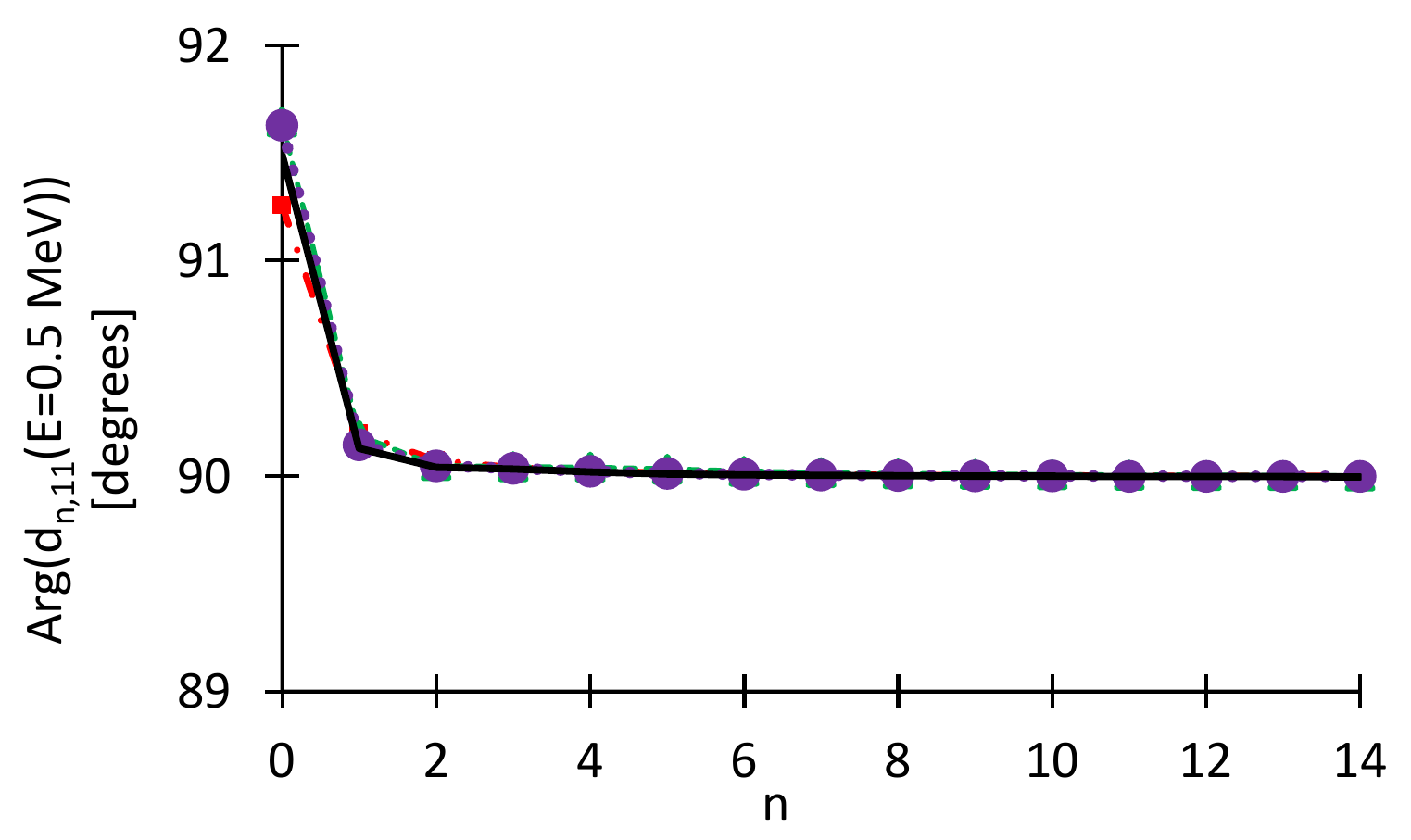}%
}\hspace{-0.1cm}
\caption[]{\clg Arguments of ${^{3}SD_{1}}$ wave function amplitudes 
at the energy of~$E=0.5$~MeV obtained  with 
different~$N$,  $N_{\textrm{max}}=12$, $\hbar \Omega=30$~MeV, and smoothing 
parameter~$a=2.5$  using the eigenfunction SRFs and $^3D_{1}$ as the first  channel 
with~$i=1$: 
a)~$\textrm{Arg}(d_{n,22}(E))$ (``strong'' $^{3}S_{1}$ component), 
b)~$\textrm{Arg}(d_{n,21}(E))$ (``weak'' $^{3}S_{1}$ component), 
c)~$\textrm{Arg}(d_{n,12}(E))$ (``weak'' $^{3}D_{1}$ component),  
d)~$\textrm{Arg}(d_{n,11}(E))$ (``strong'' $^{3}D_{1}$ component). 
}
 \label{fig:S20}
\vspace{-1ex}
\end{figure*}

 In Fig.~\ref{fig:S17}, we present the convergence of the phase shifts and mixing angle with~$N$ at $N_{\textrm{max}}=12$ obtained using
 the~$^3D_{1}$~as the first channel corresponding to~$i=1$ and setting~$l_{1}=2$~and~$l_{2}=0$
(i.\,e., {solving first the complete set of equations in}  the~$^3D_{1}$ {partial wave}
and {next switching to} the reduced {set of equations in} the~$^3S_{1}$ { partial wave).  We obtain reasonable results for $N=5$ and~$N=6$
at least up to~$E=16$~MeV for all quantities. 
The  results are still accurate in cases of~$N=7$ and~$N=8$.  
However, $\delta_{0}$ {obtained with~$N=8$}
shows some oscillation 
{around~$E\approx12$~MeV}
 and~$\delta_{2}$ and~$\clg\epsilon$ { have} 
 jumps 
 around~$ E\approx3$~MeV in cases of both~$N=7$ and~$N=8$.  \color{black} Those numerical issues are fixed by \color{black} adopting \color{black} the SRF selection \color{black} \color{black} in Fig. \ref{fig18} in the main text. \color{black}
 
 Figure \ref{fig:S18} shows the results of calculations with~${N_{\max}=20}$, with the rest of the parameters 
{ the same as}
in \color{black} Fig.~\ref{fig18} in the main text. \color{black} It is seen that we get an excellent description of all scattering observables
up to the energy of 50~MeV if we use a large enough set of SRFs
${N=11\,\{0,1,2,3,4,5,6,7,8\}}$.

{\clg
In Fig.~\ref{fig:S19} the absolute magnitudes of the wave function 
expansion coefficients~$d_{n,ij}(E)$~are shown
 at the energy of~$E=0.5$~MeV where we obtain~$\delta_{0}$, $\delta_{2}$, and~$\epsilon$
close to exact value in calculations  with 
different~$N$,  $N_{\textrm{max}}=12$, $\hbar \Omega=30$~MeV, and smoothing 
parameter~$a=2.5$  using the eigenfunction SRFs and $^3D_{1}$ as the first  channel 
with~$i=1$.  We observe a fast convergence of the amplitudes even with small~$N$ values.  The arguments of the coefficients converge in a similar manner as seen in Fig.~\ref{fig:S20}.

\section{\protect\clg\boldmath$
{S}$-matrix poles
}
\label{SectionIV}
Figure~\ref{fig:S21}~shows the convergence  {\clg with~$N_{\textrm{max}}$ of the resonance energy~$E_{r}$ and 
width}~$\Gamma$ of $n \alpha$ scattering using the Woods--Saxon interaction with parameters fitted in Ref. \cite{WoodsSaxon1979}.  The results are
obtained by  the HORSE method.  We {\clg see that at}~$N_{\textrm{max}}=10$, we obtain results very close to the exact values.

\begin{figure}[t!]
\subfloat[]{%
  \includegraphics[width=1\columnwidth]{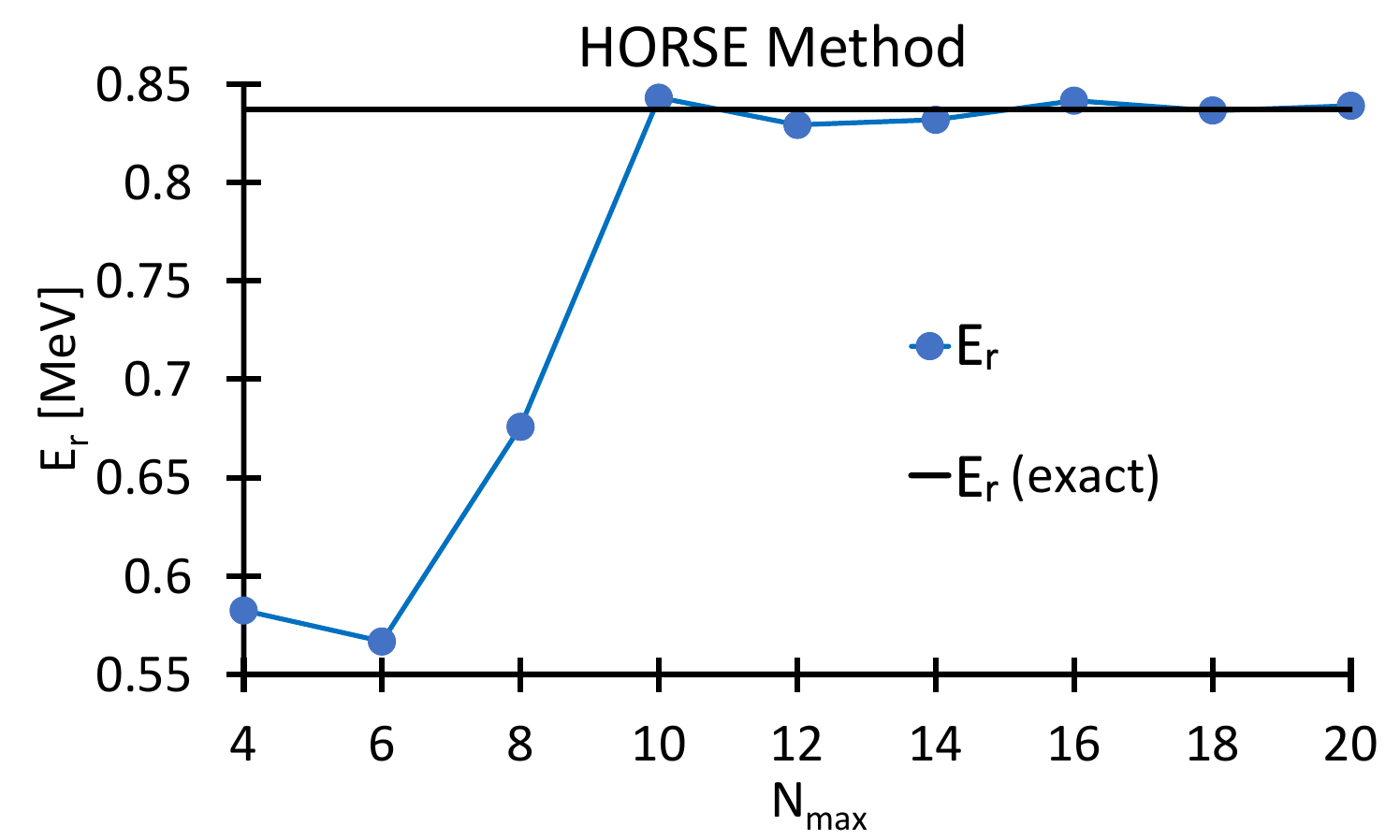}%
}\hfill
\subfloat[]{%
  \includegraphics[width=1\columnwidth]{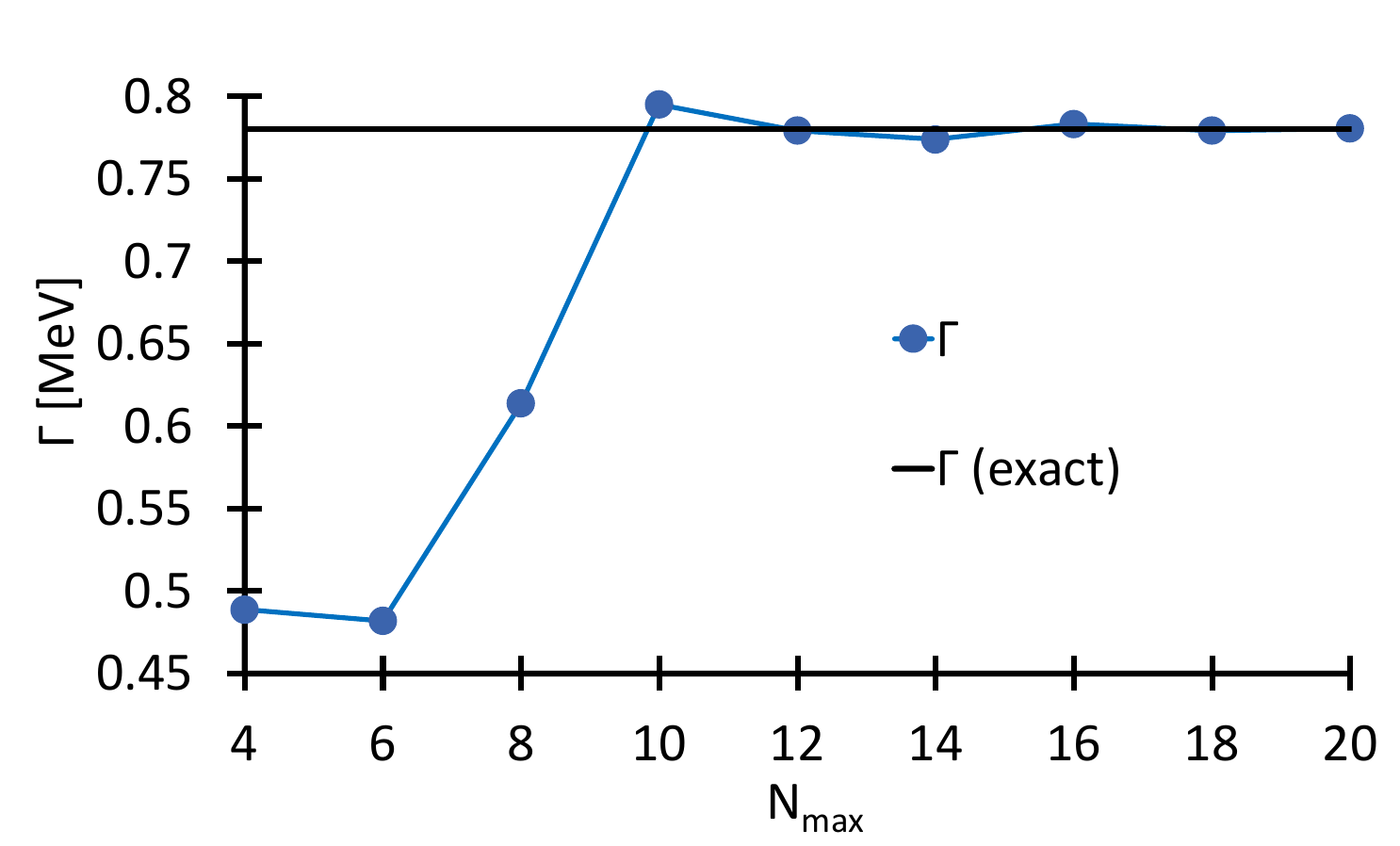}%
}\hfill
\vspace{-1ex}
\caption[]{\clg $\frac32^{-}$ resonance in~$n\alpha$ scattering supported by the 
Woods--Saxon potential: a) resonance energy~$E_{r}$ and b)~resonance width~$\Gamma$
 obtained by locating the $S$-matrix pole within the
HORSE method  using various $N_{\max}$ truncations, 
 $\hbar \Omega=30$~MeV, {\clg and} 
 smoothing 
 parameter~$a=5$. 
}
 \label{fig:S21}
\end{figure}

Figure~\ref{fig:S22}~shows \color{black} the convergence of the deuteron ground state energy~$E_{b}$ 
with~$N_{\textrm{max}}$~within the HORSE method. 
The deuteron ground state energy is seen to converge much
faster to the exact value than the variational results obtained by
  the direct diagonalization of the truncated Hamiltonian~$H^{tr}$.
Approximately~$N_{\textrm{max}}=50$ is needed to achieve three-digit accuracy for direct diagonalization as compared to~${N_{\textrm{max}}=14}$~needed for the same accuracy for the~$S$-matrix pole location via the HORSE method.

In \color{black} Fig.~\ref{fig:S23}, \color{black} we present $\breve{d}_{n,i}(E_{b})$ \color{black} of the deuteron wave function obtained by \color{black} the adapted \color{black} Efros method with different~$N$~using eigenfunction SRFs, $N_{\max}=12$, $\hbar \Omega=30$~MeV, and smoothing parameter~$a=5$.  We see \color{black} a high accuracy of expansion coefficients calculated even with small~$N$.

\begin{figure}[t!]
{%
  \includegraphics[width=1\columnwidth]{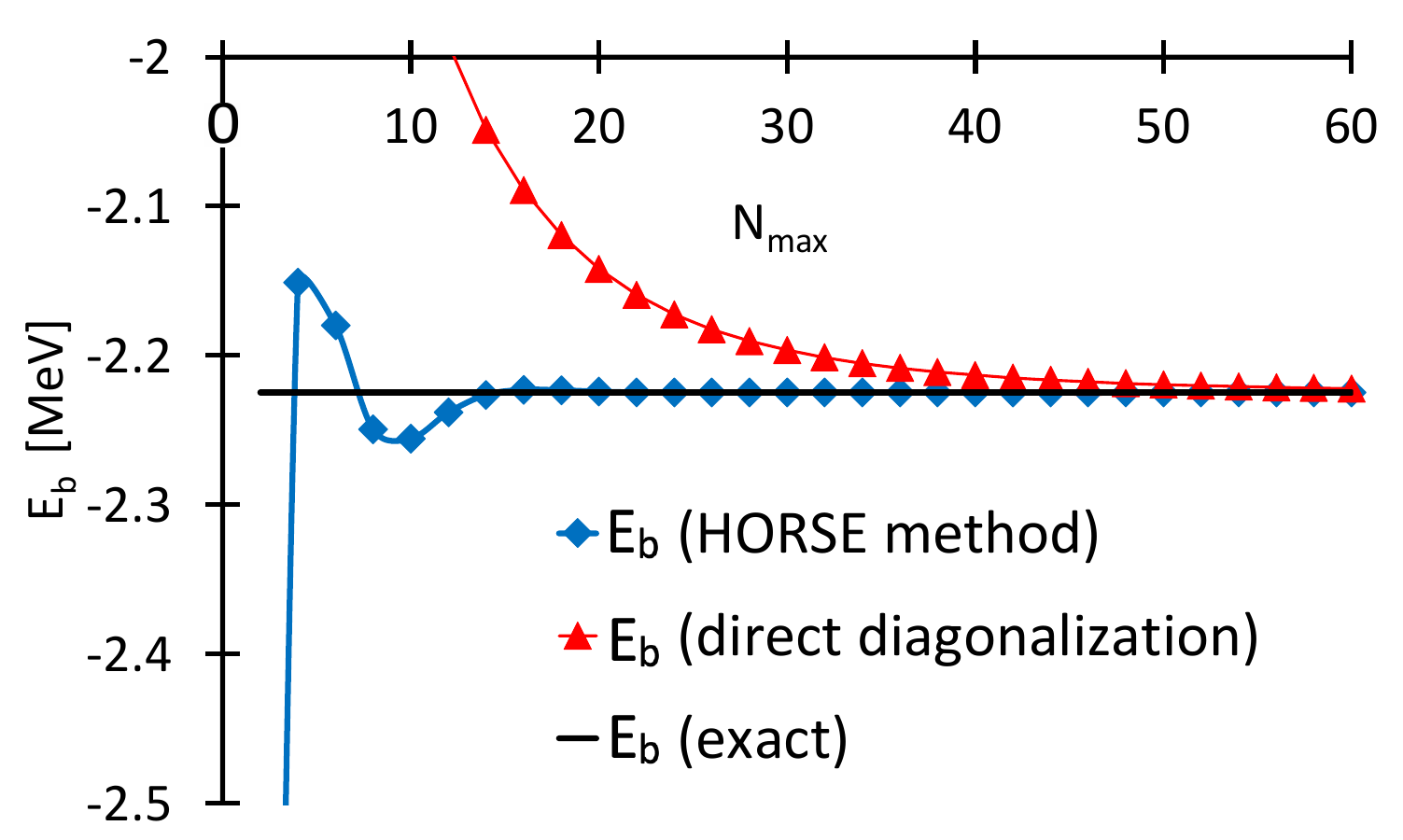}%
}
\caption[]{\clg
Deuteron bound state energy~$E_{b}$ obtained with Daejeon16 $NN$ interaction
using various~$N_{\textrm{max}}$, 
$\hbar \Omega=30$~MeV, and 
smoothing parameter~
$a=5$ by the HORSE method
and by the 
direct diagonalization of~$H^{tr}$.
}
 \label{fig:S22}
\end{figure}

\begin{figure}[t!]
\subfloat[]{%
  \includegraphics[width=1\columnwidth]{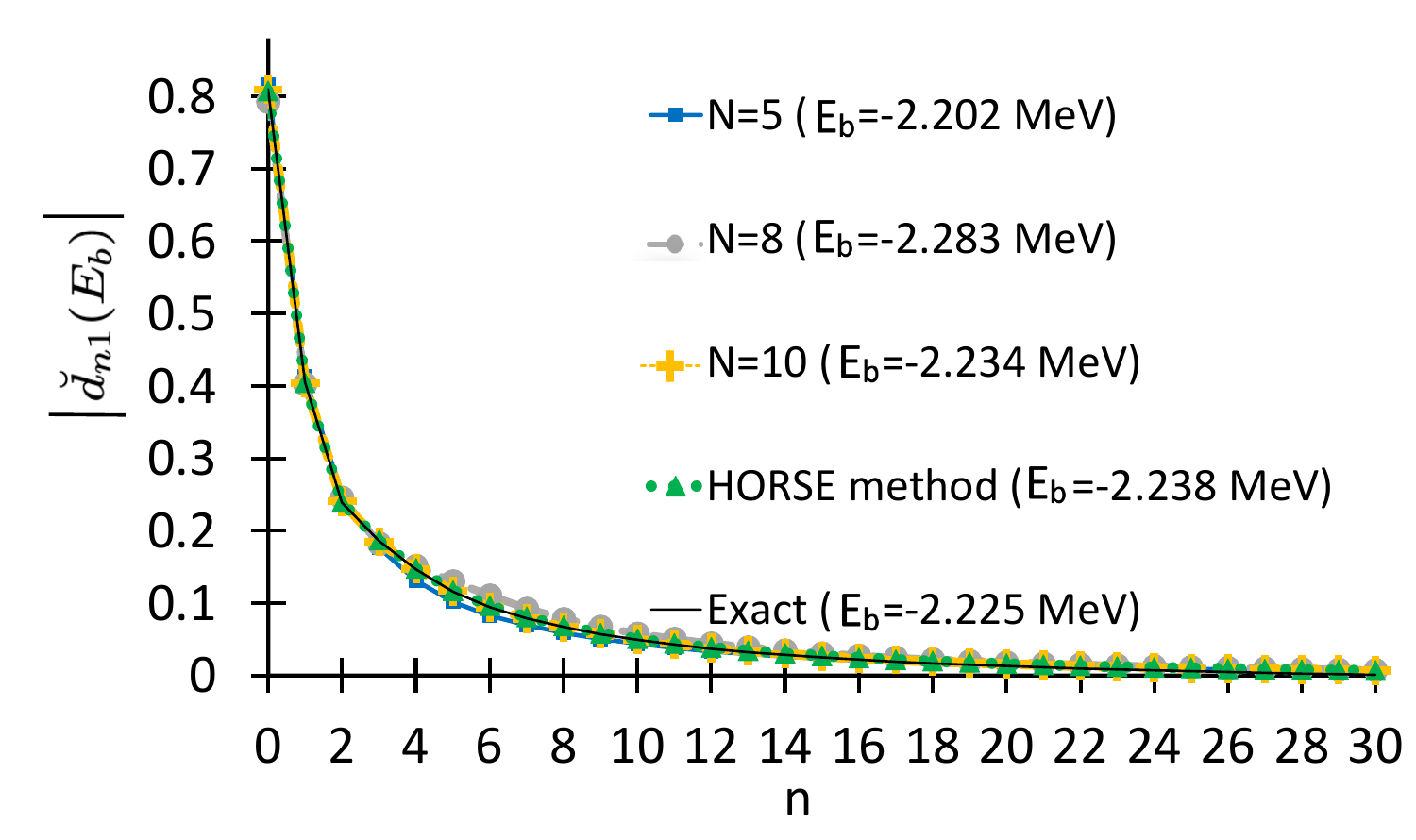}%
}\hfill
\subfloat[]{%
  \includegraphics[width=1\columnwidth]{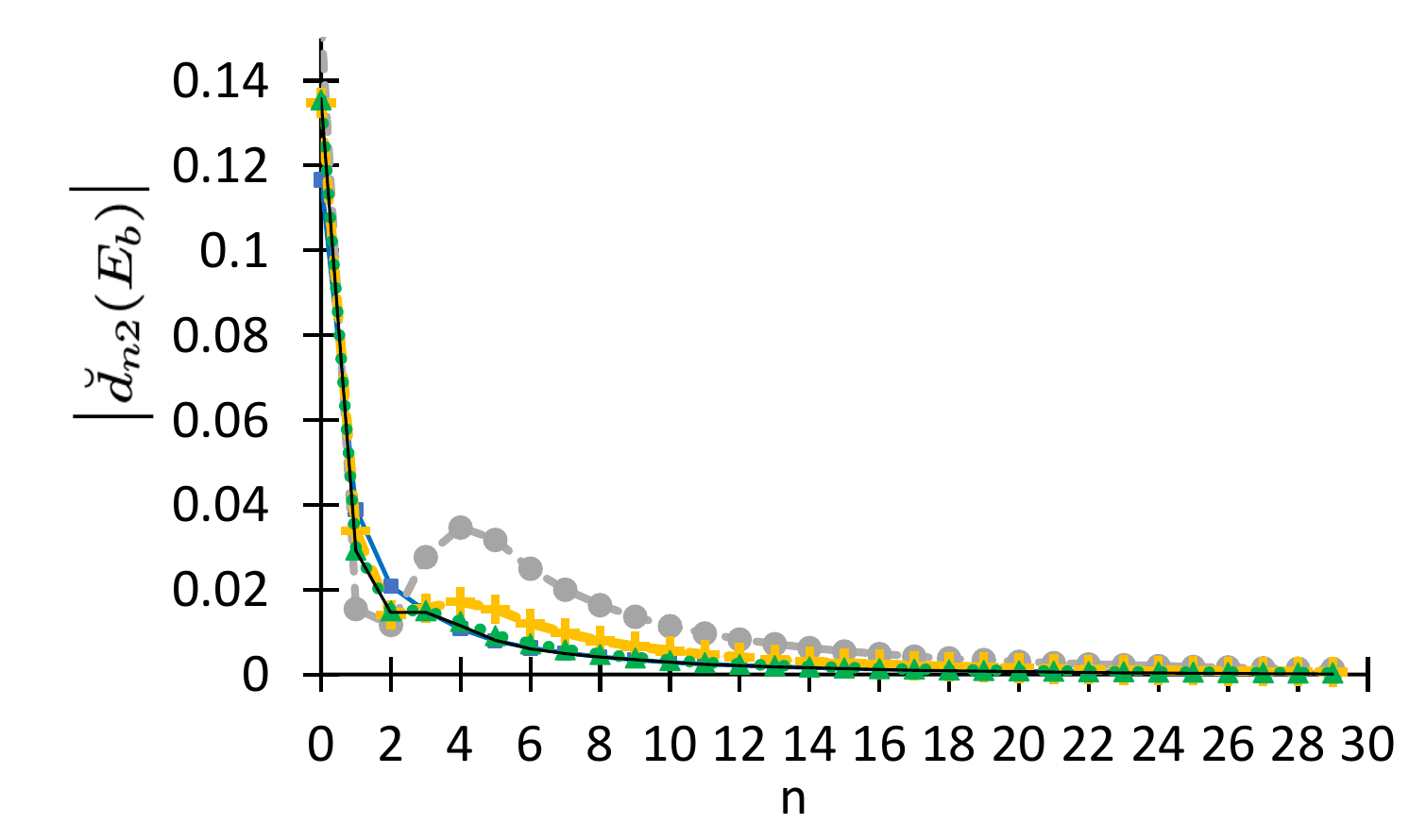}%
}
\caption[]{Absolute magnitudes of expansion coefficients~$|\breve{d}_{n,i}(E_{b})|$
of the deuteron wave function obtained by locating the $S$-matrix pole within \color{black} the adapted \color{black}
Efros method with different~$N$ using eigenfunction SRFs, 
 $N_{\max}=12$, \color{black} $\hbar \Omega=30$~MeV, and smoothing 
parameter~$a=5$: a)~$^3S_{1}$~components, b)~$^3D_{1}$~components.
The deuteron energy obtained with each~$N$ is given in the legend in parentheses.  
We present also for
comparison the exact values of~$|\breve{d}_{n,i}(E_{b})|$~($E_{b}=-2.225$~MeV) \color{black} and those values obtained by the HORSE method.  Symbols are often obscured by other symbols especially in panel (a).\color{black}
}
 \label{fig:S23}
\end{figure}

\begin{figure}[b!]
\subfloat[]{%
  \includegraphics[width=1\columnwidth]{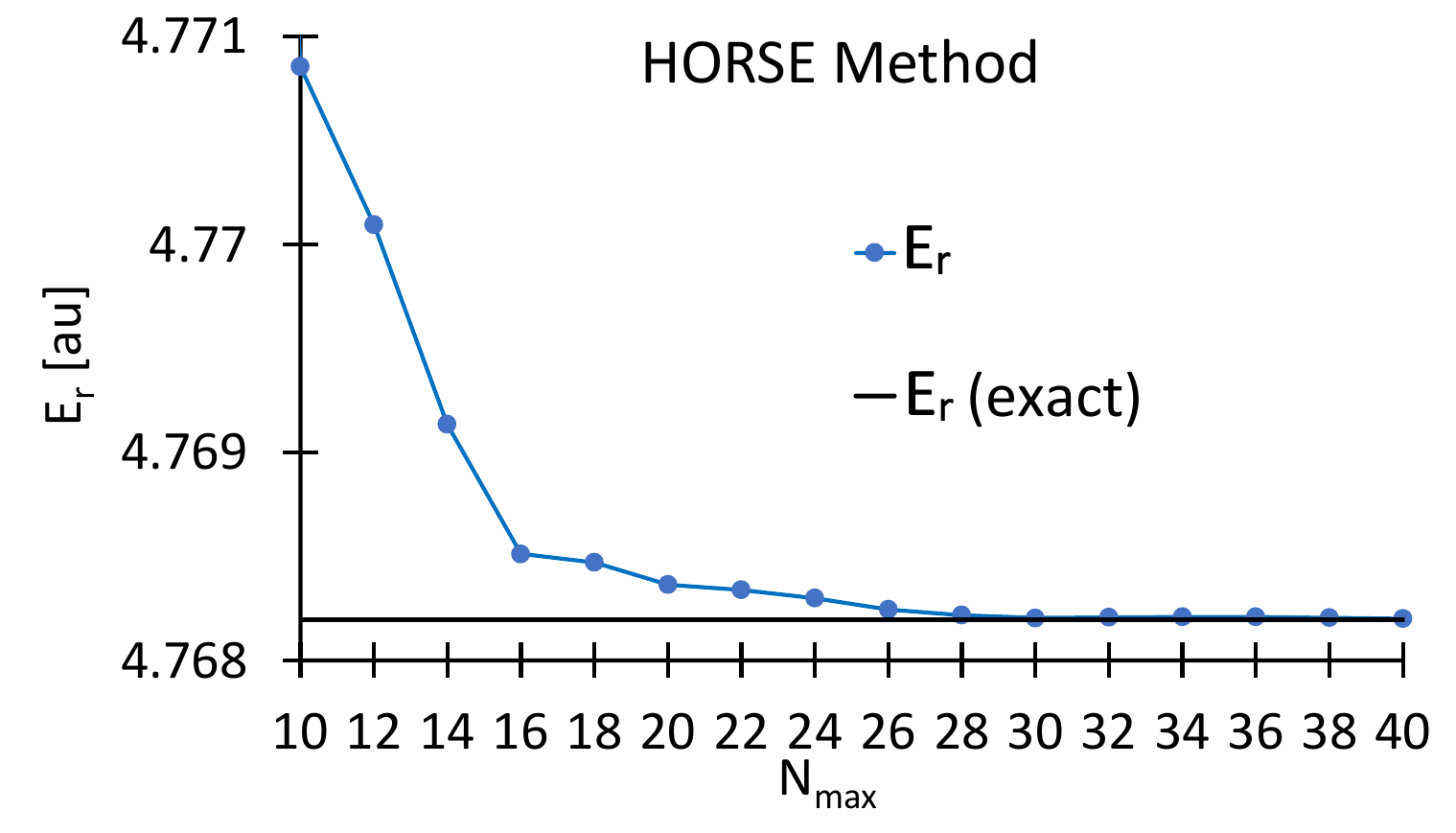}%
}\hfill
\subfloat[]{%
  \includegraphics[width=1\columnwidth]{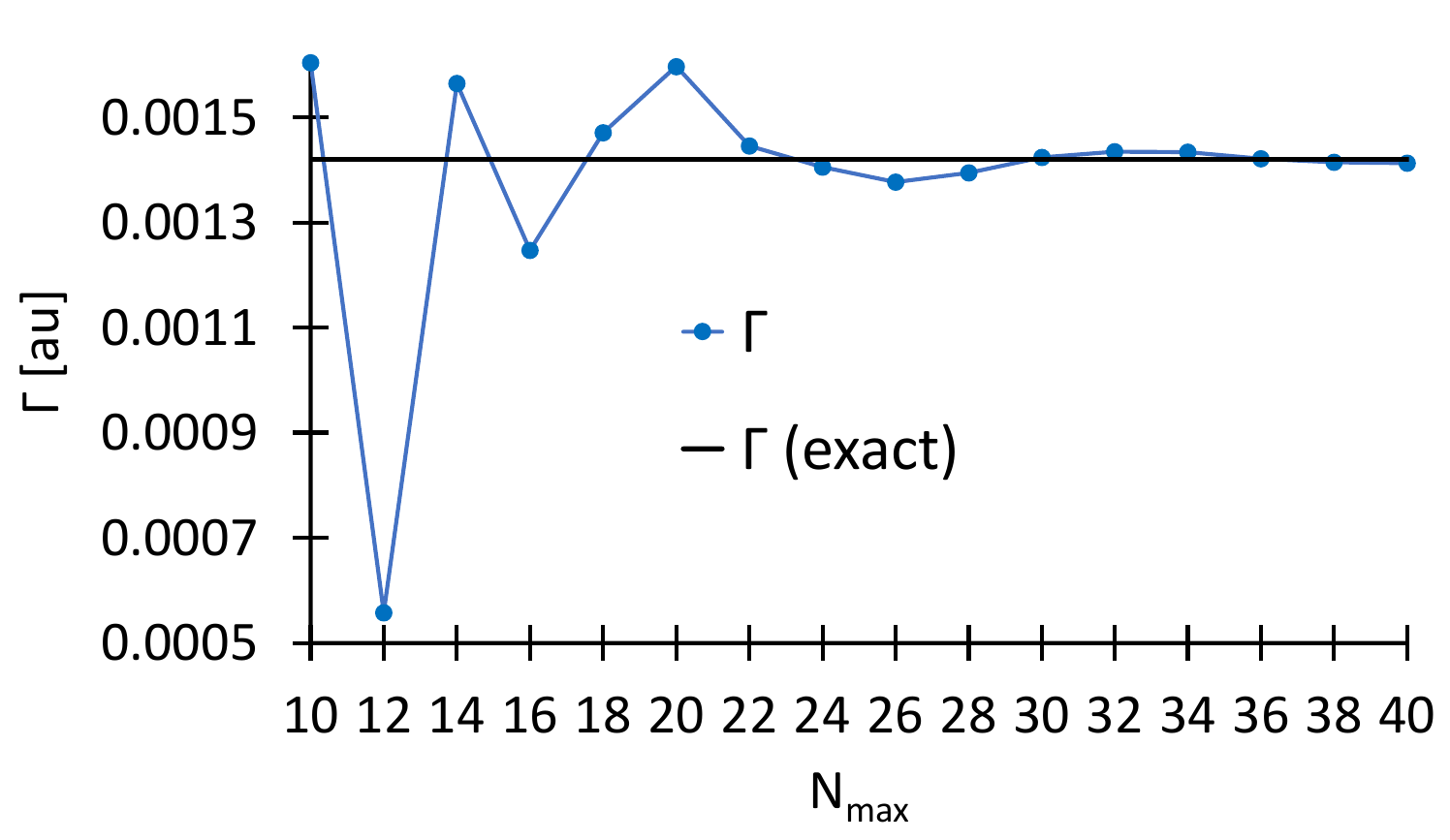}%
}
\caption[]{\clg a)~Energy~$E_{r}$ and b)~width~$\Gamma$ of the 
resonance in the two-channel Noro--Taylor model (see Ref. \cite{Norro1980}) in HORSE calculations 
with~$\Omega=1.5$~au, smoothing parameter~$a=5$, and various~$N_{\max}$.
}
 \label{fig:S24}
\end{figure}

Finally, we consider 
{ a resonance in the Noro--Taylor two-channel}
model originally presented in Ref. \cite{Norro1980} {which
includes two~$s$-wave channels ($l_{1}=l_{2}=0$) and the potential
\renewcommand{\theequation}{S\arabic{equation}} 
\setcounter{equation}{9} 

\begin{figure}[t!]
\subfloat[]{%
  \includegraphics[width=\columnwidth]{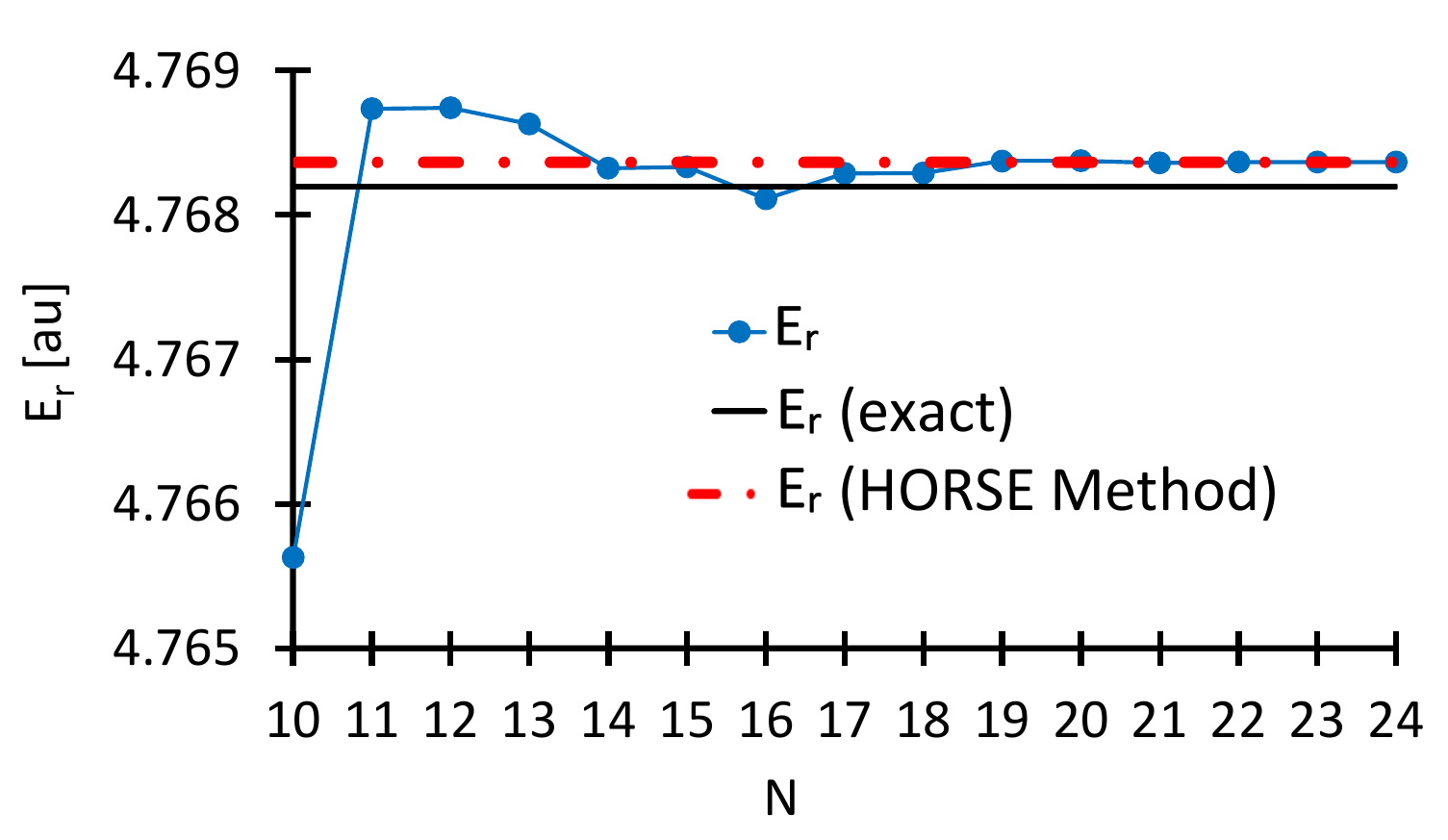}%
}\hfill
\subfloat[]{%
  \includegraphics[width=\columnwidth]{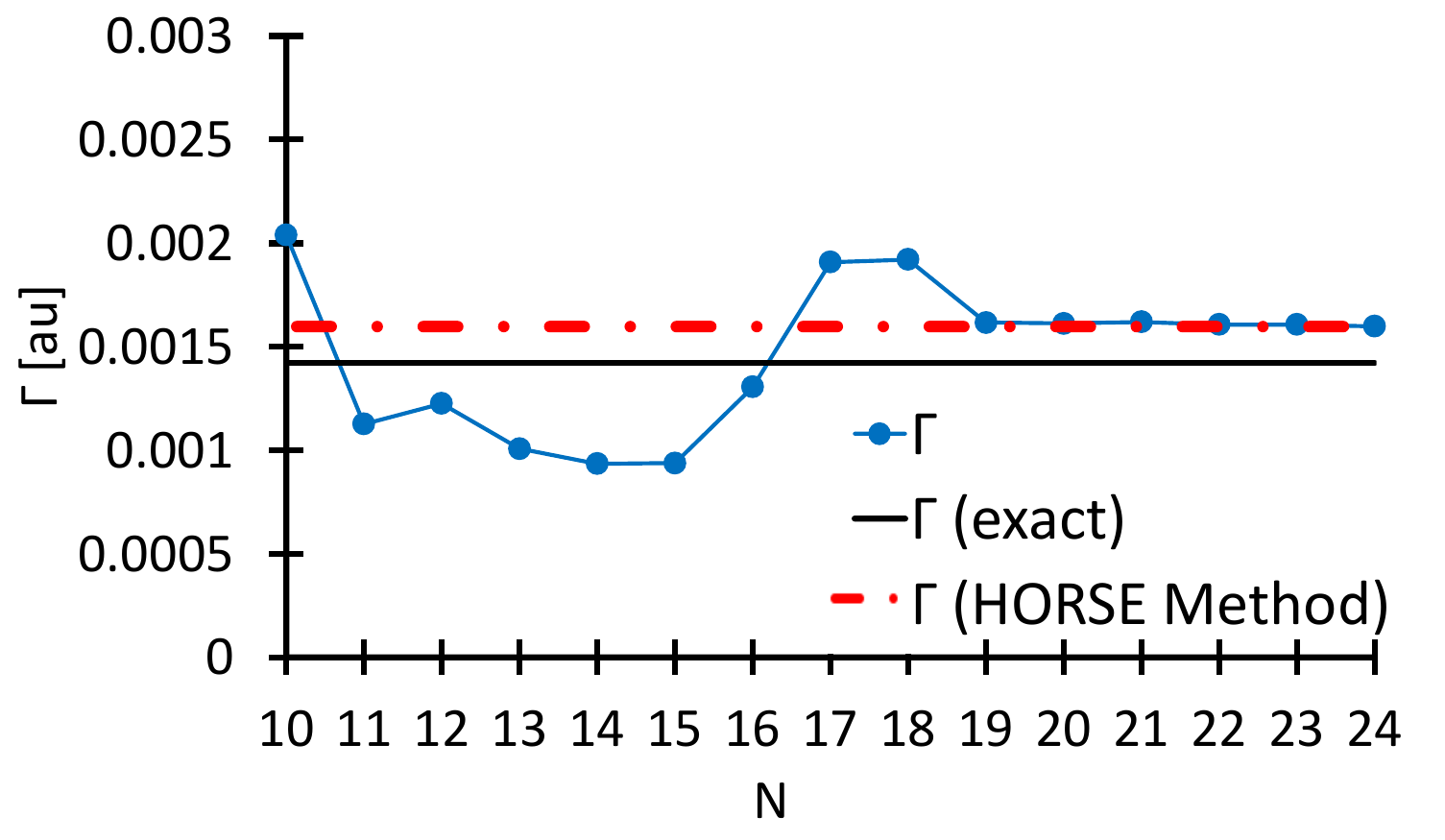}%
}
\caption[]{\clg a)~Energy~$E_{r}$ and b)~width~$\Gamma$ of the 
resonance in the two-channel Noro--Taylor model (see Ref. \cite{Moiseyev1990}) in calculations 
within \color{black} the adapted \color{black} Efros method with eigenfunction SRFs, $\Omega=1.5$~au, { $N_{\max}=20$}, 
smoothing parameter~$a=5$, and various~$N$~as compared with the exact results and with the HORSE method results. \color{black}
}
 \label{fig:S25}
\end{figure}

\begin{subequations}\label{Noro-Taylor}
\renewcommand{\theequation}{S\arabic{equation}\alph{equation}} 
\begin{gather}
\tag{S10a}\label{Noro-Taylor-V}
V_{ij} = \lambda_{ij} r^2 e^{-r} + E_{i}^{\textrm{thresh.}} \delta_{ij},
\end{gather}
\begin{gather}
\tag{S10b}\label{Noro-Taylor-lam}
\lambda_{11}=1,\quad \lambda_{12}=\lambda_{21}=-7.5,\quad \lambda_{22}=7.5,
\end{gather}
\begin{gather}
\tag{S10c}\label{Noro-Taylor-E}
E_{1}^{\textrm{thresh.}}=0,\quad E_{2}^{\textrm{thresh.}}=0.1.
\end{gather}
\end{subequations}
The atomic units are used here.} \color{black} The interaction~{\eqref {Noro-Taylor}} \color{black}
has a rich spectrum containing four bound states and { a number of} 
resonances, and is therefore a good test problem for the numerical validity of a particular approach.  We explore the lowest-lying narrow resonance in this system at the 
energy ${E_{r}=4.7682}$~au with the width~$\Gamma=0.00142$~au.

\color{black} Figure~\ref{fig:S24} \color{black} shows the convergence with $N_{\textrm{max}}$~of the resonant energy~$E_{r}$ and width~$\Gamma$~in the Noro--Taylor two-channel
model obtained by the HORSE method.  $N_{\textrm{max}}=10$  is sufficient for obtaining~$E_{r}$ { with an accuracy
of~1\%.} 
For { the width}~$\Gamma$, approximately~$N_{\textrm{max}}=50$~is needed to achieve a similar level of accuracy. {Note, however, that the resonance is very narrow and
the corresponding $\Gamma$
is  small; thus the accuracy of~1\% for the width means  fixing it within a very small interval 
of absolute values. For both of these observables, }we see a good level of stability beyond the truncation of~{$N_{\max}=22$.}

 The~results~of~calculations~of~$E_{r}$ and~$\Gamma$ within \color{black} the adapted \color{black} Efros method
with eigenfunction SRFs,~$\Omega=1.5$ au, ${N_{\max}=20}$, and smoothing parameter~$a=5$ at various~$N$ are presented in
\color{black} Fig.~\ref{fig:S25} \color{black} and are 
compared with the HORSE method result. 
 We see that { the}
level of accuracy within better than {0.1\%  of the exact result  is} 
achieved for~$E_{r}$ { at~$N=10$}.  
{For the width~$\Gamma$, a relative accuracy of better than 15\%
is achieved at around~$N=19$.}
{ We note that both the resonance energy~$E_{r}$ and its width~$\Gamma$ converge with~$N$ to the HORSE results which require larger~$N_{\max}$ values (see \color{black} Fig.~\ref{fig:S24} \color{black})
to reproduce accurately the width. With $N=19$ we exactly reproduce the HORSE width; thus one
needs to perform calculations
with larger~$N_{\max}$ to describe the~$\Gamma$ value with higher accuracy.  
}

\end{document}